\documentclass[12pt,eqno,epsf]{article}
\usepackage{amsmath,amssymb,graphicx,slashbox}
\numberwithin{equation}{section}

\def\mydate{January 8, 2014}
\def\ignore#1{{}}

\newcounter{sxn}

\newcounter{axn}

\date{}

\newdimen\mybaselineskip
\renewcommand{\thefootnote}{\arabic{footnote}}

\newcommand{\beeq}{\begin{equation}}
\newcommand{\eneq}{\end{equation}}
\newcommand{\beqn}{\begin{eqnarray}}
\newcommand{\eeqn}{\end{eqnarray}}

\newcommand{\alp}{\alpha}
\newcommand{\bt}{\beta}
\newcommand{\gm}{\gamma}
\newcommand{\Gm}{\Gamma}
\newcommand{\dlt}{\delta}
\newcommand{\Dlt}{\Delta}
\newcommand{\ep}{\epsilon}
\newcommand{\tht}{\theta}

\newcommand{\vth}{\vartheta}

\newcommand{\lmd}{\lambda}
\newcommand{\Lmd}{\Lambda}
\newcommand{\sgm}{\sigma}
\newcommand{\Sgm}{\Sigma}
\newcommand{\Ups}{\Upsilon}
\newcommand{\vph}{\varphi}
\newcommand{\omg}{\omega}
\newcommand{\Omg}{\Omega}
\newcommand{\dalp}{\dot{\alpha}}
\newcommand{\dbt}{\dot{\beta}}

\newcommand{\be}{\begin{equation}}
\newcommand{\ee}{\end{equation}}
\newcommand{\bea}{\begin{eqnarray}}
\newcommand{\eea}{\end{eqnarray}}
\newcommand{\eql}{\!\!\!&=\!\!\!&}

\newcommand{\defa}{\!\!\!&\equiv\!\!\!&}

\newcommand{\toa}{\!\!\!&\to\!\!\!&}

\newcommand{\tl}[1]{\tilde{#1}}
\newcommand{\bdm}[1]{{\mbox{\boldmath $#1$}}}
\newcommand{\sbdm}[1]{\mbox{\scriptsize \boldmath $#1$}}
\newcommand{\tr}{{\rm tr}\,}
\newcommand{\str}{{\rm str}}
\newcommand{\diag}{{\rm diag}}
\newcommand{\der}{\partial}
\newcommand{\dr}{\!\!d}
\newcommand{\hc}{{\rm h.c.}}
\newcommand{\ie}{{i.e.}}
\newcommand{\id}{\mbox{\boldmath $1$}}

\newcommand{\vev}[1]{\langle #1 \rangle}
\newcommand{\Lvev}[1]{\left\langle #1 \right\rangle}
\newcommand{\brkt}[1]{\left( #1 \right)}
\newcommand{\brc}[1]{\left\{ #1 \right\}}
\newcommand{\sbk}[1]{\left[ #1 \right]}
\newcommand{\abs}[1]{\left| #1 \right|}

\renewcommand{\Re}{{\rm Re}\,}
\renewcommand{\Im}{{\rm Im}\,}

\newcommand{\cA}{{\cal A}}
\newcommand{\cB}{{\cal B}}
\newcommand{\cC}{{\cal C}}
\newcommand{\cD}{{\cal D}}

\newcommand{\cF}{{\cal F}}
\newcommand{\cG}{{\cal G}}
\newcommand{\cH}{{\cal H}}

\newcommand{\cK}{{\cal M}}
\newcommand{\cL}{{\cal L}}
\newcommand{\cM}{{\cal K}}
\newcommand{\cN}{{\cal N}}
\newcommand{\cO}{{\cal O}}
\newcommand{\cP}{{\cal P}}
\newcommand{\cQ}{{\cal Q}}

\newcommand{\cT}{{\cal T}}
\newcommand{\cV}{{\cal V}}
\newcommand{\cU}{{\cal U}}
\newcommand{\cW}{{\cal W}}
\newcommand{\cX}{{\cal X}}
\newcommand{\cY}{{\cal Y}}
\newcommand{\cZ}{{\cal Z}} 
\newcommand{\bE}{{\mathbb E}}

\newcommand{\bP}{{\mathbb P}}

\newcommand{\dscp}{\delta_{\rm sc}^{(1)}}
\newcommand{\dscq}{\delta_{\rm sc}^{(2)}}
\newcommand{\dgt}{\delta_{\rm sg}}

\newcommand{\Io}{I_{\rm o}}
\newcommand{\Ie}{I_{\rm e}}
\newcommand{\Jo}{J_{\rm o}}
\newcommand{\Je}{J_{\rm e}}
\newcommand{\Ko}{K_{\rm o}}
\newcommand{\Ke}{K_{\rm e}}
\newcommand{\Dlp}{\Delta^{\rm 1loop}}
\newcommand{\Omglp}{\Omg_{\rm eff}^{\rm 1loop}}
\newcommand{\Tr}{{\rm Tr}}
\newcommand{\Istr}{{\rm Istr}}
\newcommand{\avh}{r_s}

\begin{document}
\thispagestyle{empty}

\baselineskip=12pt

{\small \noindent \mydate    
\hfill }

{\small \noindent \hfill  KEK-TH-1609}

\baselineskip=35pt plus 1pt minus 1pt

\vskip 1.5cm

\begin{center}
{\LARGE\bf One-loop K\"ahler potential in 5D gauged} \\
{\LARGE\bf supergravity with generic prepotential}

\vspace{1.5cm}
\baselineskip=20pt plus 1pt minus 1pt

\normalsize

{\large\bf Yutaka\ Sakamura}$\!${\def\thefootnote{\fnsymbol{footnote}}
\footnote[1]{\tt e-mail address: sakamura@post.kek.jp}}

\vspace{.3cm}
{\small \it KEK Theory Center, Institute of Particle and Nuclear Studies, 
KEK, \\ Tsukuba, Ibaraki 305-0801, Japan} \\ \vspace{3mm}
{\small \it Department of Particles and Nuclear Physics, \\
The Graduate University for Advanced Studies (Sokendai), \\
Tsukuba, Ibaraki 305-0801, Japan} 
\end{center}

\vskip 1.0cm
\baselineskip=20pt plus 1pt minus 1pt

\begin{abstract}
We calculate one-loop contributions to the K\"ahler potential 
in 4D effective theory of 5D gauged supergravity (SUGRA) 
on $S^1/Z_2$ 
with {\it a generic form of the prepotential} and arbitrary 
boundary terms. 
Our result is applicable to a wide class of 5D SUGRA models. 
The derivation is systematically performed by means of 
an $N=1$ superfield formalism 
based on the superconformal formulation of 5D SUGRA. 
As an illustrative example, 
we provide an explicit expression of the K\"ahler potential 
in the case of 5D flat spacetime. 
\end{abstract}


\newpage

\section{Introduction}
Higher-dimensional supergravities (SUGRA) have been attracted much attention 
and extensively studied in various aspects, such as 
the model building in the context of the brane-world scenario, 
effective theories of the superstring theory or M-theory, 
AdS/CFT correspondence, etc. 
Among them, five-dimensional (5D) SUGRA compactified on an orbifold~$S^1/Z_2$ 
has been thoroughly investigated since 
it is the simplest setup for supersymmetric (SUSY) brane-world models, and 
it is shown to appear as an effective theory of the strongly coupled heterotic 
string theory~\cite{Horava-Witten} 
compactified on a Calabi-Yau 3-fold~\cite{Lukas:1998}. 
Besides, SUSY extensions of the Randall-Sundrum model~\cite{Randall:1999ee}
are also constructed in 5D SUGRA 
on $S^1/Z_2$~\cite{Gherghetta:2000qt,Falkowski:2000er,Altendorfer:2000rr}. 

Models with an extra dimension can easily realize the large hierarchy 
between the electroweak and the Planck scales or 
among the fermion masses in the standard model. 
The former is obtained by the warped geometry 
along the extra dimension~\cite{Randall:1999ee}, 
and the latter is by the wave function localization of matter fields 
in the extra dimension~\cite{ArkaniHamed:1999dc,Kaplan:2000av}. 
In both mechanisms, some mass scales have to be introduced in the 5D bulk. 
The warped geometry is induced by the 5D cosmological constant, and 
the wave function profiles are controlled by 5D masses of the matters. 
In SUGRA context, these mass scales are introduced by gauging some isometries 
with some 5D vector multiplets. 
Namely, we have to consider the gauged SUGRA. 
When the extra dimension is compactified on $S^1/Z_2$, 
the four-dimensional (4D) vector components in such vector multiplets 
must be $Z_2$-odd. 
Every 5D SUGRA model has this type of vector field, 
\ie, the graviphoton.\footnote{
In this paper, the terminology ``graviphoton" denotes 
a vector field in the gravitational multiplet 
of the on-shell formulation. 
It should be distinguished from the off-diagonal components of the 5D metric. }
Therefore, most models based on 5D gauged SUGRA assume that 
the vector multiplet that gauges the isometries to induce the mass scales 
is the graviphoton multiplet. 
However this is not the only possibilities. 
There can be other vector multiplets whose 4D vector components are $Z_2$-odd.  
The 5D mass scales can also be obtained by gauging with these multiplets. 
Such 5D vector multiplets contain $Z_2$-even real scalar fields. 
These scalar fields have 4D zero-modes, and do not have any potential terms
at least at tree level. 
Thus we refer to them as moduli in this paper.\footnote{
These moduli are actually identified with the shape moduli of 
the compactified space for a 5D effective theory of the heterotic 
M-theory on the Calabi-Yau manifold~\cite{Lukas:1998}, for example. } 
In fact, one linear combination of these moduli corresponds to 
the size modulus of the fifth dimension, \ie, the radion, 
which belongs to the same 5D supermultiplet as the graviphoton.  

In the case that a model has more than one moduli, they generically mix with 
each other. 
Such mixing is characterized by a cubic polynomial, which is referred to as 
the norm function in this paper. 
This corresponds to the prepotential in 4D $N=2$ SUSY gauge theories. 
As mentioned above, most models based on 5D SUGRA implicitly assumed 
a special form of the norm function such that the radion does not mix 
with the other moduli. 
In our previous works~\cite{Abe:2008an,Abe:2011rg}, 
we derived 4D effective theory of 5D SUGRA with more than one moduli at tree level, 
and found that some terms appear in the K\"ahler potential, 
which do not exist in the single modulus case. 
We also showed those terms can significantly affect the flavor structure 
of the effective theory when the fermion mass hierarchy is realized 
by the wave function localization, and pointed out a possibility 
that the SUSY flavor problem is avoided. 
This indicates an importance of considering 
arbitrary form of the norm function with multi moduli 
when we construct a realistic model based on 5D SUGRA. 

\ignore{
In order to construct a phenomenologically viable model with the extra 
dimension, we have to stabilize the radion to some finite value. 
Such modulus stabilization can occur at tree level 
due to a vacuum configuration of a bulk scalar field that has nontrivial 
potentials at the orbifold boundaries~\cite{Goldberger:1999uk}. 
This mechanism is simple, but an extra bulk scalar field has to be 
introduced just for the modulus stabilization. 
The modulus can also be stabilized by the vacuum energy. 
This mechanism has an advantage that it does not require 
any additional fields for the stabilization, 
in contrast to the above mechanism. 
}

For a construction of realistic 5D SUGRA models, 
mediation of SUSY-breaking effects to our observable sector 
and stabilization of the radion to some finite value are 
indispensable issues. 
In some of the mechanisms for them, 
one-loop quantum corrections to the K\"ahler potential 
in 4D effective theory are relevant. 
For example, SUSY breaking at one of the boundaries of $S^1/Z_2$ 
can be transmitted to the other boundary where we live 
by the quantum loop effects of the bulk 
fields~\cite{Kaplan:1999ac}-\cite{Rattazzi:2003rj}, 
and the radion can be stabilized by the vacuum energy 
through the Casimir effect~\cite{Fabinger:2000jd}-\cite{Sakamura:2010ju}. 
The soft SUSY-breaking parameters and the radion mass 
are induced from the one-loop K\"ahler potential 
after taking into account the SUSY-breaking effects. 
These contributions are finite in spite of 
the non-renormalizability of 5D SUGRA. 
This is because each relevant loop diagram must touch both boundaries 
and cannot shrink to a point. 
Thus the inverse of the size of the extra dimension provides 
an effective cutoff in the momentum integral. 

The one-loop corrections to the effective K\"ahler potential 
in the context of 5D SUGRA have already been discussed 
in Refs.~\cite{Gherghetta:2001sa,Rattazzi:2003rj,Buchbinder:2003qu,
Gregoire:2004nn,Falkowski:2005fm}. 
However these works assume that the graviphoton multiplet 
(or the radion multiplet) is the only moduli multiplet which is relevant 
to the gauging of the isometries to induce the 5D mass scales. 
As mentioned above, this is only a special case among generic 5D SUGRA. 
Thus we extend the above works to more general class of theories 
in this paper. 
We calculate the one-loop K\"ahler potential for 5D SUGRA on $S^1/Z_2$ 
with {\it an arbitrary form of the norm function}. 
Our derivation is performed in an $N=1$ superfield formalism 
based on the superconformal formulation of 
5D SUGRA~\cite{Kugo:2000af}-\cite{Kugo:2002js}, 
which is developed in our previous works~\cite{Abe:2004ar,Sakamura:2012bj}. 
This makes it possible to deal with general 5D SUGRA 
in a systematic and transparent manner. 
Thus the result is applicable to a wide class of models based on 5D SUGRA. 

The paper is organized as follows. 
In the next section, we briefly review our previous works, 
which provide an $N=1$ superfield description of 5D SUGRA on $S^1/Z_2$ 
with an arbitrary prepotential. 
In Sec.~\ref{1loop_K}, we derive an expression of one-loop contributions 
to the 4D effective K\"ahler potential by means of 
the background field method and the superfield formalism. 
In Sec.~\ref{Flat_case}, we apply the formula obtained in Sec.~\ref{1loop_K} 
in the case that 5D spacetime is flat as an illustrative example. 
Sec.~\ref{summary} is devoted to the summary. 
In Appendix~\ref{trf:SC}, we list the 5D superconformal transformation laws 
in terms of the $N=1$ superfields. 
In Appendix~\ref{superspinPi}, we collect the definitions of useful 
projection operators in the $N=1$ superspace and their properties. 
In Appendix~\ref{tree:action}, we review the derivation of 
the effective K\"ahler potential at tree level. 
We show some detailed calculations to pick up 
quadratic terms for the bulk fluctuation superfields 
in Appendix~\ref{quad_flct}, and to derive the boundary conditions for them 
in Appendix~\ref{BC:flct_md}. 
In Appendix~\ref{BS_component}, we provide an explicit expression 
of the one-loop Lagrangian in a simple case 
in terms of the bosonic components of the superfields.

\section{Superfield description of 5D SUGRA} \label{SF:5DSUGRA} 
In this paper, we consider 5D SUGRA compactified on an orbifold~$S^1/Z_2$. 
We take the fundamental region of $S^1/Z_2$ as $0\leq y\leq L$, 
where $y$ is the coordinate of the extra dimension. 
The most general metric for the background spacetime 
that has the 4D Poincar\'e symmetry has a form of 
\be
 ds^2 = e^{2\sgm(y)}\eta_{\mu\nu}dx^\mu dx^\nu-\Lvev{e_y^{\;\;4}}^2dy^2, 
 \label{warp:metric}
\ee
where $\eta_{\mu\nu}=\diag(1,-1,-1,-1)$, 
$e^{\sgm(y)}$ is the warp factor, which is determined 
by solving 5D Einstein equation, and $\Lvev{e_y^{\;\;4}}$ is the background value 
of the component of the f\"unfbein~$e_y^{\;\;4}$.\footnote{
We can always choose the coordinate~$y$ so that $\Lvev{e_y^{\;\;4}}=1$, 
but we leave it to be an arbitrary positive value in this paper. }
Notice that we can always absorb the warp factor in (\ref{warp:metric}) 
by making use of the dilatation symmetry. 
In fact, the warp factor does not appear explicitly in our calculations 
since our formalism keeps the superconformal symmetries manifest. 
The information of the warped geometry is encoded 
in the gauging for the compensator hypermultiplets~\cite{Abe:2007zv}. 

In this section, we review our previous works~\cite{Abe:2004ar,Sakamura:2012bj} 
that complete an $N=1$ superfield description of 5D SUGRA on $S^1/Z_2$ 
(see also Refs.~\cite{Paccetti:2004ri}-\cite{Kuzenko:2008wr}).  
Our superfield description is based on the superconformal formulation 
developed in Refs.~\cite{Kugo:2000af}-\cite{Kugo:2002js}, 
and is considered as an extension of Ref.~\cite{Linch:2002wg} 
to a generic system of vector multiplets and hypermultiplets.

\subsection{Decomposition into $N=1$ superfields}
The 5D superconformal transformations are divided into 
two parts~$\dscp$ and $\dscq$, where $\dscp$ forms an $N=1$ subalgebra, 
and $\dscq$ is the rest part. 
As shown in Ref.~\cite{Kugo:2002js}, 
each 5D superconformal multiplet 
can be decomposed into $N=1$ superconformal multiplets, 
which only respect $\dscp$ manifestly. 
We have explicitly shown in Ref.~\cite{Sakamura:2011df} 
how each $N=1$ superconformal multiplet is expressed 
by an $N=1$ superfield with the aid of the fields in 
the gravitational multiplet. 
We will consider the following three types of 5D superconformal multiplets 
in this paper.\footnote{
We do not consider the tensor multiplets, which are discussed 
in Ref.~\cite{Gunaydin:1999zx,Kugo:2002vc}, for simplicity. }

\begin{description}
\item[Hypermultiplet] \mbox{}\\
A hypermultiplet~$\mathbb H^a$ $(a=1,2,\cdots,n_C+n_H)$ 
is decomposed into two chiral superfields~$(\Phi^{2a-1},\Phi^{2a})$, 
which have opposite $Z_2$-parities. 
We can always label the chiral superfields so that they have 
the $Z_2$-parities listed in Table~\ref{Z2_parity}. 
The hypermultiplets are divided into two classes. 
One is the compensator multiplets~$a=1,2,\cdots,n_C$ 
and the other is the physical matter multiplets~$a=n_C+1,\cdots,n_C+n_H$. 
The former is auxiliary degrees of freedom and eliminated by 
the superconformal gauge fixing.\footnote{ 
The number of the compensator multiplets~$n_C$ characterizes the hyperscalar manifold.
For example, it is $USp(2,2n_H)/USp(2)\times USp(2n_H)$ 
for $n_C=1$, and $SU(2,n_H)/SU(2)\times SU(n_H)$ for $n_C=2$. } 
The Weyl and the chiral weights of the superfields are also listed 
in Table~\ref{Z2_parity}.\footnote{
The Weyl and the chiral weights are the charges of the dilatation 
and of $U(1)_A\subset SU(2)_U$, respectively.  
These weights of a superfield denote those of the lowest component 
in the superfield. } 

\item[Vector multiplet] \mbox{}\\
A vector multiplet~$\mathbb V^I$ $(I=1,2,\cdots,n_V)$
is decomposed into $N=1$ vector and chiral superfields~$(V^I,\Sgm^I)$, 
which have opposite $Z_2$-parities. 
The vector multiplets are also divided into two classes 
according to their $Z_2$-parities. 
One is a class of the gauge multiplets, 
which are denoted as ${\mathbb V}^{\Ie}$ ($\Ie=1,\cdots,n_{V_{\rm e}}$). 
In this class, $V^{\Ie}$ are $Z_2$-even and have zero-modes 
that are identified with the gauge superfields in 4D effective theory. 
The other is a class of the moduli multiplets,  
which are denoted as ${\mathbb V}^{\Io}$ ($\Io=1,\cdots,n_{V_{\rm o}}$). 
In this class, $V^{\Io}$ are $Z_2$-odd and have no zero-modes. 
Instead, the chiral multiplets~$\Sgm^{\Io}$ have zero-modes~$T^{\Io}$  
whose scalar components do not have any potential terms 
at tree level. 
Thus we refer to $T^{\Io}$ as the moduli superfields in this paper. 
At least one vector multiplet belongs to the latter class. 
In the single modulus case ($n_{V_{\rm o}}=1$), 
the vector component of such a multiplet is identified with the graviphoton. 

\item[Weyl multiplet (Gravitational multiplet)] \mbox{}\\
The 5D Weyl multiplet~${\mathbb E}_W$ is also decomposed into 
six real superfields~$U^\mu$ ($\mu=0,1,2,3$), $U^y$ and $V_E$,\footnote{
The superfield~$U^y$ is related to $U^4$ in Ref.~\cite{Sakamura:2012bj} 
by $U^y=U^4/\vev{V_E}$, where $\vev{V_E}$ is the background value of $V_E$ 
and was assumed to be 1 in Ref.~\cite{Sakamura:2012bj}. 
} and a complex spinor superfield~$\Psi^\alp$, 
which include components of the f\"unfbein, 
$\tl{e}_\mu^{\;\;\underline{\nu}}$, $e_\mu^{\;\;4}$, $e_y^{\;\;4}$, 
and $e_y^{\;\;\underline{\nu}}$, respectively. 
Here, $\tl{e}_\mu^{\;\;\underline{\nu}}\equiv 
e_\mu^{\;\;\underline{\nu}}-\dlt_\mu^{\;\;\nu}$ is the fluctuation mode 
around the background~$\vev{e_\mu^{\;\:\underline{\nu}}}=\dlt_\mu^{\;\;\nu}$. 
Since the Weyl multiplet is the gauge multiplet for 
5D superconformal symmetry, these superfields transform nonlinearly 
under $\dscp$ and $\dscq$ as shown in Appendix~\ref{trf:SC}. 
Hence we cannot assign the Weyl and the chiral weights for them, 
except for $V_E$. 
In fact, $V_E$ transforms under $\dscp$ in a similar way to the vector superfields~$V^I$ 
because its components do not have 4D Lorentz indices. 
\end{description}

\begin{table}[t]
\begin{center}
\begin{tabular}{|c||c|c|c|c|c|c|c|c|c|c|} \hline
\rule[-2mm]{0mm}{7mm} 5D multiplet & \multicolumn{2}{c|}{Hypermultiplet} 
& \multicolumn{4}{c|}{Vector multiplet} & 
\multicolumn{4}{c|}{Weyl multiplet} \\ \hline
$N=1$ superfield & $\Phi^{2a-1}$ & $\Phi^{2a}$ & $V^{\Io}$ & $\Sgm^{\Io}$ & 
$V^{\Ie}$ & $\Sgm^{\Ie}$ & $U^\mu$ & $U^y$ & $V_E$ & $\Psi^\alp$ \\ \hline
$Z_2$-parity & $-$ & $+$ & $-$ & $+$ & $+$ & $-$ & $+$ & $-$ & $+$ & $-$ \\ \hline
Weyl weight & $3/2$ & $3/2$ & 0 & 0 & 0 & 0 &  &  & $-1$ &  \\ \hline
Chiral weight & $3/2$ & $3/2$ & 0 & 0 & 0 & 0 &  &  & 0 &  \\ \hline
\end{tabular}
\end{center}
\caption{The decomposition of 5D superconformal multiplets into $N=1$ superfields. 
The orbifold $Z_2$-parities, the Weyl and the chiral weights of 
the $N=1$ superfields are also shown. }
\label{Z2_parity}
\end{table}

\subsection{5D SUGRA Lagrangian}
5D SUGRA action is determined by 5D superconformal 
transformations~$\dscp$, $\dscq$ and the supergauge 
transformation~$\dgt$~\cite{Sakamura:2012bj}. 
In the following, we keep terms up to linear order 
in the gravitational superfields for each interaction terms. 
Basically we use the two-component spinor notations of Ref.~\cite{Wess:1992cp}, 
except for the metric and the spinor derivatives. 
We take the convention of the 4D metric as $\eta_{\mu\nu}=\diag(1,-1,-1,-1)$ 
so as to match it to that of Ref.~\cite{Kugo:1982cu}, 
and define the spinor derivatives~$D_\alp$ and $\bar{D}_{\dalp}$ as 
\be
 D_\alp \equiv \frac{\der}{\der\tht^\alp}-i\brkt{\sgm^\mu\bar{\tht}}_\alp\der_\mu, 
 \;\;\;\;\;
 \bar{D}_{\dalp} \equiv -\frac{\der}{\der\bar{\tht}^{\dalp}}
 +i\brkt{\tht\sgm^\mu}_{\dalp}\der_\mu, 
\ee
which satisfy $\brc{D_\alp,\bar{D}_{\dalp}}=2i\sgm^\mu_{\alp\dalp}\der_\mu$. 
The spinor derivatives are understood as the left-derivatives. 
It is convenient to define the following differential operators. 
\bea
 \hat{\der}_y \defa \der_y-\brkt{\frac{1}{4}\bar{D}^2\Psi^\alp D_\alp
 +\frac{1}{2}\bar{D}^{\dalp}\Psi^\alp\bar{D}_{\dalp}D_\alp
 +\frac{w+n}{24}\bar{D}^2D^\alp\Psi_\alp+\hc}, \nonumber\\
 \Dlt_\mu \defa \frac{1}{4}\bar{\sgm}_\mu^{\dalp\alp}
 \brkt{D_\alp\bar{D}_{\dalp}-\bar{D}_{\dalp}^RD_\alp^R}, 
\eea
where $w$ and $n$ are the Weyl and the chiral weights of a superfield 
which $\hat{\der}_y$ acts on, and $(w+n)^\dagger=w-n$. 
The spinor derivatives~$D_\alp^R$ and $\bar{D}_{\dalp}^R$ 
are defined by the right-derivatives. 
Then $\Dlt_\mu$ satisfies the Leibniz rule on a product of bosonic superfields. 
On (anti-)chiral superfields, $\Dlt_\mu=-i\der_\mu$ ($\Dlt_\mu=i\der_\mu$). 
It should be noted that , for a chiral superfield~$\Phi$, 
$\der_y\Phi$ is not a chiral superfield in a superconformal sense 
because its $\dscp$-transformation law is no longer 
that of a chiral superfield~\cite{Sakamura:2012bj}. 
Instead, $\hat{\der}_y\Phi$ transforms as a chiral superfield under $\dscp$. 
Thus $\hat{\der}_y$ is understood as a covariant derivative for $\dscp$. 
Similarly, $D_\alp$ and $\bar{D}_{\dalp}$ do not preserve the $\dscp$-transformation 
law of the $N=1$ superfields, either. 
For them, however, there are no corresponding covariant derivatives for $\dscp$. 

In the $d^4\tht$-integral, which corresponds to the $D$-term formula 
in Ref.~\cite{Kugo:1982cu}, a chiral superfield~$\Phi$ must appear 
through the combination of
\be
 \cU(\Phi) \equiv \brkt{1+iU^\mu\der_\mu+iU^y\der_y}\Phi. 
\ee
The first two terms correspond to an embedding of a chiral multiplet 
into a general multiplet in 4D superconformal formulation~\cite{Kugo:1982cu}, 
and the third term 
is necessary for the $\dscq$-invariance of the action. 

5D SUGRA is characterized by a cubic polynomial for the vector multiplets, 
which is referred to as the norm function 
in Refs.~\cite{Kugo:2000af}-\cite{Kugo:2002js}, 
\be
 \cN(\Sgm) \equiv C_{IJK}\Sgm^I\Sgm^J\Sgm^K, 
\ee
where a real constant tensor~$C_{IJK}$ is completely symmetric for the indices. 
This corresponds to the prepotential of $N=2$ SUSY gauge theories. 
For $C_{IJK}$, there is a set of 
normalized anti-hermitian matrices~$\brc{t_I}$, 
which satisfies~\cite{Kugo:2000af} 
\be
 C_{IJK} = \frac{ic^3}{6}\tr\brkt{t_I\brc{t_J,t_K}}. 
\ee
where $\tr(t_It_J)=-\frac{1}{2}\dlt_{IJ}$, and 
a real constant~$c$ can take different values for each simple or Abelian group. 
Some of the gauge symmetries are broken by 
the orbifold projection, and $t_{\Io}$ and $t_{\Ie}$ are the broken 
and the unbroken generators, respectively. 

The supergauge transformation is expressed as 
\bea
 &&e^V \to e^{\cU(\Lmd)}e^V e^{\cU(\Lmd)^\dagger}, \;\;\;\;\;
 \Sgm \to e^\Lmd\brkt{\Sgm-\hat{\der}_y}e^{-\Lmd}, \nonumber\\
 &&\Phi_{\rm odd} \to \brkt{e^{-\Lmd}}^t\Phi_{\rm odd}, \;\;\;\;\;
 \Phi_{\rm even} \to e^\Lmd\Phi_{\rm even},  \label{trf:gauge}
\eea
where the transformation parameter~$\Lmd$ is a chiral superfield, 
and $\Phi_{\rm odd}$ and $\Phi_{\rm even}$ 
are $(n_C+n_H)$-dimensional column vectors that consist of 
$\Phi^{2a-1}$ and $\Phi^{2a}$, respectively.  
We have used a matrix notation~$(V,\Sgm)\equiv 2ig(V^I,\Sgm^I)t_I$. 
The gauge coupling~$g$ can take different values for each simple or 
Abelian factor of the gauge group.  
The gauge-invariant field strength superfields are defined as\footnote{
Note that $\cV$ is not hermitian, but $e^{-\frac{V}{2}}\cV e^{\frac{V}{2}}$ is. 
} 
\bea
 \cW_\alp \eql \frac{1}{4}\bar{D}^2\left\{e^V D_\alp e^{-V}
 -\frac{1}{2}\bar{\sgm}_\mu^{\dbt\bt}D_\alp U^\mu\bar{D}_{\dbt}
 \brkt{e^V D_\bt e^{-V}} \right. \nonumber\\
 &&\hspace{10mm}\left.
 +iD_\alp U^\mu e^V\der_\mu e^{-V}-iU^\mu\der_\mu\brkt{e^V D_\alp e^{-V}}
 \right\},  \nonumber\\
 \cV \eql e^V\tl{\der}_ye^{-V}
 +\cU(\Sgm)+e^V\cU(\Sgm)^\dagger e^{-V} \nonumber\\
 &&+i\der_y U^y\brkt{\Sgm-e^V\Sgm^\dagger e^{-V}}
 -\frac{i\vev{V_E}^2}{2}\brkt{D^\alp U^y\cW_\alp
 -\bar{D}_{\dalp}U^y e^V(\cW^\dagger)^{\dalp}e^{-V}}, 
\eea
where
\bea
 \tl{\der}_y \defa \der_y-\frac{1}{4}\bar{D}^2\Psi^\alp D_\alp
 -\frac{1}{4}D^2\bar{\Psi}_{\dalp}\bar{D}^{\dalp}
 -\frac{i}{2}\sgm^\mu_{\alp\dalp}
 \brkt{\bar{D}^{\dalp}\Psi^\alp+D^\alp\bar{\Psi}^{\dalp}}\der_\mu \nonumber\\
 &&+\brc{\der_y U^\mu+\frac{1}{2}\sgm^\mu_{\alp\dalp}
 \brkt{\bar{D}^{\dalp}\Psi^\alp-D^\alp\bar{\Psi}^{\dalp}}}\Dlt_\mu. 
\eea
They transform under (\ref{trf:gauge}) as 
\be
 \cW_\alp \to e^\Lmd\cW_\alp e^{-\Lmd}, \;\;\;\;\;
 \cV \to e^{\cU(\Lmd)}\cV e^{-\cU(\Lmd)}. 
\ee
We can check that 
these field strength superfields follow the correct $\dscp$-transformation laws. 
The Weyl weights of $\cW_\alp$ and $\cV$ are $3/2$ and $0$, respectively.

\begin{description}
\item[Matter Lagrangian] \mbox{}\\
The 5D SUGRA Lagrangian is expressed as 
\bea
 \cL \eql \cL_{\rm kin}^{{\mathbb E}_W}
 -\int\dr^4\tht\;\brkt{1+\frac{\Dlt_\mu U^\mu}{3}}
 \brkt{2V_E\Omg_{\rm h}+V_E^{-2}\Omg_{\rm v}}  \nonumber\\
 &&+\sbk{\int\dr^2\tht\;\brkt{W_{\rm h}+W_{\rm v}}+\hc}
 +2\sum_{y_*=0,L}\cL^{(y_*)}_{\rm bd}\dlt(y-y_*), \label{5D_action}
\eea
where $\cL_{\rm kin}^{{\mathbb E}_W}$ denotes kinetic terms 
for the Weyl multiplet, 
$\cL^{(y_*)}_{\rm bd}$ ($y_*=0,L$) 
are the boundary localized Lagrangians at $y=y_*$, and 
\bea
 \Omg_{\rm h} \defa \cU(\Phi_{\rm odd})^\dagger\tl{d}(e^V)^t\cU(\Phi_{\rm odd})
 +\cU(\Phi_{\rm even})^\dagger\tl{d}e^{-V}\cU(\Phi_{\rm even}), \nonumber\\
 \tl{d} \defa \diag(\id_{n_C},-\id_{n_H}), \nonumber\\
 \Omg_{\rm v} \defa \cN(\cV) 
 = -\frac{c^3}{24g^3}\tr\brkt{\cV^3}, \nonumber\\
 W_{\rm h} \defa \Phi_{\rm odd}^t\tl{d}\brkt{\hat{\der}_y-\Sgm}\Phi_{\rm even}
 -\Phi_{\rm even}^t\tl{d}\brkt{\hat{\der}_y+\Sgm^t}\Phi_{\rm odd}, \nonumber\\
 W_{\rm v} \defa \frac{c^3}{16g^3}\tr\sbk{\Sgm\cW^2
 -\frac{1}{24}\bar{D}^2\brkt{\cZ^\alp}\brkt{\cW_\alp-\frac{1}{4}\cW_\alp^{(2)}}}
 +\cdots. 
 \label{def:Omgs}
\eea
Here, $W_{\rm v}$ represents the supersymmetric 
Chern-Simons terms,\footnote{
The counterpart in the global 5D SUSY theory is shown 
in Refs.~\cite{ArkaniHamed:2001tb,Hebecker:2008rk}. }  
and a part of it provides the kinetic term for the vector superfield~$V$ 
after the superconformal gauge fixing. 
The ellipsis in $W_{\rm v}$ denotes terms that vanish 
in the Wess-Zumino gauge. 
$\cW^{(2)}_\alp$ is a quadratic part of $\cW_\alp$ in $V$, and 
\be
 \cZ_\alp \equiv \brc{X,\der_y D_\alp X}_{\mathbb E}
 -\{\hat{\der}_y X,D_\alp X\}_{\mathbb E}, 
\ee
where $X\equiv\brkt{1+U^\mu \Dlt_\mu}V-iU^y (\Sgm-e^V\Sgm^\dagger e^{-V})$, and 
\bea
 \brc{\cX,\cY_\alp}_{\mathbb E} \defa \brc{\cX,\sbk{\cY_\alp}_{\mathbb E}}
 -\frac{1}{2}\bar{\sgm}_\mu^{\dbt\bt}\brkt{
 U^\mu\brc{D_\bt\bar{D}_{\dbt}\cX,\cY_\alp}
 +D_\alp U^\mu\brc{\bar{D}_{\dbt}\cX,\cY_\bt}}, \nonumber\\
 \sbk{\der_y D_\alp X}_{\mathbb E} \defa 
 D_\alp\hat{\der}_y X-\frac{1}{2}\bar{\sgm}_\mu^{\dbt\bt}U^\mu
 D_\alp D_\bt\bar{D}_{\dbt}\der_y X \nonumber\\
 &&+\frac{1}{4}\brkt{\sgm_{\alp\dbt}^\mu\der_y U_\mu
 +\bar{D}_{\dbt}\Psi_\alp-D_\alp\bar{\Psi}_{\dbt}}D^2\bar{D}^{\dbt}X, \nonumber\\
 \sbk{D_\alp X}_{\mathbb E} \defa D_\alp X
 -\frac{1}{2}\bar{\sgm}_\mu^{\dbt\bt}U^\mu D_\alp D_\bt\bar{D}_{\dbt}X. 
\eea

\item[Kinetic terms for $\bdm{\bE_W}$] \mbox{}\\
In contrast to the matter sector, 
$\cL_{\rm kin}^{{\mathbb E}_W}$ is quadratic in the gravitational superfields. 
It should be identified from the invariance of the action 
up to linear order in the gravitational superfields. 
This requires an extension of the 5D superconformal transformations~(\ref{dscp}) 
and (\ref{dscq}) by including linear terms in the gravitational superfields. 
For the purpose of this paper, we only need terms in $\cL^{\bE_W}_{\rm kin}$ 
that are independent of the quantum fluctuation 
of the matter superfields. 
Hence, we can treat the matter superfields in 
the corrections to (\ref{dscp}) and (\ref{dscq}) as the background values.  
The corrected transformations involving $U^\mu$ are listed 
in (\ref{md:dscq}) in Appendix~\ref{trf:SC}. 
By requiring the invariance of the action 
under the corrected transformations, we find 
\be
 \cL_{\rm kin}^{{\mathbb E}_W} = \int\dr^4\tht\;\left\{
 \Lvev{\frac{2V_E\Omg_{\rm h}+V_E^{-2}\Omg_{\rm v}}{3}}E_2
 +\Lvev{\frac{V_E^{-1}\Omg_{\rm h}-4V_E^{-4}\Omg_{\rm v}}{3}}\cC^\mu\cC_\mu
 \right\},  \label{cL_kin^EW}
\ee
where the symbol~$\Lvev{\cdots}$ denotes the background value, and 
\bea
 E_2 \defa -\frac{1}{8}U_\mu D^\alp\bar{D}^2D_\alp U^\mu
 +\frac{1}{3}(\Dlt_\mu U^\mu)^2-\brkt{\der_\mu U^\mu}^2, \nonumber\\
 \cC^\mu \defa \der_y U^\mu+\frac{1}{2}\sgm_{\alp\dalp}^\mu
 \brkt{\bar{D}^{\dalp}\Psi^\alp-D^\alp\bar{\Psi}^{\dalp}}
 +\vev{V_E}^2\der^\mu U^y. 
\eea
In addition to the above terms, the following term is expected 
to appear in the 5D Lagrangian. 
\be
 \cL_{\rm add} = -\Lvev{\frac{\Omg_{\rm v}}{V_E^4}}\bar{D}^{\dalp}\Psi^\alp
 D_\alp\bar{\Psi}_{\dalp}. \label{cL_add}
\ee
This term is necessary to obtain the correct kinetic terms for 
the vector superfields~(\ref{cL_kin^vec}). 
In order to justify the existence of this term, 
we need to modify $\dscp$ and $\dscq$ further by including 
$\Psi_\alp$-dependent terms in the right-hand sides of (\ref{dscp}) 
and (\ref{dscq}). 
Here we leave this task for future works, 
and just assume (\ref{cL_add}).

\item[Boundary localized terms] \mbox{}\\
We can introduce terms localized on the 4D boundaries of $S^1/Z_2$. 
The boundary actions are described by the action formulae of 
4D superconformal formulation~\cite{Kugo:1982cu}, and expressed 
in terms of the superfields as~\cite{Sakamura:2011df} 
\bea
 \cL_{\rm bd}^{(y_*)} \eql \int\dr^4\tht\;
 \brc{-\frac{2}{3}\Lvev{\Omg_{\rm bd}^{(y_*)}}E_2+2\brkt{1+\frac{\Dlt_\mu U^\mu}{3}}
 \Omg_{\rm bd}^{(y_*)}} \nonumber\\
 &&+\sbk{\int\dr^2\tht\;\brc{\phi^3P^{(y_*)}(\chi)
 -\frac{1}{2}\tr\brkt{f^{(y_*)}(\chi)\cW^\alp\cW_\alp}}+\hc}, 
 \label{cL_bd}
\eea
where 
\be
 \Omg_{\rm bd}^{(y_*)} = -\frac{3}{2}\abs{\cU(\phi_C)}^2
 \exp\brc{-\frac{K^{(y_*)}(\cU(\chi),V_{\rm 4D})}{3}}. 
\ee
Chiral superfields~$\phi_C$ and $\chi^a$ ($a=1,2,\cdots)$ are the 4D compensator 
and the physical matter superfields, and $V_{\rm 4D}^I$ are 4D vector superfields.  
A real function~$K^{(y_*)}$ is the K\"ahler potential, 
and holomorphic functions~$P^{(y_*)}$ and 
$f^{(y_*)}$ are the superpotential and the gauge kinetic functions, respectively.  
Note that $\cU(\phi)=\brkt{1+iU^\mu\der_\mu}\phi$ in the above Lagrangian 
since $U^y$ is $Z_2$-odd and vanishes on the boundaries. 
In general, $\chi^a$ and $V_{\rm 4D}^I$ can be either boundary values of 
the $Z_2$-even bulk superfields or 
additional 4D superfields localized on the boundaries. 
In contrast to the 5D bulk action, we have only one compensator 
chiral multiplet. 
Thus, one combination of $Z_2$-even 5D compensators~$\Phi^{2a}$ 
($a=1,\cdots,n_C$) plays its role.\footnote{
Since the gravity is unique in the whole system, 
the boundary compensator multiplets must be the boundary values 
of the bulk compensator multiplets. }

In the case of $n_C=1$, $\Phi^2$ is the only $Z_2$-even compensator superfield. 
Hence, the 4D chiral compensator superfield~$\phi_C$ 
in $\cL_{\rm bd}^{(y_*)}$ ($y_*=0,L$) is identified as 
\be
 \phi_C = \left.\brkt{\Phi^2}^{2/3}\right|_{y=y_*},  \label{def:phi_C1}
\ee
because $\phi_C$ must have $w=n=1$. 
The bulk physical matter superfields can appear in $\cL_{\rm bd}^{(y_*)}$ 
in the forms of 
\be
 \chi^a = \left.\frac{\Phi^{2a+2}}{\Phi^2}\right|_{y=y_*}, 
 V_{\rm 4D} = V|_{y=y_*}, 
 \label{def:chi1}
\ee
because the physical matter superfields must have zero Weyl (chiral) weight 
in the 4D superconformal formulation~\cite{Kugo:1982cu}.

In the case of $n_C=2$, there are two $Z_2$-even 
compensator superfields~$\Phi^2$ and $\Phi^4$. 
In this case, we have to eliminate 
one combination of the 5D compensator multiplets. 
In Ref.~\cite{Fujita:2001bd}, this is done by 
introducing a nondynamical (auxiliary) Abelian 
vector multiplet~${\mathbb V}_T=(V_T,\Sgm_T)$, 
and gauging a $U(1)$ subgroup of the isometries, which is referred to 
as $U(1)_T$, by it. 
The $U(1)_T$ charges~$Q_T$ are chosen as 
$Q_T(\Phi^1)=Q_T(\Phi^4)=Q_T(\Phi^{2a+4})=+1$ and 
$Q_T(\Phi^2)=Q_T(\Phi^3)=Q_T(\Phi^{2a+3})=-1$ ($a\geq 1$). 
Since the 4D superfields must be neutral for $U(1)_T$, they are identified as 
\be
 \phi_C = \left.\brkt{\Phi^2\Phi^4}^{1/3}\right|_{y=y_*}, \;\;\;\;\;
 \chi^a = \left.\frac{\Phi^{2a+4}}{\Phi^4}
 \right|_{y=y_*}.  \label{def:phichi2}
\ee
\end{description}

As pointed out in Ref.~\cite{Correia:2006pj}, $V_E$ does not have a kinetic term 
and can be integrated out. 
From (\ref{5D_action}), $V_E$ is expressed as 
\be
 V_E = \brkt{\frac{\Omg_{\rm v}}{\Omg_{\rm h}}}^{1/3}. 
\ee
After integrating it out, the 5D Lagrangian becomes 
\bea
 \cL \eql \int\dr^4\tht\;\left\{\Lvev{\Omg_{\rm v}^{1/3}\Omg_{\rm h}^{2/3}}E_2
 -\Lvev{\Omg_{\rm v}^{-1/3}\Omg_{\rm h}^{4/3}}\brkt{\cC^\mu\cC_\mu
 +\bar{D}^{\dalp}\Psi^\alp D_\alp\bar{\Psi}_{\dalp}} \right. \nonumber\\
 &&\hspace{15mm}\left. 
 -3\brkt{1+\frac{\Dlt_\mu U^\mu}{3}}\Omg_{\rm v}^{1/3}\Omg_{\rm h}^{2/3}\right\} 
 \nonumber\\
 &&+\sbk{\int\dr^2\tht\;\brkt{W_{\rm h}+W_{\rm v}}+\hc}
 +2\sum_{y_*=0,L}\cL^{(y_*)}_{\rm bd}\dlt(y-y_*).  \label{5D_action2}
\eea
In our previous paper~\cite{Sakamura:2012bj}, we implicitly assumed 
that $\vev{\Omg_{\rm v}}=\vev{\Omg_{\rm h}}=1$ 
(in the unit of the 5D Planck mass), 
but we need their explicit dependences on 
the background superfields of the matters for the derivation of 
the one-loop effective K\"ahler potential. 

In order to obtain the Poincar\'e SUGRA, we have to impose 
the superconformal gauge-fixing conditions to eliminate the extra symmetries. 
For example, the dilatation symmetry will be fixed by the condition, 
$\Omg_{\rm v}|_0=\Omg_{\rm h}|_0=1$ in the 5D Planck unit,\footnote{
This condition must be consistent with the orbifold projection, 
which indicates that 
$C_{\Io J_{\rm o}K_{\rm e}}=C_{\Ie J_{\rm e}K_{\rm e}}=0$. \label{zero:C_IJK}
} 
where the symbol~$|_0$ denotes the lowest component of the superfield. 
However, these gauge-fixing conditions are incompatible with 
the $N=1$ off-shell structure. 
Thus we will add the gauge-fixing terms in the calculations in Sec.~\ref{1loop_K}, 
instead of imposing such conditions. 

\ignore{
The most general metric that has the 4D Poincar\'e symmetry has a form of
\be
 ds^2 = e^{2\sgm(y)}\eta_{\mu\nu}dx^\mu dx^\nu-dy^2, \label{warp:metric}
\ee
where $e^{\sgm(y)}$ is the warp factor, which is determined 
by solving 5D Einstein equation. 
Notice that (\ref{warp:metric}) is conformally flat. 
Thus we can always move to a coordinate system where 
the metric is the 5D flat one by using the dilatation and the coordinate change 
of $y$. 
Since our formalism keeps the superconformal symmetries manifest, 
the warp factor does not appear explicitly in our calculations. 
The information of the warped geometry is encoded 
in the gauging for the compensator hypermultiplets~\cite{Abe:2007zv}. 
}

\section{One-loop effective K\"ahler potential} \label{1loop_K}
In our previous works~\cite{Abe:2008an,Abe:2011rg,Abe:2006eg}, 
we derived the 4D effective action at tree level. 
We provide a brief review of the derivation in Appendix~\ref{tree:action}. 
In this section, we calculate the one-loop contributions to 
the effective K\"ahler potential. 

\subsection{Background field method}
We calculate the one-loop effective K\"ahler potential 
by using the background field method~\cite{Grisaru:1996ve}.\footnote{
The first calculation of the one-loop effective K\"ahler potential 
by means of the $N=1$ superfield technique was provided 
in Ref.~\cite{Buchbinder:1994iw}. 
} 
First we split each superfield into the background 
and the fluctuation parts. 
Since we are interested in the effective theory for the zero-modes 
of the matter superfields, we only consider the background values 
of the $Z_2$-even matter superfields~$\Phi_{\rm even}$, $V^{\Ie}$, 
and $\Sgm^{\Io}$. 
We move to a gauge where $\vev{\Sgm^I}$ are zero 
by the supergauge transformation for the background superfields. 
This is accomplished by choosing the transformation parameter 
as (\ref{def:Lmd_Sgm}) (or (\ref{def:Lmd_Sgm2}) in the Abelian case). 
Then $\vev{V^{\Io}}$ become discontinuous at $y=L$. 
In fact, in the case that the gauge group is Abelian, 
their boundary conditions are 
\bea
 \lim_{y\to 0}\vev{V^{\Io}} \eql 0, \;\;\;\;\;
 \lim_{y\to L}\vev{V^{\Io}} = -T^{\Io}-\bar{T}^{\Io},  \nonumber\\
 \left.\vev{V^{\Io}}\right|_{y=0} \eql \left.\vev{V^{\Io}}\right|_{y=L} = 0,  
 \label{bdcd:V^I}
\eea
where the limits are taken from the bulk region~$0<y<L$, and 
\be
 T^{\Io} \equiv \int_0^L\dr y\;\vev{\Sgm^{\Io}}.  \label{def:T^I}
\ee
We refer to the chiral superfields~$T^{\Io}$ 
as the moduli superfields in this paper. 
In order to take them into account, we also keep the background values of $V^{\Io}$ 
in addition to those of the $Z_2$-even matter superfields. 
Thus, each matter superfield is split as 
\bea
 \Phi_{\rm odd} \eql \tl{\Phi}_{\rm odd}, \;\;\;\;\;
 \Phi_{\rm even} = \bdm{\Phi}+\tl{\Phi}_{\rm even}, \nonumber\\
 V \eql \bdm{V}+\tl{V}, \;\;\;\;\;
 \Sgm = \tl{\Sgm}, 
\eea
where $\bdm{\Phi}$ and $\bdm{V}$ are the background values 
and the quantities with tilde denote the fluctuation parts. 
We neglect derivative terms in the effective K\"ahler potential, 
and thus we treat $\bdm{\Phi}$ and $\bdm{V}$ 
as functions of only $y$ in the following calculations. 
The gravitational superfields~$U^\mu$, $U^y$, and $\Psi_\alp$ 
are considered as the fluctuation modes. 
($V_E$ has already been integrated out.) 
As we have pointed out in Ref.~\cite{Sakamura:2012bj}, 
$U^y$ can be gauged away by $\dscq$ given in (\ref{dscq}) in Appendix~\ref{trf:SC}. 
So we take the gauge where $U^y=0$ in the following.\footnote{
In this gauge, we do not need to consider contributions from the ghost 
for $\dscq$ because it is decoupled 
from the background superfields~$\bdm{\Phi}$ and $\bdm{V}$. } 

We expand the 5D Lagrangian~(\ref{5D_action2}) and pick up quadratic terms 
in the fluctuation superfields. 
\be
 \cL = \sum_{F}\int\dr^4\tht\;F^\dagger\cO_F F+\cdots, 
 \label{def:cO_F}
\ee
where $F$ runs over the fluctuation superfields, 
and $\cO_F$ are differential operators that depend on 
$\bdm{\Phi}$ and $\bdm{V}$. 
Then the one-loop contribution to the effective action~$\Dlp S$ 
is calculated as 
\be
 \Dlp S = \frac{i}{2(2\pi)^4}\sum_F
 \int\dr^4p\;\Tr\brkt{\str\,\ln\cO_F}, 
\ee
where $\str$ is the supertrace   
over the functional space on the 16-dimensional graded vector space 
built from all combinations of $\tht$ and $\bar{\tht}$, and 
$\Tr$ is the trace over the remaining space including 
the functional space of $y$. 
Here we denote an integrand of the $d^4\tht$-integral 
for $\str$ as ${\rm Istr}$~\cite{Flauger:2012ie}. 
Namely, it follows that 
\be
 \str\,\ln\cO_F \equiv \int\dr^4\tht\;\Istr\ln\cO_F. 
\ee
Then the one-loop contribution to 
the K\"ahler potential~$\Omg_{\rm eff}=-3e^{-K_{\rm eff}/3}$ 
is expressed as
\be
 \Omglp = \frac{i}{2(2\pi)^4}\sum_F\int\dr^4p\;
 \Tr\brkt{\Istr\ln\cO_F}. 
 \label{formula:Omglp}
\ee
Since $\Omglp$ is a function of the background superfields 
whose dependences on $x^\mu$ and $\tht$ ($\bar{\tht}$) are now neglected, 
$\Istr$ is calculated by 
\be
 \Istr\ln\cO_F 
 = \sbk{\ln\cO_F\brkt{\tht^2\bar{\tht}^2}}_{\tht=\bar{\tht}=0}.   
 \label{def:Istr}
\ee
Its values for various operators are collected in (\ref{Istr:formulae}).

\subsection{Quadratic terms for fluctuation modes}
Here we pick up the quadratic terms in the fluctuation superfields, 
and find explicit forms of $\cO_F$ in (\ref{def:cO_F}). 
Detailed calculations are shown in Appendix~\ref{quad_flct}. 

\subsubsection{Bulk sector}
Using the superspin projectors defined by (\ref{def:Pis}) 
in Appendix~\ref{superspinPi}, 
$U^\mu$ is decomposed as~\cite{Gregoire:2004nn,Gates:1983nr,Gates:2003cz} 
\be
 U^\mu = \sum_s\Pi_s^{\mu\nu}U_\nu \equiv \sum_sU_s^\mu, 
 \label{U:decompose}
\ee
where $s=0,1/2,1,3/2$. 
We choose the gauge-fixing term for the superconformal symmetry~$\dscp$ as 
\be
 \cL_{\rm gf}^{\rm sc} = \int\dr^4\tht\;
 \frac{\Lvev{\Omg_{\rm v}^{1/3}\Omg_{\rm h}^{2/3}}}{\xi_{\rm sc}}
 \hat{U}_\mu\Box_4\Pi_{\rm gf}^{\mu\nu}(\xi_{\rm sc})
 \hat{U}_\nu,  \label{cL_gf^sc}
\ee
where $\xi_{\rm sc}$ is the gauge-fixing parameter, $\Box_4\equiv\der^\mu\der_\mu$, 
and 
\bea
 \Pi_{\rm gf}^{\mu\nu}(\xi_{\rm sc}) 
 \defa \eta^{\mu\nu}-\Pi_{3/2}^{\mu\nu}
 -\frac{2\xi_{\rm sc}}{3}\Pi_0^{\mu\nu}, \nonumber\\
 \hat{U}_\mu \defa U_\mu+\frac{3i\xi_{\rm sc}}{(3-2\xi_{\rm sc})\Box_4}
 \der_\mu\brkt{\cT+\tl{\Phi}_C-\bar{\cT}-\bar{\tl{\Phi}}_C} \nonumber\\
 &&+\frac{\xi_{\rm sc}}{2\Box_4}\brkt{\eta_{\mu\nu}
 +\frac{2\xi_{\rm sc}}{3-2\xi_{\rm sc}}\Pi_{0\mu\nu}}
 \Dlt^\nu\brkt{\tl{V}_{\rm v}+\tl{V}_{\rm h}}. 
\eea
Here, $\cT$, $\tl{\Phi}_C$, $\tl{V}_{\rm v}$ and $\tl{V}_{\rm h}$ 
are defined as~\footnote{ 
$\cT$ and $V_{\cT}$ correspond to the 5D radion and 
the graviphoton superfields, respectively. 
}
\bea
 \cT \defa \frac{\cN_I}{3\cN}(\vev{\cV})\tl{\Sgm}^I, \;\;\;\;\;
 V_{\cT} \equiv \frac{\cN_I}{3\cN}(\vev{\cV})\tl{V}^I, \;\;\;\;\;
 \tl{\Phi}_C \equiv \frac{2}{3}\Ups^\dagger\tl{\Phi}_{\rm even}, \;\;\;\;\;
 \Ups \equiv \frac{1}{\Lvev{\Omg_{\rm h}}}
 \tl{d}e^{-\sbdm{V}}\bdm{\Phi},  \nonumber\\
 \tl{V}_{\rm v} \defa -\frac{\cN_I}{3\cN}(\vev{\cV})\der_y\tl{V}^I,  \;\;\;\;\;
 \tl{V}_{\rm h} \equiv \frac{2}{3}\bdm{\Phi}^\dagger\Ups_I\tl{V}^I, \;\;\;\;\;
 \Ups_I \equiv \frac{1}{\Lvev{\Omg_{\rm h}}}\frac{\der}{\der\bdm{V}^I}
 \tl{d}e^{-\sbdm{V}}\bdm{\Phi},  
 \label{def:cT}
\eea
where $\cN_I\equiv\der\cN/\der\cV^I$. 
Then the cross terms between $U_\mu$ and the other superfields are 
canceled, and we obtain 
\bea
 \cL+\cL_{\rm gf}^{\rm sc} 
 \eql \int\dr^4\tht\;
 \brc{-U_{3/2}^\mu\cO_{3/2}U_{3/2\mu}
 +\bar{U}^\mu\cO_{\bar{U}}\bar{U}_\mu}
 +\cO(\xi_{\rm sc}) \nonumber\\
 &&+\int\dr^4\tht\;\frac{\cN_I\cN_J}{2\cN}(\vev{\cV})\tl{V}^I
 \Box_4P_T\tl{V}^J+\cdots,   \label{quad:SUGRA}
\eea
where $P_T$ is a projection operator defined in (\ref{def:P_CT}), and 
\bea
 \bar{U}^\mu \defa U^\mu-U_{3/2}^\mu 
 = \brkt{\Pi_0^{\mu\nu}+\Pi_{1/2}^{\mu\nu}+\Pi_1^{\mu\nu}}U_\nu, \nonumber\\
 \cO_{3/2} \defa \Lvev{\Omg_{\rm v}^{1/3}\Omg_{\rm h}^{2/3}}
 \brkt{\Box_4+\cD_U}, \;\;\;\;\;
 \cO_{\bar{U}} \equiv \Lvev{\Omg_{\rm v}^{1/3}\Omg_{\rm h}^{2/3}}
 \brkt{\frac{1}{\xi_{\rm sc}}\Box_4+\cD_U}, \nonumber\\
 \cD_U \defa -\frac{\der_y\brkt{\Lvev{\Omg_{\rm v}^{-1/3}\Omg_{\rm h}^{4/3}}\der_y}}
 {\Lvev{\Omg_{\rm v}^{1/3}\Omg_{\rm h}^{2/3}}}. 
 \label{def:cD_U}
\eea
The last term in (\ref{quad:SUGRA}) is combined with the quadratic terms 
in the vector sector shown in (\ref{quad_flct:grv}), 
and provides the kinetic terms for $\tl{V}^I$, 
\be
 \cL_{\rm kin}^{\rm vec} = \int\dr^4\tht\;
 \Lvev{\Omg_{\rm v}}a_{IJ}\tl{V}^I\Box_4P_T\tl{V}^J,  \label{cL_kin^vec}
\ee
where 
\be
 a_{IJ} \equiv -\frac{1}{2\cN}\brkt{\cN_{IJ}-\frac{\cN_I\cN_J}{\cN}}, \;\;\;\;\;
 \cN_{IJ} \equiv \frac{\der^2\cN}{\der\cV^I\der\cV^J}. 
 \label{def:a}
\ee
The arguments of the norm function and its derivatives are 
understood as $\vev{\cV^I}$ in this and the next subsections. 
These kinetic terms are consistent with those in Ref.~\cite{Kugo:2000af}. 

In the following, we consider a case that the gauge group is Abelian 
for simplicity. 
We choose the gauge-fixing term for the supergauge symmetry~$\dgt$ as 
\be
 \cL_{\rm gf}^{\rm sg} = \int\dr^4\tht\;
 \frac{\Lvev{\Omg_{\rm v}}a_{IJ}}{\xi_{\rm sg}}
 \hat{V}^I\Box_4 P_C\hat{V}^J,  \label{GF:gg}
\ee
where $\xi_{\rm sg}$ is the gauge-fixing parameter, $P_C$ is 
the chiral projection operator defined in (\ref{def:P_CT}), and 
\bea
 \hat{V}^I \defa \tl{V}^I
 +\frac{\xi_{\rm sg}a^{IJ}}{\Lvev{\Omg_{\rm v}}\Box_4}
 \brkt{\Xi_J+\bar{\Xi}_J}. 
\eea
The definition of $\Xi_I$ is given in (\ref{def:Xi_I}). 
Then the cross terms between $\tl{V}$ and the chiral superfields 
are canceled. 

As a result, the quadratic terms 
for the fluctuation superfields in the 5D Lagrangian are summarized as 
\bea
 \cL_{\rm quad} \eql \int\dr^4\tht\;
 U_\mu\brc{-\cO_{3/2}\Pi_{3/2}^{\mu\nu}
 +\cO_{\bar{U}}\brkt{\eta^{\mu\nu}-\Pi_{3/2}^{\mu\nu}}}U_\nu \nonumber\\
 &&+\int\dr^4\tht\;
 \tl{V}^I\brc{(\cO_T)_{IJ}P_T+(\cO_C)_{IJ}P_C}\tl{V}^J \nonumber\\
 &&+\int\dr^4\tht\;(\vph^\dagger,\vph^t)\begin{pmatrix} \cM & 
 \bar{W}\frac{\bar{D}^2}{4\Box_4} \\
 W\frac{D^2}{4\Box_4} & \cM^t \end{pmatrix}
 \begin{pmatrix} \vph \\ \bar{\vph} \end{pmatrix},  \label{quad:cL_bulk}
\eea
where $\vph\equiv(\tl{\Sgm}^I,\tl{\Phi}_{\rm even},\tl{\Phi}_{\rm odd})^t$, 
and 
\bea
 (\cO_T)_{IJ} \defa \Lvev{\Omg_{\rm v}}a_{IK}\brc{
 \dlt^K_{\;\;J}\Box_4+(\cD_V)^K_{\;\;J}}+\cO(\xi_{\rm sg}), \nonumber\\
 (\cO_C)_{IJ} \defa \Lvev{\Omg_{\rm v}}a_{IK}\brc{
 \frac{\dlt^K_{\;\;J}}{\xi_{\rm sg}}\Box_4+(\cD_V)^K_{\;\;J}}
 +\cO(\xi_{\rm sg}), \nonumber\\
 (\cD_V)^I_{\;\:J} \defa -\frac{a^{IK}}{\Lvev{\Omg_{\rm v}}}
 \der_y\brc{\Lvev{\Omg_{\rm v}^{1/3}\Omg_{\rm h}^{2/3}}
 \brkt{(a\cdot\cP_V)_{KJ}\der_y+\frac{\cN_K}{3\cN}\Ups_J^\dagger\bdm{\Phi}}} 
 \nonumber\\
 &&+\Lvev{\frac{\Omg_{\rm h}}{\Omg_{\rm v}}}^{2/3}a^{IK}
 \brkt{\frac{\cN_J}{3\cN}\Ups_K^\dagger\bdm{\Phi}\der_y
 -\Lvev{\frac{\der_I\der_J\Omg_{\rm h}}{\Omg_{\rm h}}}
 +\frac{\Ups_I^\dagger\bdm{\Phi}\bdm{\Phi}^\dagger\Ups_J}{3}}, 
 \nonumber\\
 \cM \defa \Lvev{\Omg_{\rm v}^{1/3}\Omg_{\rm h}^{2/3}}
 \begin{pmatrix} (a\cdot\cP_V)_{IJ} & -\frac{\cN_I}{3\cN}\Ups^\dagger & 0 \\
 -\frac{\cN_J}{3\cN}\Ups & -\frac{1}{\Lvev{\Omg_{\rm h}}}\tl{d}e^{-\sbdm{V}}
 +\frac{1}{3}\Ups\Ups^\dagger & 0 \\
 0 & 0 & -\frac{1}{\Lvev{\Omg_{\rm h}}}\tl{d}(e^{\sbdm{V}})^t \end{pmatrix}
 +\cO(\xi_{\rm sg}), \nonumber\\
 W \defa \begin{pmatrix} 0 & 0 & -\bdm{\Phi}^t\tl{d}\hat{t}_I^t \\
 0 & 0 & -\tl{d}\der_y \\ -\tl{d}\hat{t}_J\bdm{\Phi} & \tl{d}\der_y & 0 
 \end{pmatrix}. 
 \label{expr:cOs}
\eea
Here $\hat{t}_I\equiv 2igt_I$ are hermitian generators, and   
\be
 (\cP_V)^I_{\;\;J} \equiv \dlt^I_{\;\;J}-\frac{\vev{\cV^I}\cN_J}{3\cN}. 
 \label{def:cP_V}
\ee
is a projection operator~\cite{Kugo:2000af}, which has a property, 
\be
 \cN_I(\cP_V)^I_{\;\;J} = (\cP_V)^I_{\;\;J}\vev{\cV^J} = 0, \;\;\;\;\;
 \cP_V^2 = \id_{n_V}. 
\ee
The definitions of $\Ups$ and $\Ups_I$ are given in (\ref{def:cT}). 
For the purpose of calculating the one-loop K\"ahler potential, 
it is convenient to choose the gauge-fixing parameters as 
$\xi_{\rm sc}=\xi_{\rm sg}=0$.

\subsubsection{Boundary sector}
From (\ref{cL_bd}), the quadratic terms for the fluctuation superfields 
in the boundary Lagrangians are found to be
\bea
 \cL_{\rm boundary}^{(y_*)} 
 \eql \cL_{\rm bd}^{(y_*)}+\cL_{\rm gf}^{{\rm sc}(y_*)}
 +\cL_{\rm gf}^{{\rm sg}(y_*)} \nonumber\\
 \eql \int\dr^4\tht\;\abs{\bdm{\phi}_C}^2h^{(y_*)}\brc{
 -U_\mu\Box_4\Pi_{3/2}^{\mu\nu}U_\nu
 +\frac{1}{\zeta_{\rm sc}^{(y_*)}}
 U_\mu\Box_4\brkt{\eta^{\mu\nu}-\Pi_{3/2}^{\mu\nu}}U_\nu} 
 \nonumber\\
 &&+\int\dr^4\tht\;\left[\Re f_{\Ie\Je}^{(y_*)}
 \brc{\tl{V}^{\Ie}\Box_4\brkt{P_T+\frac{1}{\zeta_{\rm sg}^{(y_*)}}P_C}
 \tl{V}^{\Je}} 
 -\frac{3}{2}\abs{\bdm{\phi}_C}^2h_{\Ie\Je}^{(y_*)}\tl{V}^{\Ie}\tl{V}^{\Je}
 \right] \nonumber\\
 &&+\int\dr^4\tht\;\sbk{\abs{\bdm{\phi}_C}^2h_{a\bar{b}}^{(y_*)}
 \tl{\chi}^a\bar{\tl{\chi}}^b
 +\brkt{\bar{\bdm{\phi}}_Ch_{\bar{a}}^{(y_*)}
 \tl{\phi}_C\bar{\tl{\chi}}^a
 +\hc}+h^{(y_*)}|\tl{\phi}_C|^2} \nonumber\\
 &&+\sbk{\int\dr^2\tht\;\brkt{\frac{1}{2}\bdm{\phi}_C^3
 P_{ab}^{(y_*)}\tl{\chi}^a\tl{\chi}^b
 +3\bdm{\phi}_C^2P_a^{(y_*)}\tl{\phi}_C\tl{\chi}^a
 +3\bdm{\phi}_CP^{(y_*)}\tl{\phi}_C^2}+\hc} \nonumber\\
 &&+\cO\brkt{\zeta_{\rm sc}^{(y_*)},\zeta_{\rm sg}^{(y_*)}}+\cdots, 
 \label{quad:cL_bd}
\eea
where $\bdm{\phi}_C$ ($\tl{\phi}_C$) 
and $\bdm{\chi}^a$ ($\tl{\chi}^a$) are 
the background (fluctuation) parts of the compensator 
and the physical chiral superfields~$\phi_C$ and $\chi^a$, and 
$h^{(y_*)}\equiv -3\exp\brkt{-K^{(y_*)}/3}$, 
$h^{(y_*)}_a\equiv\der h^{(y_*)}/\der\chi^a$, 
$h^{(y_*)}_{\Ie}\equiv\der h^{(y_*)}/\der V^{\Ie}$, $\cdots$, 
whose arguments are $(\bdm{\chi},\bdm{V})$. 
We have chosen the boundary gauge-fixing terms 
for the superconformal and the gauge symmetries as 
\bea
 \cL_{\rm gf}^{{\rm sc}(y_*)} \eql 
 -\int\dr^4\tht\;\frac{\abs{\bdm{\phi}_C}^2h^{(y_*)}}
 {\zeta_{\rm sc}^{(y_*)}}
 \hat{U}_\mu\Box_4\Pi_{\rm gf}^{\mu\nu}(\zeta_{\rm sc})
 \hat{U}_\nu, \nonumber\\
 \hat{U}_\mu \eql U_\mu+\frac{3i\zeta_{\rm sc}^{(y_*)}}
 {(3-2\zeta_{\rm sc}^{(y_*)})\Box_4}
 \der_\mu\brkt{\frac{h_a^{(y_*)}}{h^{(y_*)}}\tl{\chi}^a
 +\frac{\tl{\phi}_C}{\bdm{\phi}_C}-\hc} \nonumber\\
 &&+\frac{\zeta_{\rm sc}^{(y_*)}}{2\Box_4}
 \brkt{\eta_{\mu\nu}+\frac{2\zeta_{\rm sc}^{(y_*)}}
 {3-2\zeta_{\rm sc}^{(y_*)}}\Pi_{0\mu\nu}}
 \brkt{\frac{h_{\Ie}^{(y_*)}}{h^{(y_*)}}\Dlt^\nu\tl{V}^{\Ie}}, 
 \nonumber\\
 \cL_{\rm gf}^{{\rm sg}(y_*)} \eql \int\dr^4\tht\;
 \frac{\Re f_{\Ie\Je}^{(y_*)}(\bdm{\chi})}{\zeta_{\rm sg}^{(y_*)}}
 \hat{V}_{y_*}^{\Ie}\Box_4 P_C\hat{V}_{y_*}^{\Je}, \nonumber\\
 \hat{V}^{\Ie}_{y_*} \defa \tl{V}^{\Ie}
 +\frac{3\zeta_{\rm sg}^{(y_*)}}{2\Box_4}
 \brc{F^{\Ie\Je}\brkt{\abs{\bdm{\phi}_C}^2h_{\Je a}^{(y_*)}\tl{\chi}^a
 +\bar{\bdm{\phi}}_Ch_{\Je}^{(y_*)}\tl{\phi}_C+\hc}}, 
\eea
where $F^{\Ie\Je}$ is an inverse matrix of 
$\Re f_{\Ie\Je}^{(y_*)}(\bdm{\chi})$. 
In the following, we will choose the gauge-fixing parameters as 
$\zeta_{\rm sc}^{(y_*)}=\zeta_{\rm sg}^{(y_*)}=0$. 

In the case that $\phi_C$ and $\chi^a$ are the boundary values of 
the bulk superfields, the relations~(\ref{def:phi_C1}) and (\ref{def:chi1}) 
(or (\ref{def:phichi2})) in $\cL_{\rm bd}^{(L)}$ must be modified
for the background superfields because we have performed 
the discontinuous gauge transformation at $y=L$. 
(See (\ref{rel:bulk-bd}).)
In the case of $n_C=1$, for example, the relations are modified as 
\be
 \bdm{\phi}_C = \left.e^{-2k_{\Io}T^{\Io}}\brkt{\bdm{\Phi}^1}^{2/3}\right|_{y=L}, 
 \;\;\;\;\;
 \bdm{\chi}^a = \left.\frac{\brc{\exp\brkt{T^{\Io}\check{t}_{\Io}}\bdm{\Phi}}^{a+1}}
 {\bdm{\Phi}^1}\right|_{y=L},  \label{modify:rel:phichi}
\ee
where $k_{\Io}$ and $\check{t}_{\Io}$ are defined in (\ref{form:generator}). 

The above boundary-localized terms affect 
the boundary conditions for the fluctuation modes of the bulk superfields,  
which are no longer determined only by the orbifold parities. 
We derive them in Appendix~\ref{BC:flct_md}.

\subsection{Integration of fluctuation modes} \label{Int:flctuation}
In this subsection, we perform the integration of the fluctuation modes, 
and obtain formal expressions of the one-loop contributions to $\Omglp$. 

\subsubsection{Contribution from gravitational superfields}
The contribution from the gravitational superfields is 
\be
 \Omg_{\rm eff}^U = \frac{i}{2(2\pi)^4}\int\dr^4p\;
 \Tr\brc{\Istr\ln\cO_U-\Istr\ln
 \brkt{\frac{1}{\xi_{\rm sc}}
 \Lvev{\Omg_{\rm v}^{1/3}\Omg_{\rm h}^{2/3}}
 \Pi_{\rm gf}}}, \label{expr:Omg^U}
\ee
where
\be
 \cO_U^{\mu\nu} \equiv -\cO_{3/2}\Pi_{3/2}^{\mu\nu}
 +\cO_{\bar{U}}\brkt{\eta^{\mu\nu}-\Pi_{3/2}^{\mu\nu}}. 
\ee
The second term in (\ref{expr:Omg^U}) is a contribution 
from the ghost for $\dscp$. 
(See the gauge-fixing term~(\ref{cL_gf^sc}).) 
Since 
\be
 \ln\cO_U = \Pi_{3/2}^{\mu\nu}\ln(-\cO_{3/2})
 +\brkt{\eta^{\mu\nu}-\Pi_{3/2}^{\mu\nu}}\ln\cO_{\bar{U}}, 
\ee
it follows that 
\bea
 &&\Tr\brc{\Istr\ln\cO_U-\Istr\ln\brkt{\frac{1}{\xi_{\rm sc}}
 \Lvev{\Omg_{\rm v}^{1/3}\Omg_{\rm h}^{2/3}}\Pi_{\rm gf}}} \nonumber\\
 \eql \Istr\Pi_{3/2}\Tr\ln(-\cO_{3/2})
 +\Istr(\eta-\Pi_{3/2})\Tr\ln\cO_{\bar{U}} 
 -\Istr\Pi_{\rm gf}\Tr\ln\brkt{\Lvev{\Omg_{\rm v}^{1/3}\Omg_{\rm h}^{2/3}}}
 \nonumber\\
 \eql \frac{4}{\Box_4}\Tr\ln\cO_{3/2}
 -\frac{4\xi_{\rm sc}}{3\Box_4}\Tr\ln\brkt{
 \Lvev{\Omg_{\rm v}^{1/3}\Omg_{\rm h}^{2/3}}}+\cdots, 
\eea
where the ellipsis denotes terms independent of 
the background superfields. 
Thus, when $\xi_{\rm sc}=0$, $\Omg_{\rm eff}^U$ is calculated as 
\bea
 \Omg_{\rm eff}^U \eql \frac{i}{2(2\pi)^4}\int\dr^4p\;
 \frac{4}{-p^2}\Tr\ln\cO_{3/2}(p^2)+\cdots \nonumber\\
 \eql -\int\frac{d^4p_E}{(2\pi)^4}\;\frac{2}{p_E^2}
 \ln{\rm Det}\cO_{3/2}(-p_E^2)+\cdots, 
 \label{Omg_eff^U}
\eea
where $p_E^\mu\equiv(p^1,p^2,p^3,-ip^0)$ is the Wick-rotated Euclidean momentum, 
and ${\rm Det}$ is the functional determinant, which is expressed as 
\be
 \brkt{{\rm Det}\,\cO_{3/2}}^{-1/2} = \int\cD F_U\;
 \exp\brc{-\int_0^L\dr y\;F_U\cO_{3/2}F_U}.  \label{Det:O_32}
\ee
The integral variable~$F_U$ is a function of $y$, and can be expanded as 
\be
 F_U(y) = \sum_k f_U(y;\mu_U^{(k)})F_U^{(k)}, 
\ee
where $f_U(y;\mu_U)$ is an eigenfunction of $\cD_U$ defined in (\ref{def:cD_U}) 
with an eigenvalue~$\mu_U^2$, \ie, 
\be
 \cD_U f_U(y;\mu_U) = \mu_U^2 f_U(y;\mu_U).  \label{md_eq:U}
\ee
This has a form of the Sturm-Liouville equation. 
Thus the eigenfunctions satisfy the orthonormal condition, 
\be
 \int_0^L\dr y\;\Lvev{\Omg_{\rm v}^{1/3}\Omg_{\rm h}^{2/3}}
 f_U(y;\mu_U^{(k)})f_U(y;\mu_U^{(l)}) = \dlt_{kl}. 
\ee
Then, (\ref{Det:O_32}) is rewritten as 
\bea
 \brkt{{\rm Det}\,\cO_{3/2}}^{-1/2} \eql 
 \int\prod_k\cD F_U^{(k)}\;
 \exp\brc{-\sum_k F_U^{(k)}\brkt{p_E^2+\mu_U^{(k)2}}F_U^{(k)}} \nonumber\\
 \eql \prod_k\brkt{p_E^2+\mu_U^{(k)2}}^{-1/2}, 
\eea
up to an irrelevant normalization constant. 
Therefore, (\ref{Omg_eff^U}) becomes 
\bea
 \Omg_{\rm eff}^U \eql \int\frac{d^Dp_E}{(2\pi)^D}\;\frac{2}{p_E^2}
 \sum_k\ln\brkt{p_E^2+\mu_U^{(k)2}}+\cdots \nonumber\\
 \eql -\frac{2\Gm(1-\frac{D}{2})}{(4\pi)^{\frac{D}{2}}
 \brkt{\frac{D}{2}-1}}\sum_{k}\mu_U^{(k)D-2}+\cdots,   \label{Omg_eff^U2}
\eea
where $\Gm(z)$ is the gamma function. 
We have used the dimensional reduction~\cite{Siegel:1979wq} 
to regularize the divergent momentum integral.\footnote{
Since our formalism respects the superconformal symmetry, 
a momentum cutoff should not be introduced in contrast to Ref.~\cite{Brignole:2000kg}. 
}

\subsubsection{Contribution from vector superfields}
The contribution from the vector superfields is 
\be
 \Omg_{\rm eff}^V = \frac{i}{2(2\pi)^4}\int\dr^4p\;
 \Tr\brc{\Istr\ln\cO_V-\Istr\ln\brkt{\Lvev{\Omg_{\rm v}}a P_C}}, 
 \label{expr:Omg^vec}
\ee
where $a$ is the matrix defined in (\ref{def:a}), and 
\be
 \cO_V \equiv \cO_TP_T+\cO_CP_C. 
\ee
The second term in (\ref{expr:Omg^vec}) is a contribution 
from the ghost for $\dgt$. 
(See the gauge-fixing term~(\ref{GF:gg}).) 
Since 
\be
 \ln\cO_V = P_T\ln\cO_T+P_C\ln\cO_C, 
\ee
it follows that 
\bea
 &&\Tr\brc{\Istr\ln\cO_V-\Istr\ln\brkt{\Lvev{\Omg_{\rm v}}a P_C}} \nonumber\\
 \eql (\Istr P_T)\Tr\ln\cO_T+(\Istr P_C)\Tr\ln\cO_C
 -(\Istr P_C)\Tr\brkt{\Lvev{\Omg_{\rm v}}a} \nonumber\\
 \eql \frac{2}{\Box_4}\Tr\brc{\ln\cO_T-\ln\cO_C
 +\ln\brkt{\Lvev{\Omg_{\rm v}}a}}. 
\eea
When $\xi_{\rm sg}\to 0$, 
\be
 \ln\cO_C \to \ln\brkt{\frac{\Lvev{\Omg_{\rm v}}}{\xi_{\rm sg}}
 a\Box_4} = \ln\brkt{\Lvev{\Omg_{\rm v}}a}+\cdots, 
\ee
where the ellipsis denotes terms independent of 
the background superfields. 
Therefore, $\Omg_{\rm eff}^V$ is calculated as 
\bea
 \Omg_{\rm eff}^V \eql \frac{i}{(2\pi)^4}\int\dr^4p\;
 \frac{1}{-p^2}\Tr\ln\cO_T(p^2)+\cdots \nonumber\\
 \eql -\int\frac{d^4p_E}{(2\pi)^4}\;\frac{1}{p_E^2}\ln{\rm Det}\,\cO_T(-p_E^2)+\cdots. 
 \label{Omg_eff^vec}
\eea
Similarly to the derivation of (\ref{Omg_eff^U2}), this can be rewritten as 
\bea
 \Omg_{\rm eff}^V \eql -\int\frac{d^Dp_E}{(2\pi)^D}\;\frac{1}{p_E^2}
 \sum_k\ln\brkt{p_E^2+\mu_V^{(k)2}}+\cdots \nonumber\\
 \eql \frac{\Gm(1-\frac{D}{2})}{(4\pi)^{\frac{D}{2}}
 \brkt{\frac{D}{2}-1}}\sum_{k}\mu_V^{(k)D-2}+\cdots, \label{Omg_eff^V2}
\eea
where $\mu_V^{(k)2}$ are eigenvalues of $\cD_V$ defined in (\ref{expr:cOs}), \ie, 
\be
 (\cD_V)^I_{\;\;J} f_V^J(y;\mu_V) = \mu_V^2 f_V^I(y;\mu_V),  \label{md_eq:V}
\ee
and the eigenfunctions satisfy the orthonormal condition, 
\be
 \int_0^L\dr y\;\Lvev{\Omg_{\rm v}}a_{IJ}f_V^I(y;\mu_V^{(k)})f_V^J(y;\mu_V^{(l)}) 
 = \dlt_{kl}.  
\ee

\subsubsection{Contribution from chiral superfields}
The contribution from the chiral superfields is
\be
 \Omg_{\rm eff}^{\rm ch} = \frac{i}{2(2\pi)^4}\int\dr^4p\;
 \Tr\Istr\brkt{\bP\ln\cO_{\rm ch}},  \label{expr:Omg^ch}
\ee
where 
\be
 \bP \equiv \begin{pmatrix} P_+ & \\ & P_- \end{pmatrix}, \;\;\;\;\;
 \cO_{\rm ch} \equiv \begin{pmatrix} \cM & \bar{W}\frac{\bar{D}^2}{4\Box_4} \\
 W\frac{D^2}{4\Box_4} & \cM^t \end{pmatrix}. 
\ee
The chiral projection operators~$P_\pm$ are defined in (\ref{def:bP}). 
In (\ref{expr:Omg^ch}), $\bP$ is necessary 
because we have integrated the chiral fluctuation modes. 
Here, $\cO_{\rm ch}$ is rewritten as 
\be
 \cO_{\rm ch} = \begin{pmatrix} \cM & 0 \\ 0 & \cM^t \end{pmatrix}
 \brkt{\id+M_{\rm ch}}, 
\ee
where
\be
 M_{\rm ch} \equiv \begin{pmatrix} 0 & 
 \cM^{-1}\bar{W}\frac{\bar{D}^2}{4\Box_4} \\
 (\cM^t)^{-1}W\frac{D^2}{4\Box_4} & 0 \end{pmatrix}. 
\ee
Notice that $\cM$ is a normal matrix when $\xi_{\rm sg}=0$ 
(see (\ref{expr:cOs})), and 
\be
 \cM^{-1} \equiv \Lvev{\Omg_{\rm v}^{-1/3}\Omg_{\rm h}^{-2/3}}
 \begin{pmatrix} a^{IJ} & -\vev{\cV^I}\bdm{\Phi}^\dagger & 0 \\
 -\bdm{\Phi}\vev{\cV^J} & -\Lvev{\Omg_{\rm h}}e^{\sbdm{V}}\tl{d} 
 +\bdm{\Phi}\bdm{\Phi}^\dagger & 0 \\
 0 & 0 & -\Lvev{\Omg_{\rm h}}(e^{-\sbdm{V}})^t\tl{d} \end{pmatrix}, 
\ee
where $\vev{\cV^I}=-\der_y\bdm{V}^I$. 
When $\xi_{\rm sg}\neq 0$, $\cM$ becomes a differential operator matrix and 
$\cM^{-1}$ must be understood as the Green's function for it. 

Since only even powers of $M_{\rm ch}$ contribute to the trace, 
it follows that 
\bea
 \Tr\,\Istr\brkt{\bP\ln\cO_{\rm ch}} \eql \Tr\,\Istr
 \begin{pmatrix} P_+\ln\cM & 0 \\ 0 & P_-\ln\cM^t \end{pmatrix}
 +\Tr\,\Istr\brc{\bP\ln\brkt{\id+M_{\rm ch}}} \nonumber\\
 \eql -\frac{2}{\Box_4}\Tr\ln\cM
 +\frac{1}{2}\Tr\,\Istr\brc{\bP\ln\brkt{\id-M_{\rm ch}^2}} \nonumber\\
 \eql -\frac{1}{\Box_4}\Tr\brc{2\ln\det\cM
 +\tr\ln\brkt{1+\frac{\cM^{-1}\bar{W}(\cM^t)^{-1}W}{\Box_4}}}, 
\eea
where $\Tr$ in the third line denotes the trace over only the functional space 
of $y$. 
Therefore, (\ref{expr:Omg^ch}) is calculated as 
\bea
 \Omg_{\rm eff}^{\rm ch} \eql \frac{i}{2(2\pi)^4}\int\dr^4p\;
 \frac{1}{p^2}\Tr\brc{2\ln\det\cM
 +\tr\ln\brkt{-p^2+\cD_{\rm ch}}}+\cdots \nonumber\\
 \eql \int\frac{d^4p_E}{2(2\pi)^4}\;\frac{1}{p_E^2}\Tr\brc{2\ln\det\cM
 +\tr\ln\brkt{p_E^2+\cD_{\rm ch}}}+\cdots. 
 \label{Omg_eff^ch}
\eea
where 
\be
 \cD_{\rm ch} \equiv \cM^{-1}\bar{W}(\cM^t)^{-1}W.  \label{def:cD_ch}
\ee
Similarly to the derivation of (\ref{Omg_eff^U2}) or (\ref{Omg_eff^V2}), 
this can be rewritten as 
\bea
 \Omg_{\rm eff}^{\rm ch} \eql \int\frac{d^Dp_E}{2(2\pi)^D}\;
 \frac{1}{p_E^2}\sum_k\ln\brkt{p_E^2+\mu_{\rm ch}^{(k)2}}+\cdots \nonumber\\
 \eql -\frac{\Gm(1-\frac{D}{2})}{2(4\pi)^{\frac{D}{2}}
 \brkt{\frac{D}{2}-1}}\sum_{k}\mu_{\rm ch}^{(k)D-2}+\cdots, 
 \label{Omg_eff^ch2}
\eea
where $\mu_{\rm ch}^{(k)2}$ are eigenvalues of $\cD_{\rm ch}$, \ie, 
\be
 \cD_{\rm ch}f_{\rm ch}(y;\mu_{\rm ch}) = \mu_{\rm ch}^2f_{\rm ch}(y;\mu_{\rm ch}). 
 \label{md_eq:ch}
\ee

\ignore{
Here, $\det\cM$ is calculated as 
\bea
 \det\cM \eql \Lvev{\Omg_{\rm v}^{1/3}\Omg_{\rm h}^{2/3}}^{n_V+2n_C+2n_H}
 \det \begin{pmatrix} (a\cdot\cP_V)_{IJ} & -\frac{\cN_I}{3\cN}\Ups^\dagger \\
 -\frac{\cN_J}{3\cN}\Ups & -\frac{\tl{d}e^{-V}}{\Lvev{\Omg_{\rm h}}}
 +\frac{1}{3}\Ups\Ups^\dagger \end{pmatrix}
 \det\brkt{-\frac{\tl{d}(e^V)^t}{\Lvev{\Omg_{\rm h}}}}  \nonumber\\
 \eql \Lvev{\Omg_{\rm v}^{1/3}\Omg_{\rm h}^{2/3}}^{n_V+2n_C+2n_H}
 \det\brkt{-\frac{\tl{d}e^{-V}}{\Lvev{\Omg_{\rm h}}}
 +\frac{1}{3}\Ups\Ups^\dagger}
 \det\brkt{-\frac{\tl{d}(e^V)^t}{\Lvev{\Omg_{\rm h}}}} \nonumber\\
 &&\times\det\brc{(a\cdot\cP_V)_{IJ}-\frac{\cN_I}{3\cN}\Ups^\dagger
 \brkt{-\frac{\tl{d}e^{-V}}{\Lvev{\Omg_{\rm h}}}
 +\frac{1}{3}\Ups\Ups^\dagger}^{-1}\frac{\cN_J}{3\cN}\Ups} \nonumber\\
 \eql \Lvev{\Omg_{\rm v}^{1/3}\Omg_{\rm h}^{2/3}}^{n_V+2n_C+2n_H}
 \det\brkt{\frac{\id_{n_C+n_H}}{\Lvev{\Omg_{\rm h}}^2}}\cdot
 \det\brkt{\id-\frac{1}{3}\Phi_{\rm even}\Ups^\dagger}\cdot\det a 
 \nonumber\\
 \eql \frac{2}{3}\Lvev{\Omg_{\rm v}^{1/3}\Omg_{\rm h}^{2/3}}^{n_V+n_C+n_H}
 \Lvev{\Omg_{\rm v}^{1/3}\Omg_{\rm h}^{-4/3}}^{n_C+n_H}\det a, 
\eea
and 
\be
 \cM^{-1}\bar{W}(\cM^t)^{-1}W = \begin{pmatrix} 
 \cM_1^{-1}\bar{w}\bar{\cM}_2^{-1}w^t & \\ & \cM_2^{-1}\bar{w}^t\bar{\cM}_1^{-1}w 
 \end{pmatrix}, 
\ee
where
\bea
  \cM_1^{-1} \defa \Lvev{\Omg_{\rm v}^{-1/3}\Omg_{\rm h}^{-2/3}}
 \begin{pmatrix} a^{IJ} & -\vev{\cV^I}\bdm{\Phi}^\dagger \\
 -\bdm{\Phi}\vev{\cV^J} & -\Lvev{\Omg_{\rm h}}e^{\sbdm{V}}\tl{d} 
 +\bdm{\Phi}\bdm{\Phi}^\dagger \end{pmatrix}, \nonumber\\
 \cM_2^{-1} \defa -\Lvev{\frac{\Omg_{\rm v}}{\Omg_{\rm h}}}^{1/3}
 (e^{-\sbdm{V}})^t\tl{d}, \;\;\;\;\;
 w \equiv \begin{pmatrix} -\bdm{\Phi}^t\tl{d}\hat{t}_I^t \\ -\tl{d}\der_y 
 \end{pmatrix}. 
\eea
Thus, (\ref{expr:Omg^ch}) is calculated as 
\bea
 \Omg_{\rm eff}^{\rm ch} \eql \frac{i}{2(2\pi)^4}\int\dr^4p\;\frac{1}{p^2}
 \Tr\brc{2\ln\det\cM
 +\tr\ln\brkt{\Box_4+\cM^{-1}\bar{W}\bar{\cM}^{-1}W^t}}+\cdots \nonumber\\
 \eql \Tr\int\frac{d^4p_E}{32\pi^4}\frac{1}{p_E^2}\left[
 2\ln\det a+\tr\ln\brc{\Lvev{\Omg_{\rm v}^{1/3}\Omg_{\rm h}^{2/3}}^2
 \brkt{p_E^2+\cM_1^{-1}\bar{w}\bar{\cM}_2^{-1}w^t}} \right. \nonumber\\
 &&\hspace{30mm}\left.
 +\tr\brc{\Lvev{\Omg_{\rm v}^{1/3}\Omg_{\rm h}^{-4/3}}^2
 \brkt{p_E^2+\cM_2^{-1}\bar{w}^t\bar{\cM}_1^{-1}w}}\right]+\cdots \nonumber\\
 \eql \Tr\int\frac{d^4p_E}{32\pi^4}\frac{1}{p_E^2}
 \left[2\ln a+\frac{\ln\brkt{p_E^2+\cD_{12}}}
 {\Lvev{\Omg_{\rm v}^{2/3}\Omg_{\rm h}^{4/3}}} 
 +\Lvev{\frac{\Omg_{\rm h}^{8/3}}{\Omg_{\rm v}^{2/3}}}
 \ln\brkt{p_E^2+\cD_{21}} \right]+\cdots 
 \label{Omg_eff^ch}
\eea
where we have dropped terms independent of the background superfields, and 
\be
 \cD_{12} \equiv  \Lvev{\Omg_{\rm v}^{2/3}\Omg_{\rm h}^{4/3}}
 \cM_1^{-1}\bar{w}\bar{\cM}_2^{-1}w^t, \;\;\;\;\;
 \cD_{21} \equiv \Lvev{\frac{\Omg_{\rm v}^{2/3}}{\Omg_{\rm h}^{8/3}}}
 \cM_2^{-1}\bar{w}^t\bar{\cM}_1^{-1}w. 
\ee
At the last step in (\ref{Omg_eff^ch}), we have rescaled the Euclidean momentum~$p_E$, 
and $\Tr$ includes the trace over the matrix indices~$\tr$. 
Note that the first terms in (\ref{Omg_eff^vec}) and (\ref{Omg_eff^ch}), 
which are quadratically divergent, are canceled. 
In the global SUSY theories, 
In Ref.~\cite{Brignole:2000kg}, there are quadratic terms 
in the momentum cutoff scale~$\Lmd$, which stem from 
the non-minimal kinetic terms. 
In contrast, 
no appropriate cutoff can be introduced in the momentum integral 
because our formalism respects the superconformal symmetry. 
%
}

\subsubsection{Contribution from boundary actions}
Here we calculate the contributions to $\Omglp$ 
from the boundary Lagrangians~(\ref{quad:cL_bd}). 

The contribution from the gravitational superfields is 
\bea
 \Omg_{\rm eff}^{(y_*)U} \eql 
 \frac{i}{2(2\pi)^4}\int\dr^4p\;\lim_{\zeta_{\rm sc}^{(y_*)}\to 0}
 \brc{\Istr\ln\cO_U^{\rm bd}
 -\Istr\ln\brkt{\frac{\abs{\bdm{\phi}_C}^2h^{(y_*)}}
 {\zeta_{\rm sc}^{(y_*)}}\Pi_{\rm gf}(\zeta_{\rm sc}^{(y_*)})}}
 \nonumber\\
 \eql -\int\frac{d^4p_E}{8\pi^4}\frac{1}{p_E^2}\ln\brkt{\abs{\phi_C}^2h^{(y_*)}}
 +\cdots \nonumber\\
 \eql -\int\frac{dp_E^2}{8\pi^2}\ln\brkt{\abs{\phi_C}^2h^{(y_*)}}+\cdots, 
 \label{Omg_eff^bdU}
\eea
where the ellipsis denotes terms independent of the background superfields, and 
\be
 \cO_U^{\rm bd} = \abs{\bdm{\phi}_C}^2h^{(y_*)}\Box_4\brc{
 -\Pi_{3/2}^{\mu\nu}+\frac{1}{\zeta_{\rm sc}^{(y_*)}}
 \brkt{\eta^{\mu\nu}-\Pi_{3/2}^{\mu\nu}}}. 
\ee
Recall that our formalism respects the superconformal symmetry. 
Thus (\ref{Omg_eff^bdU}) is independent of the background superfields 
because their dependences can be absorbed by rescaling the momentum 
as $p_E^2\to p_E^2/\ln(\abs{\phi_C}^2h^{(y_*)})$. 

The contribution from the vector superfields is 
\bea
 \Omg_{\rm eff}^{(y_*)V} \eql \frac{i}{2(2\pi)^4}
 \int\dr^4p\;\lim_{\zeta_{\rm sg}^{(y_*)}\to 0}
 \tr\brc{\Istr\ln\brkt{\cO_T^{(y_*)}P_T
 +\cO_C^{(y_*)}P_C}-\Istr\ln\brkt{\Re f_{\Ie\Je}^{(y_*)}P_T}} 
 \nonumber\\
 \eql -\int\frac{d^4p_E}{16\pi^4}\frac{1}{p_E^2}
 \tr\brc{\ln\Re f^{(y_*)}+\ln\brkt{p_E^2+\cK_V^{2(y_*)}}}+\cdots, 
 \label{Omg_eff^bdV}
\eea
where  
\bea
 (\cO_T^{(y_*)})_{\Ie\Je} \defa \Re f^{(y_*)}_{\Ie\Ke}\brc{\dlt^{\Ke}_{\;\;\Je}\Box_4 P_T
 +\brkt{\cK_V^{2(y_*)}}^{\Ke}_{\;\;\Je}}, \nonumber\\
 (\cO_C^{(y_*)})_{\Ie\Je} \defa \Re f^{(y_*)}_{\Ie\Ke}
 \brc{\frac{\dlt^{\Ke}_{\;\;\Je}}{\zeta_{\rm sg}^{(y_*)}}\Box_4 P_T
 +\brkt{\cK_V^{2(y_*)}}^{\Ke}_{\;\;\Je}}, \nonumber\\
 \brkt{\cK_V^{2(y_*)}}^{\Ie}_{\;\;\Je} \defa -\frac{3}{2}F^{(y_*)\Ie\Ke}
 \abs{\bdm{\phi}_C}^2h^{(y_*)}_{\Ke\Je},  \label{def:cO_T:bd}
\eea
and $F^{(y_*)\Ie\Je}$ is an inverse matrix of $\Re f^{(y_*)}_{\Ie\Je}$. 
The first term in the second line of (\ref{Omg_eff^bdV}) is 
independent of the background superfields because they can be absorbed 
by the momentum rescaling. 
Thus, (\ref{Omg_eff^bdV}) becomes 
\bea
 \Omg_{\rm eff}^{(y_*)V} \eql -\int\frac{d^Dp_E}{(2\pi)^D}\;
 \frac{1}{p_E^2}\tr\ln\brkt{p_E^2+\cK_V^{2(y_*)}}+\cdots \nonumber\\
 \eql \frac{\Gm(1-\frac{D}{2})}{(4\pi)^{\frac{D}{2}}\brkt{\frac{D}{2}-1}}
 \tr\brkt{\cM_V^{2(y_*)}}^{\frac{D}{2}-1}+\cdots. 
\eea

Since the boundary Lagrangian in the chiral sector is written as
\be
 \cL_{\rm bd}^{(y_*)} = \int\dr^4\tht\;
 \phi_{(y_*)}^\dagger\begin{pmatrix} \cM_{(y_*)} & 
 \bar{W}_{(y_*)}\frac{\bar{D}^2}{4\Box_4} \\
 W_{(y_*)}\frac{D^2}{4\Box_4} & \cM_{(y_*)}^t \end{pmatrix}
 \phi_{(y_*)}+\cdots, 
\ee
where $\phi_{(y_*)}\equiv (\tl{\phi}_C,\tl{\chi}^a)$, and 
\bea
 \cM_{(y_*)} \defa -3\begin{pmatrix} h^{(y_*)} & 
 \bdm{\phi}_Ch_a^{(y_*)} \\ \bar{\bdm{\phi}}_Ch_{\bar{b}}^{(y_*)} & 
 \abs{\bdm{\phi}_C}^2h_{a\bar{b}}^{(y_*)} \end{pmatrix}, 
 \nonumber\\
 W_{(y_*)} \defa \begin{pmatrix} 3\bdm{\phi}_CP^{(y_*)} & 
 \frac{3}{2}\bdm{\phi}_C^2P_a^{(y_*)} \\
 \frac{3}{2}\bdm{\phi}_C^2P_b^{(y_*)} & 
 \frac{1}{2}\bdm{\phi}_C^3P_{ab}^{(y_*)} \end{pmatrix}, 
\eea
the contribution from the chiral superfields is 
\bea
 \Omg_{\rm eff}^{(y_*){\rm ch}} \eql \frac{i}{2(2\pi)^4}\int\dr^4p\;
 \frac{2}{p^2}\tr\brc{\ln\cM_{(y_*)}
 +\frac{1}{2}\ln\brkt{1-\frac{\cK^{2(y_*)}_{\rm ch}}{p^2}}} 
 \nonumber\\
 \eql \int\frac{d^Dp_E}{2(2\pi)^D}\;\frac{1}{p_E^2}
 \tr\ln\brkt{p_E^2+\cK_{\rm ch}^{2(y_*)}}+\cdots \nonumber\\
 \eql -\frac{\Gm(1-\frac{D}{2})}{2(4\pi)^{\frac{D}{2}}
 \brkt{\frac{D}{2}-1}}\tr\brkt{\cK_{\rm ch}^{2(y_*)}}^{\frac{D}{2}-1}+\cdots, 
 \label{Omg_eff^bdch}
\eea
where 
\be
 \cK_{\rm ch}^{2(y_*)} \equiv \cM_{(y_*)}^{-1}\bar{W}_{(y_*)}
 \brkt{\cM_{(y_*)}^t}^{-1}W_{(y_*)}.  \label{def:cM_ch:bd}
\ee
We have dropped the first term in the first line of (\ref{Omg_eff^bdch}) 
at the second equality because it can be absorbed by the momentum rescaling.

\subsection{Eigenvalues of differential operators} \label{Eigenvalues}
Here we derive equations satisfied by 
the eigenvalues of $\cD_F$ ($F=U,V,{\rm ch}$), which appear 
in (\ref{Omg_eff^U2}), (\ref{Omg_eff^V2}) and (\ref{Omg_eff^ch2}). 
\ignore{
The contributions from the bulk 
sector~(\ref{Omg_eff^U}), (\ref{Omg_eff^vec}) and (\ref{Omg_eff^ch}) 
contain the traces of the differential 
operators~$\cD_U$, $\cD_V$ and $\cD_{\rm ch}$ 
defined in (\ref{def:cD_U}), (\ref{expr:cOs}) and (\ref{def:cD_ch}), 
respectively. 
They can be calculated by summing up their eigenvalues. 
Let us denote the eigenfunctions of $\cD_F$ ($F=U,V,{\rm ch}$) as $f_F(y;\mu_F)$, 
which satisfy 
\bea
 \cD_U f_U(y;\mu_U) \eql \mu_U^2 f_U(y;\mu_U), \nonumber\\
 (\cD_V)^I_{\;\;J}f_V^J(y;\mu_V) \eql \mu_V^2 f_V^I(y;\mu_V), \nonumber\\
 \cD_{\rm ch}f_{\rm ch}(y;\mu_{\rm ch}) \eql \mu_{\rm ch}^2f_{\rm ch}(y;\mu_{\rm ch}), 
 \label{md_eq1}
\eea
where $\mu_F^2$ are the eigenvalues.\footnote{
Although $\mu_{\rm ch}$ is generally complex, 
we can always make it real and positive 
by redefining the phase of $f_{\rm ch}(y;\mu_{\rm ch})$. }
The last equation can be rewritten as 
\be
 \begin{pmatrix} \cM_1^{-1}\bar{w}\bar{\cM}_2^{-1}w^t & \\
 & \cM_2^{-1}\bar{w}^t\bar{\cM}_1^{-1}w \end{pmatrix}
 \begin{pmatrix} f_{\rm 12} \\ f_{\rm 21} \end{pmatrix}
 = \mu_{\rm ch}^2 \begin{pmatrix} f_{12}
 \\ f_{21} \end{pmatrix},  \label{ch:md_eq}
\ee
where $f_{\rm ch}=(f_{12},f_{21})^t$, and 
\bea
 \cM_1^{-1} \eql \Lvev{\Omg_{\rm v}^{-1/3}\Omg_{\rm h}^{-2/3}}
 \begin{pmatrix} a^{IJ} & -\vev{\cV^I}\bdm{\Phi}^\dagger \\
 -\bdm{\Phi}\vev{\cV^J} & -\Lvev{\Omg_{\rm h}}e^{\sbdm{V}}\tl{d}
 +\bdm{\Phi}\bdm{\Phi}^\dagger \end{pmatrix}, \nonumber\\
 \cM_2^{-1} \eql -\Lvev{\frac{\Omg_{\rm h}}{\Omg_{\rm v}}}^{1/3}
 (e^{-\sbdm{V}})^t\tl{d}, \;\;\;\;\;
 w \equiv \begin{pmatrix} -\bdm{\Phi}^t\tl{d}\hat{t}_I^t \\
 -\tl{d}\der_y \end{pmatrix}. 
\eea
}
Since we have already integrated out the fluctuation superfields, 
we rewrite the background superfields~$\bdm{\Phi}$ and $\bdm{V}$ 
as $\Phi_{\rm even}$ and $V$ in the following. 
From the procedure summarized in Appendix~\ref{tree:action},
we see that $\Phi_{\rm even}$ and $V^{\Ie}$ are independent of $y$ 
while $V^{\Io}$ have nontrivial $y$-dependences. 
As explained in Appendix~\ref{tree:Kahler}, 
such $y$-dependences cannot be determined 
by the equations of motion~\cite{Abe:2006eg}. 
Instead, their functional forms are determined 
when they are regarded as functions of 
$V_s$ defined by 
\be
 V_s \equiv s_{\Io}V^{\Io}, \label{def:V_s}
\ee
where $s_{\Io}$ are arbitrarily chosen constants~\cite{Abe:2008an,Abe:2011rg}. 
This has the following boundary conditions. 
\be
 V_s|_{y=0} = 0, \;\;\;\;\;
 \lim_{y\to L}V_s = \bar{V}_s \equiv -2s_{\Io}\Re T^{\Io}. 
\ee
As we will explicitly see in the next section, 
the $s_{\Io}$-dependences are canceled in the final result. 

In order to rewrite the eigenvalue equations as differential equations for $V_s$, 
we rescale $\tl{\Sgm}^{\Io}$ as 
\be
 \tl{\Sgm}^I \to \hat{\Sgm}^I \equiv \frac{\tl{\Sgm}^I}
 {s_{\Io}\vev{\cV^{\Io}}} 
 = -\frac{\tl{\Sgm}^I}{\der_y V_s}.  \label{rescale:Sgm}
\ee
Then, ~(\ref{md_eq:U}), (\ref{md_eq:V}) and (\ref{md_eq:ch}) are rewritten as
\bea
 \tl{D}_U\tl{f}_U(V_s;\mu_U) \eql \mu_U^2\tl{f}_U(V_s;\mu_U), \nonumber\\
 \tl{\cD}_V\tl{f}_V(V_s;\mu_V) \eql \mu_V^2\tl{f}_V(V_s;\mu_V), \nonumber\\
 \tl{\cD}_1\bar{\tl{\cD}}_2\tl{f}_{12}(V_s;\mu_{\rm ch}) 
 \eql \mu_{\rm ch}^2\tl{f}_{12}(V_s;\mu_{\rm ch}), \nonumber\\
 \tl{\cD}_2\bar{\tl{\cD}}_1\tl{f}_{21}(V_s;\mu_{\rm ch}) 
 \eql \mu_{\rm ch}^2\tl{f}_{21}(V_s;\mu_{\rm ch}),  \label{md_eq2}
\eea
where 
\bea
 f_U(y;\mu_U) \eql \tl{f}(V_s(y);\mu_U), \;\;\;\;\;
 f_V(y;\mu_V) = \tl{f}(V_s(y);\mu_V), \nonumber\\
 f_{\rm ch}(y;\mu_{\rm ch}) \eql 
 \begin{pmatrix} \tl{f}_{12}(V_s(y);\mu_{\rm ch}) \\
 \tl{f}_{21}(V_s(y);\mu_{\rm ch}) \end{pmatrix}, 
\eea
and
\bea
 \tl{\cD}_U \defa -\frac{1}{\Lvev{
 \Omg_{\rm v}^{1/3}\Omg_{\rm h}^{2/3}}}
 \der_{V_s}\brkt{\Lvev{
 \frac{\Omg_{\rm h}^{4/3}}{\Omg_{\rm v}^{1/3}}}\der_{V_s}}, 
 \nonumber\\
 (\tl{\cD}_V)^I_{\;\;J} \defa 
 -\frac{a^{IK}}{\Lvev{\Omg_{\rm v}}}\der_{V_s}
 \brc{\Lvev{\Omg_{\rm v}^{1/3}\Omg_{\rm h}^{2/3}}
 \brkt{(a\cdot\cP_V)_{KJ}\der_{V_s}
 -\frac{\cN_K}{3\cN}\Ups_J^\dagger\Phi_{\rm even}}} \nonumber\\
 &&-\Lvev{\frac{\Omg_{\rm h}}{\Omg_{\rm v}}}^{2/3}a^{IK}
 \brkt{\frac{\cN_J}{3\cN}\Ups_K^\dagger\Phi_{\rm even}\der_{V_s}
 +\Lvev{\frac{\der_K\der_J\Omg_{\rm h}}{\Omg_{\rm h}}}
 -\frac{\Ups_K^\dagger\bdm{\Phi}\bdm{\Phi}^\dagger\Ups_J}{3}}, \nonumber\\
 \tl{\cD}_1 \defa \frac{1}{\Lvev{\Omg_{\rm v}^{1/3}\Omg_{\rm h}^{2/3}}}
 \begin{pmatrix} -a^{IJ}\Phi_{\rm even}^\dagger\tl{d}\hat{t}_J
 -v^I\Phi^\dagger_{\rm even}\tl{d}\der_{V_s} \\
 \Phi_{\rm even}\Phi_{\rm even}^\dagger\tl{d}v
 -\brkt{\Lvev{\Omg_{\rm h}}e^V-\Phi_{\rm even}\Phi_{\rm even}^\dagger\tl{d}}
 \der_{V_s} \end{pmatrix}, \nonumber\\
 \tl{\cD}_2 \defa \Lvev{\frac{\Omg_{\rm h}}{\Omg_{\rm v}}}^{1/3}
 (e^{-V})^t\brkt{\hat{t}_I^t\bar{\Phi}_{\rm even},\der_{V_s}}, 
 \;\;\;\;\;
 v^{\Io} \equiv \frac{\der_y V^{\Io}}{\der_y V_s}, \;\;\;\;\;
 v \equiv v^{\Io}\hat{t}_{\Io}. 
\eea
Here, $\tl{\cD}_1$ and $\tl{\cD}_2$ are $(n_V+n_C+n_H)\times(n_C+n_H)$ 
and $(n_C+n_H)\times (n_V+n_C+n_H)$ matrices, respectively, 
and the arguments of the norm function and its derivatives 
are $(\bdm{0}_{n_{V_{\rm e}}},v^{\Io})$. 
As explained in Appendix~\ref{tree:Kahler}, 
the $V_s$-dependences of $v^{\Io}$ and $V^{\Io}$ are determined 
by the equations of motion for the background superfields. 
Therefore, the $V_s$-dependences of $\tl{\cD}_F$ ($F=U,V,1,2$) 
are already known after deriving the tree-level K\"ahler potential. 

The boundary conditions are obtained from (\ref{bdcd:UV}) 
and (\ref{bdcd:chiral2}) as
\be
 \left.\brc{\cA_F^{(y_*)}\der_{V_s}-\cB_F^{(y_*)}}\tl{f}_F\right|_{y=y_*} 
 = 0, \;\;\;\;\; (F=U,V,{\rm ch})  \label{BC:tl}
\ee
where $\tl{f}_{\rm ch}\equiv (\tl{f}_{12},\tl{f}_{21})^t$, 
$\cA_{\rm ch}^{(y_*)}$ and $\cB_{\rm ch}^{(y_*)}$ are 
defined in (\ref{def:AB}), and 
\bea
 \cA_U^{(y_*)} \defa \Lvev{\Omg_{\rm v}^{-1/3}\Omg_{\rm h}^{4/3}}, \;\;\;\;\;
 \cB_U^{(y_*)} \equiv \eta_{y_*}\abs{\phi_C}^2h^{(y_*)}\mu_U^2, \nonumber\\
 \brkt{\cA_V^{(y_*)}}_{\Ie\Je} \defa 
 \Lvev{\Omg_{\rm v}^{1/3}\Omg_{\rm h}^{2/3}}a_{\Ie\Je}, \;\;\;\;\;
 \brkt{\cB_V^{(y_*)}}_{\Ie\Je} \equiv \eta_{y_*}\brkt{\Re f_{\Ie\Je}^{(y_*)}\mu_V^2
 +\frac{2}{3}\abs{\phi_C}^2h_{\Ie\Je}^{(y_*)}}, \nonumber\\
 \brkt{\cA_V^{(y_*)}}_{\Ie\Jo} \defa \brkt{\cB_V^{(y_*)}}_{\Ie\Jo} 
 = \brkt{\cA_V^{(y_*)}}_{\Io J} = 0, \;\;\;\;\;
 \brkt{\cB_V^{(y_*)}}_{\Io J} \equiv \dlt_{\Io J},  \label{def:cAB}
\eea
where $\eta_0=1$ and $\eta_L=-1$. 
We have used $p^2=\mu_F^2$ ($F=U,V$), which follows 
from the bulk equations of motion. 
The arguments of the norm function and its derivatives are understood as 
$(\bdm{0}_{n_{V_{\rm e}}},v^{\Io}|_{y=y_*})$, 
and $h^{(y_*)}$, $P^{(y_*)}$, $f^{(y_*)}$ and their derivatives 
are functions of $\bdm{\chi}^a$, which can be either the boundary values 
of the bulk superfields or 4D superfields localized on the boundaries. 

Note that $\tl{\cD}_V$, $\tl{\cD}_1\tl{\cD}_2$ and 
$\tl{\cD}_2\tl{\cD}_1$ can be expressed in the following forms. 
\bea
 (\tl{\cD}_V)^I_{\;\;J} \eql -\Lvev{\frac{\Omg_{\rm h}}{\Omg_{\rm v}}}^{2/3}
 \brc{\cP_V\der_{V_s}^2
 +A_V\der_{V_s}+B_V}^I_{\;\;J}, \nonumber\\
 \tl{\cD}_1\bar{\tl{\cD}}_2 \eql -\Lvev{\frac{\Omg_{\rm h}}{\Omg_{\rm v}}}^{2/3}
 \brc{\begin{pmatrix} 0 & v^I\Ups^\dagger \\
 0 & \cP_{\!{\rm ch}} \end{pmatrix}\der_{V_s}^2
 +A_{12}\der_{V_s}+B_{12}}, \nonumber\\
 \tl{\cD}_2\bar{\tl{\cD}}_1 \eql 
 -\Lvev{\frac{\Omg_{\rm h}}{\Omg_{\rm v}}}^{2/3}
 \brc{(e^{-V})^t\bar{\cP}_{\!{\rm ch}}(e^V)^t\der_{V_s}^2
 +A_{21}\der_{V_s}+B_{21}}^a_{\;\;b}, 
\eea
where matrices~$A_F$ and $B_F$ ($F=V,12,21$) 
are functions of the background superfields, and 
\be
 \cP_{\!{\rm ch}} \equiv \id_{n_C+n_H}-\Phi_{\rm even}\Ups^\dagger  
\ee
is a projection operator that satisfies 
\be
 \cP_{\!{\rm ch}}\Phi_{\rm even} = 0, \;\;\;\;\;
 \Ups^\dagger\cP_{\!{\rm ch}} = 0, \;\;\;\;\;
 \cP_{\!{\rm ch}}^2 = \cP_{\!{\rm ch}}. 
\ee
Hence (\ref{md_eq2}) is rewritten as
\bea
 \brc{\der_{V_s}^2+\der_{V_s}\ln\Lvev{
 \frac{\Omg_{\rm h}^{4/3}}{\Omg_{\rm v}^{1/3}}}\der_{V_s}
 +\Lvev{\frac{\Omg_{\rm v}}{\Omg_{\rm h}}}^{2/3}\mu_U^2}\tl{f}_U \eql 0, 
 \nonumber\\
 \brc{\cP_V\der_{V_s}^2+A_V\der_{V_s}+\tl{B}_V}
 \tl{f}_V \eql 0, \nonumber\\
 \brc{\cP_{12}\der_{V_s}^2+\tl{A}_{12}\der_{V_s}+\tl{B}_{12}}\tl{f}_{12} 
 \eql 0,  \nonumber\\
 \brc{(e^{-V})^t\bar{\cP}_{\!{\rm ch}}(e^V)^t\der_{V_s}^2
 +A_{21}\der_{V_s}+\tl{B}_{21}}\tl{f}_{21} \eql 0,  \label{md_eq3}
\eea
where 
\bea
 \tl{B}_V \defa B_V+\Lvev{\frac{\Omg_{\rm v}}{\Omg_{\rm h}}}^{2/3}\mu_V^2\id,  
 \;\;\;\;\; 
 \tl{B}_{21} \equiv B_{21}+\Lvev{\frac{\Omg_{\rm v}}{\Omg_{\rm h}}}^{2/3}
 \mu_{\rm ch}^2\id,   \nonumber\\
 \cP_{12} \defa \begin{pmatrix} \bdm{0}_{n_V} & 0 
 \\ 0 & \id_{n_C+n_H} \end{pmatrix},  \;\;\;\;\;
 \tl{A}_{12} \equiv \begin{pmatrix} \cP_V & 0 \\
 \Phi_{\rm even}\frac{\cN_J}{3\cN} & \id_{n_C+n_H} \end{pmatrix}A_{12}, 
 \nonumber\\
 \tl{B}_{12} \defa \begin{pmatrix} \cP_V & 0 \\
 \Phi_{\rm even}\frac{\cN_J}{3\cN} & \id_{n_C+n_H} \end{pmatrix}
 \brkt{B_{12}+\Lvev{\frac{\Omg_{\rm v}}{\Omg_{\rm h}}}^{2/3}\mu_{\rm ch}^2\id}. 
\eea
The appearance of the projection operators~$\cP_V$ 
and $\cP_{\!{\rm ch}}$ in (\ref{md_eq3}) reflects the fact that the graviphoton 
and the compensator superfields are unphysical 
in the superconformal formulation, while that of $\cP_{12}$ stems from 
the fact that $\tl{\Sgm}^I$ do not propagate 
in the super-Landau gauge~$\xi_{\rm sg}=0$. 

Solving (\ref{md_eq3}) with the boundary conditions at $y=0$ (\ie, $V_s=0$), 
we can express $\tl{f}_F(V_s)$ ($F=U,V,{\rm ch}$) in the form of 
\be
 \tl{f}_F(V_s) = \cC_F(V_s;\mu_F)\cdot N_F, 
\ee
where $\cC_F(V_s;\mu_F)$ are matrices that depend on 
the background superfields, and $N_F$ is an integration constant vector. 
(See eq.(\ref{tl:f:y0}) in the next section.)
Then the boundary conditions at $y=L$ (\ie, $V_s=\bar{V}_s$) are rewritten as 
\be
 \cQ_F(\mu_F)\cdot N_F = 0, \;\;\;\;\; (F=U,V,{\rm ch}) 
 \label{cMN}
\ee
where $\cQ_F(\mu_F)\equiv\left.\brkt{
\cB_F^{(L)-1}\cA_F^{(L)}\der_{V_s}-1}\cC_F\right|_{V_s=\bar{V}_s}$. 
Due to the presence of the projection operators in (\ref{md_eq3}), 
the constant vectors~$N_F$ ($F=V,{\rm ch}$) belong to 
projected spaces~${\rm PS}_F$. 
The eigenvalues~$\mu_F$ are determined by the conditions that 
(\ref{cMN}) has solutions with non-vanishing $N_F$, \ie, 
\be
 \cF_F(\mu_F) \equiv {\rm det}_{{\rm PS}_F}\cQ_F(\mu_F) = 0, 
 \;\;\;\;\; (F=U,V,{\rm ch})  \label{def:cG_F}
\ee
where ${\rm det}_{{\rm PS}_F}$ is the determinant 
restricted to the projected space~${\rm PS}_F$.

\subsection{Expression of One-loop K\"ahler potential} \label{Expr:Omglp}
Now we obtain the desired expression of 
the one-loop K\"ahler potential by summing up the contributions 
in Sec.~\ref{Int:flctuation}. 
\bea
 \Omglp \eql -\frac{\Gm(1-\frac{D}{2})}{(4\pi)^{\frac{D}{2}}\brkt{\frac{D}{2}-1}}
 \sum_{y_*=0,L}\sum_{F=V,{\rm ch}}g_F\tr\brkt{\cK_F^{2(y_*)}}^{\frac{D}{2}-1}
 \nonumber\\
 &&-\frac{\Gm(1-\frac{D}{2})}{(4\pi)^{\frac{D}{2}}\brkt{\frac{D}{2}-1}}
 \sum_{F=U,V,{\rm ch}}g_F\sum_k\brkt{\mu_F^{(k)}}^{D-2}+\cdots, 
 \label{expr:Omg_eff}
\eea
where $g_U=-2$, $g_V=-1$, $g_{\rm ch}=1/2$, and $\mu_F^2$ ($F=U,V,{\rm ch}$) 
are solutions of (\ref{def:cG_F}). 

The first line of (\ref{expr:Omg_eff}) is the contributions 
from the boundary actions, and is rewritten as 
\bea
 \Omglp \eql \sum_{y_*=0,L}\sum_{F=V,{\rm ch}}\frac{g_F}{16\pi^2}\tr\sbk{
 \cK_F^{2(y_*)}\brc{\frac{2}{4-D}-\gm+\ln(4\pi)
 -\ln\cK_F^{2(y_*)}+2}} \nonumber\\
 &&+\cO((D-4)^2)+\cdots,  \label{Omglp_bd}
\eea
where $\gm$ is the Euler's constant, and 
the matrices~$\cK_V^{2(y_*)}$ and $\cK_{\rm ch}^{2(y_*)}$ are defined by 
(\ref{def:cO_T:bd}) and (\ref{def:cM_ch:bd}), respectively. 
The divergence will be renormalized by local counterterms 
in the boundary Lagrangians~$\cL_{\rm bd}^{(y_*)}$ ($y_*=0,L$). 
\ignore{
Thus the renormalized K\"ahler potential in the $\overline{\rm MS}$ scheme is 
expressed as 
\bea
 \Omglp \eql -\sum_{y_*=0,L}\sum_{F=V,{\rm ch}}\frac{g_F}{16\pi^2}
 \tr\brc{\cK_F^{2(y_*)}\brkt{\ln\frac{\cK_F^{2(y_*)}}{\abs{\phi_C}^2\mu^2}-1}}+\cdots, 
\eea
where $\mu$ is an arbitrary constant that corresponds to 
the renormalization scale.\footnote{
There is no mass scale in the theory until the superconformal gauge-fixing conditions 
are imposed. }
The number~``$-1$'' in the parenthesis can be other number 
by rescaling $\mu$. 
The compensator superfield~$\phi_C$ that explicitly appears 
in the argument of the logarithm is necessary 
in order to compensate the Weyl weight. 
}

The summation of the eigenvalues in the second line of (\ref{expr:Omg_eff}) 
can be performed by the technique of 
Refs.~\cite{Garriga:2000jb}-\cite{Sakamura:2010ju}. 
\bea
 \sum_{k}\brkt{\mu_F^{(k)}}^{D-2} 
 \eql \oint_{C_F}\frac{dz}{2\pi i}\;\frac{\cF_F'(z)}{\cF_F(z)}z^{D-2} \nonumber\\
 \eql -\frac{D-2}{\pi}\sin\frac{\pi D}{2}
 \int_0^\infty\dr\lmd\;\lmd^{D-3}\ln\frac{\cF_F(i\lmd)}{\cF_F^{\rm asp}(i\lmd)} 
 \nonumber\\
 \eql -\frac{(D-2)}{\Gm(1-\frac{D}{2})\Gm(\frac{D}{2})}
 \int_0^\infty\dr\lmd\;\lmd^{D-3}\ln\frac{\cF_F(i\lmd)}{\cF_F^{\rm asp}(i\lmd)}, 
\eea
where the functions~$\cF_F$ are defined by (\ref{def:cG_F}), 
$C_F$ are contours that enclose the zeros of $\cF_F(z)$, 
and $\cF_F^{\rm asp}(z)$ are some analytic functions that satisfy 
\be
 \frac{\cF_F(z)}{\cF_F^{\rm asp}(z)} = 1+\cO(z^{-1}), 
\ee
for $\Im z\gg 1$. 
We can rescale $z$ by a superfield-dependent factor~$C_{\rm rs}$ 
so that $\tl{\cF}_F^{\rm asp}(z)\equiv\cF_F^{\rm asp}(C_{\rm rs}z)$ 
become independent of the superfields. 
(See (\ref{eg:cF_F^asp}).)

Therefore, $\Omglp$ is expressed as  
\bea
 \Omglp \eql \Omg_0+\Omg_L
 +\sum_{F=U,V,{\rm ch}}\frac{g_F C_{\rm rs}^2}{8\pi^2}
 \int_0^\infty\dr\lmd\;\lmd\ln\frac{\tl{\cF}_F(i\lmd)}{\tl{\cF}_F^{\rm asp}(i\lmd)},  
 \label{expr:Omg_eff^2}
\eea
where $\tl{\cF}_F(z)\equiv\cF_F(C_{\rm rs}z)$, and $\Omg_0$ and $\Omg_L$ 
are the contributions from the boundary actions at $y=0$ and $y=L$, 
which are expressed in (\ref{Omglp_bd}). 
This is our main result. 
The explicit forms of $\tl{\cF}_F(z)$, $\tl{\cF}_F^{\rm asp}(z)$, 
and $C_{\rm rs}$ are highly model-dependent, 
but we can easily find them once a model is specified. 
We will show their explicit forms in a specific case in the next section. 
The bulk contribution in (\ref{expr:Omg_eff^2}) 
also contains divergent terms. 
Such terms originate from one-loop diagrams localized on the boundaries, 
and should be absorbed into $\Omg_0$ and $\Omg_L$. 
(See the next section.)

\section{Case of flat spacetime} \label{Flat_case}
In this section, we consider a case where the spacetime geometry is flat 
and $n_C=1$ as an illustrative example. 
Namely, the compensator multiplet is neutral for the gauge symmetries, 
and the generators have the following form, 
\be
 \hat{t}_I = \begin{pmatrix} 0 & \\ & \check{t}_I \end{pmatrix}. 
\ee
Notice that $\check{t}_I$ contain the gauge coupling. 
In this case, from (\ref{fcn:v}) and (\ref{fcn:Vv}) 
in Appendix~\ref{tree:action}, we find 
\be
 v^{\Io}(V_s) = \bar{v}^{\Io}+\cO(\chi^2), \;\;\;\;\;
 V^{\Io}(V_s) = \bar{v}^{\Io}V_s+\cO(\chi^2), 
\ee
where 
\be
 \bar{v}^{\Io} \equiv \frac{\Re T^{\Io}}{s\cdot\Re T} 
 = -\frac{2\Re T^{\Io}}{\bar{V}_s}, \;\;\;\;\;
 \chi^a \equiv \frac{\Phi^{a+1}_{\rm even}}{\Phi^1_{\rm even}}. 
\ee
Therefore, we obtain 
\bea
 \Lvev{\Omg_{\rm v}} \eql -\frac{\cN(2\Re T)}{\bar{V}_s^3}
 +\cO(\chi^2),  \;\;\;\;\;
 \Lvev{\Omg_{\rm h}} = \abs{\phi_C}^3+\cO(\chi^2), \nonumber\\
  \hat{t}_I\Phi_{\rm even} \eql \cO(\chi^2), \;\;\;\;\;
 \cP_{\!{\rm ch}} = \begin{pmatrix} 0 & \\ & \id_{n_H} \end{pmatrix}
 +\cO(\chi^2), 
\eea
where $\phi_C \equiv \brkt{\Phi^1_{\rm even}}^{2/3}$. 

In the following, we do not see the dependences on $\chi^a$ coming from 
the bulk hypermultiplets, for simplicity. 
Then (\ref{md_eq3}) becomes simple. 
\bea
 &&\brc{\der_{V_s}^2+\avh^2\mu_U^2}\tl{f}_U = 0, \nonumber\\
 &&\brc{\cP_V\der_{V_s}^2+\avh^2\mu_V^2}\tl{f}_V = 0, 
 \nonumber\\
 &&\begin{pmatrix} \avh^2\mu_{\rm ch}^2(\cP_V)^I_{\;\;J} & 0 \\ 
 \avh^2\mu_{\rm ch}^2\Phi_{\rm even}^a\frac{\cN_J}{3\cN} & 
 \brkt{\der_{V_s}^2-\bar{v}\der_{V_s}+\avh^2\mu_{\rm ch}^2}\dlt^a_{\;\;b} 
 \end{pmatrix}
 \begin{pmatrix} \tl{f}_{12}^J \\ \tl{f}_{12}^b \end{pmatrix} 
 = 0, \nonumber\\
 &&\brc{\bar{\cP}_{\!{\rm ch}}\der_{V_s}^2
 +\bar{v}^t\der_{V_s}+\avh^2\mu_{\rm ch}^2}\tl{f}_{21} = 0,  
 \label{eg:md_eq}
\eea
where $\bar{v} \equiv \bar{v}^{\Io}\hat{t}_{\Io}$ and 
\be
 \avh \equiv \Lvev{\frac{\Omg_{\rm v}}{\Omg_{\rm h}}}^{1/3}
 = -\frac{\cN^{1/3}(2\Re T)}{\abs{\phi_C}\bar{V}_s}. 
\ee
Solutions of (\ref{eg:md_eq}) that satisfy the boundary conditions at $y=0$ 
are found to be 
\bea
 \tl{f}_U(V_s;\mu_U) \eql \brc{
 r_s\mu_U\cos\brkt{r_s\mu_U V_s}\cB_U^{(0)-1}\cA_U^{(0)}
 +\sin\brkt{r_s\mu_U V_s}}N_U,  \nonumber\\
 \tl{f}_V(V_s\;\mu_V) \eql 
 \cP_V\brc{r_s\mu_V\cos\brkt{r_s\mu_V V_s}\cB_V^{(0)-1}\cA_V^{(0)}
 +\sin\brkt{r_s\mu_V V_s}}N_V, 
 \nonumber\\
 \tl{f}_{\rm ch}(V_s;\mu_{\rm ch}) \eql P_{\rm ch}e^{UV_s}\left\{
 \omg_{\rm ch}\cos\brkt{\omg_{\rm ch}V_s}\cB_{\rm ch}^{(0)-1}\cA_{\rm ch}^{(0)} 
 \right. \nonumber\\
 &&\hspace{15mm}\left. 
 +\sin\brkt{\omg_{\rm ch}V_s}\brkt{\id-
 U\cB_{\rm ch}^{(0)-1}\cA_{\rm ch}^{(0)}}\right\}N_{\rm ch}, 
 \label{tl:f:y0}
\eea
where $N_F$ ($F=U,V,{\rm ch}$) are constant vectors, 
$\cA^{(0)}_F$ and $\cB^{(0)}_F$ are defined in (\ref{def:cAB}) 
and (\ref{def:AB}), and 
\bea
 P_{\rm ch} \defa \begin{pmatrix} \bdm{0}_{n_V} & & \\
 & \cP_{\!{\rm ch}} & \\
 & & \bar{\cP}_{\!{\rm ch}} \end{pmatrix}, \;\;\;\;\;
 U \equiv \frac{1}{2}\begin{pmatrix} \bdm{0}_{n_V} & & \\ & \bar{v} & \\
 & & -\bar{v}^t \end{pmatrix}, \nonumber\\
 \omg_{\rm ch} \defa \brc{\avh^2\mu_{\rm ch}^2 P_{\rm ch}-U^2}^{1/2}. 
\eea
Notice that $\cP_V$ and $P_{\rm ch}\omg_{\rm ch}$ commute with 
$\cB_V^{(0)-1}\cA_V^{(0)}$, and 
$\cB_{\rm ch}^{(0)-1}\cA_{\rm ch}^{(0)}$, respectively, 
and $P_{\rm ch}\omg_{\rm ch}U=U\omg_{\rm ch}$, 
in the present case. 
The explicit forms of $\cB_{\rm ch}^{(0)-1}\cA_{\rm ch}^{(0)}$ 
and $\cB_{\rm ch}^{(L)-1}\cA_{\rm ch}^{(L)}$ are calculated 
from (\ref{def:AB}) as 
\bea
 \cB_{\rm ch}^{(0)-1}\cA_{\rm ch}^{(0)} 
 \eql \frac{1}{r_s\mu_{\rm ch}}\begin{pmatrix} 0 & 0 & 
 -\frac{\bar{v}^{\Io}}{\bar{\phi}_C^{3/2}}(\id_{n_H+1}-\cP_{\!{\rm ch}}) \\
 0 & -G^{(0)}\tl{d} & 0 \\ 
 0 & \id_{n_H+1} & 0 \end{pmatrix}, \nonumber\\
 \cB_{\rm ch}^{(L)-1}\cA_{\rm ch}^{(L)}
 \eql \frac{1}{r_s\mu_{\rm ch}}\begin{pmatrix} 0 & 0 & 
 -\frac{\bar{v}^{\Io}}{\bar{\phi}_C^{3/2}}(\id_{n_H+1}-\cP_{\!{\rm ch}}) \\
 0 & G^{(L)}\tl{d}e^{-\bar{v}\bar{V}_s} & 0 \\ 
 0 & e^{-\bar{v}\bar{V}_s} & 0 \end{pmatrix}. 
\eea
where 
\be
 G^{(y_*)} \equiv \brkt{\bar{\cM}_{\rm bd}^{(y_*)}\mu_{\rm ch}
 -W_{\rm bd}^{(y_*)}}^{-1}. 
\ee

Using (\ref{tl:f:y0}), we obtain the expressions of $\cQ_F(\mu_F)$ in (\ref{cMN}) as 
\bea
 \cQ_U(\mu_U) \eql r_s\mu_U
 \brkt{\cB_U^{(L)-1}\cA_U^{(L)}-\cB_U^{(0)-1}\cA_U^{(0)}}
 \cos\brkt{r_s\mu_U\bar{V}_s} \nonumber\\
 &&-\brkt{r_s^2\mu_U^2\cB_U^{(L)-1}\cA_U^{(L)}\cB_U^{(0)-1}\cA_U^{(0)}+1}
 \sin\brkt{r_s\mu_U\bar{V}_s}, \nonumber\\
 \cQ_V(\mu_V) \eql \cP_V\left\{
 r_s\mu_V
 \brkt{\cB_V^{(L)-1}\cA_V^{(L)}-\cB_V^{(0)-1}\cA_V^{(0)}}
 \cos\brkt{r_s\mu_V\bar{V}_s} \right. \nonumber\\
 &&\hspace{5mm}\left.
 -\brkt{r_s^2\mu_V^2\cB_V^{(L)-1}\cA_V^{(L)}\cB_V^{(0)-1}\cA_V^{(0)}+1}
 \sin\brkt{r_s\mu_V\bar{V}_s}\right\},  \nonumber\\
 \cQ_{\rm ch}(\mu_{\rm ch}) \eql P_{\rm ch}\left\{
 \cB_{\rm ch}^{(L)-1}\cA_{\rm ch}^{(L)}e^{U\bar{V}_s}
 \omg_{\rm ch}\cos\brkt{\omg_{\rm ch}\bar{V}_s}
 -e^{U\bar{V}_s}\omg_{\rm ch}\cos\brkt{\omg_{\rm ch}\bar{V}_s}
 \cB_{\rm ch}^{(0)-1}\cA_{\rm ch}^{(0)}
 \right. \nonumber\\
 &&\hspace{5mm}
 -r_s^2\mu_{\rm ch}^2\cB_{\rm ch}^{(L)-1}\cA_{\rm ch}^{(L)}
 e^{U\bar{V}_s}\sin\brkt{\omg_{\rm ch}\bar{V}_s}
 \cB_{\rm ch}^{(0)-1}\cA_{\rm ch}^{(0)}
 -e^{U\bar{V}_s}\sin\brkt{\omg_{\rm ch}\bar{V}_s} \nonumber\\
 &&\hspace{5mm}\left.
 +\cB_{\rm ch}^{(L)-1}\cA_{\rm ch}^{(L)}e^{U\bar{V}_s}
 U\sin\brkt{\omg_{\rm ch}\bar{V}_s}
 +e^{U\bar{V}_s}U\sin\brkt{\omg_{\rm ch}\bar{V}_s}
 \cB_{\rm ch}^{(0)-1}\cA_{\rm ch}^{(0)}\right\}.  
\eea
Thus, $\cF_F(\mu_F)$ in (\ref{def:cG_F}) are found to be 
\bea
 \cF_U(\mu_U) \eql \sin\brkt{\frac{\cN^{1/3}\mu_U}{\abs{\phi_C}}}
 \left\{1-\frac{\abs{\phi_C}^2}{\mu_U^2h^{(L)}h^{(0)}}
 \right. \nonumber\\
 &&\hspace{30mm}\left. 
 -\frac{\abs{\phi_C}}{\mu_U}
 \brkt{\frac{1}{h^{(L)}}+\frac{1}{h^{(0)}}}
 \cot\brkt{\frac{\cN^{1/3}\mu_U}{\abs{\phi_C}}} \right\},  \nonumber\\
 \cF_V(\mu_V) \eql \sin^{n_V-1}\brkt{\frac{\cN^{1/3}\mu_V}{\abs{\phi_C}}}
 \det\left[\id_{n_{V_{\rm e}}}
 -\frac{\abs{\phi_C}^2}{\mu_V^2}H_V^{(L)-1}H_V^{(0)-1} \right. \nonumber\\
 &&\hspace{35mm}\left.
 -\frac{\abs{\phi_C}}{\mu_V}\brkt{H_V^{(L)-1}+H_V^{(0)-1}}
 \cot\brkt{\frac{\cN^{1/3}\mu_V}{\abs{\phi_C}}} \right], \nonumber\\
 \cF_{\rm ch}(\mu_{\rm ch}) \eql \det\brkt{
 e^{\frac{T_R}{2}}\sin\omg_T}\det\left[e^{-\frac{T_R}{2}}\sin\omg_T
 -\frac{\abs{\phi_C}^2}{\mu_{\rm ch}^2}
 H_{\rm ch}^{(L)-1}e^{\frac{T_R}{2}}\sin\omg_T H_{\rm ch}^{(0)-1}
 \right. \nonumber\\
 &&\hspace{30mm}
 -\frac{\abs{\phi_C}^2H_{\rm ch}^{(L)-1}}{\cN^{1/3}\mu_{\rm ch}^2}
 e^{\frac{T_R}{2}}\brkt{\omg_T\cos\omg_T-\frac{T_R}{2}\sin\omg_T} 
 \nonumber\\
 &&\hspace{30mm}\left. 
 -\brkt{\omg_T\cos\omg_T+\frac{T_R}{2}\sin\omg_T}
 e^{-\frac{T_R}{2}}\frac{\abs{\phi_C}^2H_{\rm ch}^{(0)-1}}
 {\cN^{1/3}\mu_{\rm ch}^2}\right], 
 \label{expr:cG_F}
\eea
where 
\bea
 \brkt{H_V^{(y_*)}}^{\Ie}_{\;\;\Je} 
 \defa \frac{a^{\Ie\Ke}}{\cN^{2/3}}\brkt{\Re f_{\Ke\Je}^{(y_*)}
 +\frac{2}{3\mu_V^2}\abs{\phi_C}^2h^{(y_*)}_{\Ke\Je}},  \nonumber\\
 \brkt{H_{\rm ch}^{(y_*)}}_{ab} \defa 
 \frac{1}{2}\brkt{\bar{h}_{a\bar{b}}^{(y_*)}
 -\frac{\abs{\phi_C}}{\mu_{\rm ch}}P_{ab}^{(y_*)}}, \nonumber\\
 T_R \defa -\bar{v}^{\Io}\check{t}_{\Io}\bar{V}_s 
 = 2\Re T^{\Io}\check{t}_{\Io}, \nonumber\\
 \omg_T \defa \brkt{\frac{\cN^{2/3}\mu_{\rm ch}^2}{\abs{\phi_C}^2}
 -\frac{T_R^2}{4}}^{1/2}. 
\eea
The arguments of $a_{\Ie\Je}$ and $\cN$ are $(\bdm{0}_{n_{V_{\rm e}}},2\Re T^{\Io})$. 
The determinants in the expressions of $\cF_V(\mu_V)$ and 
$\cF_{\rm ch}(\mu_{\rm ch})$ are taken over 
the $n_{V_{\rm e}}$-dimensional space spanned by $\tl{V}^{\Ie}$ and 
the $n_H$-dimensional space projected by $\cP_{\!{\rm ch}}$, respectively. 
Notice that the $s_{\Io}$-dependences are completely canceled in (\ref{expr:cG_F}) 
as mentioned in Sec.~\ref{Eigenvalues}. 

The contributions of the bulk superfields to $\Omglp$ are 
calculated from the formula~(\ref{expr:Omg_eff^2}) 
with the functions in (\ref{expr:cG_F}). 
Here we rescale the integral variable~$\mu_F$ as $\mu_F\to C_{\rm rs}\mu_F$, 
where $C_{\rm rs}=\abs{\phi_C}/\cN^{1/3}$. 
Then the analytic functions~$\tl{\cF}_F^{\rm asp}$ in (\ref{expr:Omg_eff^2}) 
can be chosen as 
\be
 \tl{\cF}_U^{\rm asp}(z) = \frac{i}{2}e^{-iz}, \;\;\;\;\; 
 \tl{\cF}_V^{\rm asp}(z) = \brkt{\frac{i}{2}e^{-iz}}^{n_V-1}, \;\;\;\;\;
 \tl{\cF}_{\rm ch}^{\rm asp}(z) = \brkt{\frac{i}{2}e^{-iz}}^{2n_H},  
 \label{eg:cF_F^asp}
\ee
and the bulk contribution in (\ref{expr:Omg_eff^2}) 
is expressed as  
\bea
 \Omglp \eql \frac{\abs{\phi_C}^2}{\cN^{2/3}}
 \left[\frac{(n_V+1)Q_1(0)-\tr Q_1(T_R/2)}{8\pi^2}+Q_2\tr\brkt{T_R^3} 
 \right.\nonumber\\
 &&\hspace{15mm}\left.
 +\int_0^\infty\frac{d\lmd}{8\pi^2}\sum_{F=U,V,{\rm ch}}g_F\lmd
 \ln\cG_F(\lmd) \right]+\cdots, 
 \label{eg:Omg_eff}  
\eea
where $Q_1(x)\equiv -\int_{\abs{x}}^\infty\dr\lmd\;\lmd\ln(2e^{-\lmd}\sinh\lmd)$, 
$Q_2\equiv\int_0^\infty\frac{d\lmd}{64\pi^2}\;\lmd^2\brkt{\sqrt{1+\lmd^{-2}}-1}$, and 
\bea
 \cG_U(\lmd) \eql 1+\frac{\cN^{2/3}}{\lmd^2h^{(L)}h^{(0)}}
 +\frac{\cN^{1/3}}{\lmd}\brkt{\frac{1}{h^{(L)}}+\frac{1}{h^{(0)}}}\coth\lmd, 
 \nonumber\\
 \cG_V(\lmd) \eql \det\left\{
 \id_{n_{V_{\rm e}}}+\frac{\cN^{2/3}}{\lmd^2}\hat{H}_V^{(L)-1}\hat{H}_V^{(0)-1}
 +\frac{\cN^{1/3}}{\lmd}\brkt{\hat{H}_V^{(L)-1}+\hat{H}_V^{(0)-1}}\coth\lmd\right\}, 
 \nonumber\\
 \cG_{\rm ch}(\lmd) \eql \det\left\{2e^{-\frac{T_R}{2}}e^{-\lmd}\sinh\hat{\omg}_T
 +\frac{2\cN^{2/3}}{\lmd^2}\hat{H}_{\rm ch}^{(L)-1}e^{\frac{T_R}{2}}
 e^{-\lmd}\sinh\hat{\omg}_T\hat{H}_{\rm ch}^{(0)-1} \right. \nonumber\\
 &&\hspace{10mm}
 +\frac{2\cN^{1/3}}{\lmd^2}\hat{H}_{\rm ch}^{(L)-1}e^{\frac{T_R}{2}}e^{-\lmd}
 \brkt{\hat{\omg}_T\cosh\hat{\omg}_T-\frac{T_R}{2}\sinh\hat{\omg}_T} \nonumber\\
 &&\hspace{10mm}\left.
 +2e^{-\lmd}\brkt{\hat{\omg}_T\cosh\hat{\omg}_T
 +\frac{T_R}{2}\sinh\hat{\omg}_T}e^{-\frac{T_R}{2}}\frac{\cN^{1/3}}{\lmd^2}
 \hat{H}_{\rm ch}^{(0)-1}\right\} \nonumber\\
 &&\times\brc{\det\brkt{2e^{-\frac{T_R}{2}}e^{-\lmd}\sinh\hat{\omg}_T}}^{-1}. 
\eea
The argument of the norm function~$\cN$ is 
$(\bdm{0}_{n_{V_{\rm e}}},2\Re T^{\Io})$, and 
\bea
 \brkt{\hat{H}_V^{(y_*)}}^{\Ie}_{\;\;\Je} \defa 
 a^{\Ie\Ke}\brkt{\frac{\Re f_{\Ke\Je}^{(y_*)}}{\cN^{2/3}}
 -\frac{2}{3\lmd^2}h^{(y_*)}_{\Ke\Je}}, \nonumber\\
 \brkt{\hat{H}_{\rm ch}^{(y_*)}}_{ab} \defa 
 \frac{1}{2}\brkt{\bar{h}_{a\bar{b}}^{(y_*)}+\frac{i\cN^{1/3}}{\lmd}P_{ab}^{(y_*)}}, 
 \nonumber\\
 \hat{\omg}_T \defa \brkt{\lmd^2+\frac{T_R^2}{4}}^{1/2}. 
\eea
In (\ref{eg:Omg_eff}), we have used $d\lmd\lmd=d\hat{\omg}_T\hat{\omg}_T$, and 
\be
 \int_0^\infty\dr\lmd\;\lmd\ln\det\brkt{e^{\hat{\omg}_T-\lmd}} 
 = \int_0^\infty\dr\lmd\;\lmd\tr\brkt{\hat{\omg}_T-\lmd} 
 = Q_2\tr\brkt{T_R^3}. 
\ee

As mentioned in Sec.~\ref{Expr:Omglp}, Eq.(\ref{eg:Omg_eff}) 
contain divergent terms. 
The constant~$Q_2$ is divergent and will be renormalized by local counterterms. 
The last term in (\ref{eg:Omg_eff}) also diverges in the presence of 
the boundary terms. 
In order to extract a finite part, we further rewrite it as 
\bea
 \Omglp \eql \frac{\abs{\phi_C}^2}{\cN^{2/3}}\sum_{y_*=0,L}
 \int_0^\infty\frac{d\lmd}{8\pi^2}\sum_{F=U,V,{\rm ch}}
 g_F\lmd\ln\cH_F^{(y_*)}(\lmd)  \nonumber\\
 &&+\frac{\abs{\phi_C}^2}{\cN^{2/3}}
 \int_0^\infty\frac{d\lmd}{8\pi^2}\sum_{F=U,V,{\rm ch}}
 g_F\lmd\ln\frac{\cG_F(\lmd)}{\cH_F^{(L)}(\lmd)\cH_F^{(0)}(\lmd)}+\cdots, 
 \label{eg:Omg_eff2}
\eea
where 
\bea
 \cH_U^{(y_*)}(\lmd) \defa 1+\frac{\cN^{1/3}}{\lmd h^{(y_*)}}, \;\;\;\;\;
 \cH_V^{(y_*)}(\lmd) \equiv \det\brkt{\id
 +\frac{\cN^{1/3}}{\lmd}H_V^{(y_*)-1}},  \\
 \cH_{\rm ch}^{(0)}(\lmd) \defa 
 \det\brkt{\id+\frac{\cN^{1/3}}{\lmd^2}\brkt{\hat{\omg}_T+\frac{T_R}{2}}
 H_{\rm ch}^{(0)-1}}, 
 \nonumber\\
 \cH_{\rm ch}^{(L)}(\lmd) \defa \det\brkt{\id
 +\frac{\cN^{1/3}}{\lmd^2}H_{\rm ch}^{(L)-1}e^{T_R}
 \brkt{\hat{\omg}_T-\frac{T_R}{2}}}. \nonumber
\eea
Now the second line of (\ref{eg:Omg_eff2}) is finite. 
Nonlocal effects such as the brane-to-brane mediation effects are contained 
in this part. 
We can also see that the divergent part, which is the first line 
of (\ref{eg:Omg_eff2}), does not depend on the parameters  
in $\cL_{\rm bd}^{(0)}$ and those in $\cL_{\rm bd}^{(L)}$ simultaneously. 
This indicates that the divergent terms originate from 
one-loop diagrams localized on the boundaries. 
Thus they should be combined with $\Omg_{y_*}$ ($y_*=0,L$) 
as mentioned in Sec.~\ref{Expr:Omglp}. 

As a result, the one-loop K\"ahler potential is expressed as 
\bea
 &&\Omglp(\phi_C,\chi^a,T^{\Io},V^{\Ie}) \nonumber\\
 \eql \Omg_0+\Omg_L+\frac{\abs{\phi_C}^2}{\cN^{2/3}}\left[
 \frac{(n_V+1)Q_1(0)}{8\pi^2}
 -\sum_{a=1}^{n_H}\frac{Q_1(c_{a\Io}\Re T^{\Io})}{8\pi^2}-Q_2\sum_{a=1}^{n_H}
 \brkt{2c_{a\Io}\Re T^{\Io}}^3 \right.  \nonumber\\
 &&\left.\hspace{30mm}
 +\int_0^\infty\frac{d\lmd}{8\pi^2}\sum_{F=U,V,{\rm ch}}
 g_F\lmd\ln\frac{\cG_F(\lmd)}{\cH_F^{(L)}(\lmd)\cH_F^{(0)}(\lmd)}\right] 
 +\cO(\hat{\chi}^2),  \label{FinalExpr}
\eea
where $n_V-1$ and $n_H$ are the numbers of the physical vector and hypermultiplets, 
respectively, $\cN=\cN(\bdm{0}_{n_{V_{\rm e}}},2\Re T^{\Io})$, 
$Q_1\simeq 0.30$, $(g_U,g_V,g_{\rm ch})=(-2,-1,\frac{1}{2})$, 
and $\hat{\chi}\equiv \exp\brc{\frac{1}{2}V^{\Ie}\check{t}_{\Ie}}\chi$. 
Here the generators~$\check{t}_{\Io}$ are denoted as 
$\check{t}_{\Io}=-\diag(c_{1\Io},c_{2\Io},\cdots,c_{n_H\Io})$, 
where $c_{a\Io}$ are $Z_2$-odd gauge couplings corresponding to 
the bulk masses for the hypermultiplets. 
The boundary contributions~$\Omg_{y_*}$ ($y_*=0,L$) are sum of 
(\ref{Omglp_bd}) and the first line of (\ref{eg:Omg_eff2}), 
and are renormalized by local counterterms 
in the boundary Lagrangians~$\cL_{\rm bd}^{(y_*)}$. 
The renormalized value of $Q_2$ cannot be predicted within the field theory. 
The last term in (\ref{FinalExpr}) involves 
the parameters both in $\cL_{\rm bd}^{(0)}$ and $\cL_{\rm bd}^{(L)}$, 
and becomes important when one of the boundary actions possesses 
some symmetries that are not held in the whole system. 
In such a case, terms prohibited by those symmetries are induced 
through loop diagrams involving the bulk superfields, and they are finite. 
The vector superfields~$V^{\Ie}$ appear only through $\hat{\chi}$, 
just like in the tree-level effective Lagrangian~(\ref{expr:cL_eff}). 
The overall dependence on the moduli through $\cN^{-2/3}$ represents 
the volume suppression of the extra dimension, and the nontrivial dependence 
on them are induced through the gaugings accompanied by 
the hypermultiplet bulk masses~$c_{a\Io}$. 
The bosonic component expression of $\int\dr^4\tht\;\Omglp$ is shown 
in Appendix~\ref{BS_component} in the absence of the boundary terms.

\ignore{
Expanding (\ref{eg:Omg_eff}) in terms of the superfields~$\phi_C$, 
$\chi^a$, $T^{\Io}$ and $V^{\Ie}$, we can read off various interaction terms 
in the effective K\"ahler potential. 
Notice that terms involving both boundary quantities are finite. 
This is what we expected because  
such terms are induced through one-loop diagrams 
that cannot shrink to a point and the inverse of the size of the extra dimension 
provides an effective cutoff in the momentum integral~\cite{Gregoire:2004nn}. 
Finally, we provide an explicit form of $\Omglp$ 
in the case that the boundary actions are absent, \ie, 
$h^{(y_*)}=\hat{H}_V^{(y_*)}=\hat{H}_{\rm ch}^{(y_*)}=0$, 
and the generators are denoted as 
$\check{t}_{\Io}=2\diag(c_{1\Io},c_{2\Io},\cdots,c_{n_H\Io})$, 
where $c_{a\Io}$ are $Z_2$-odd gauge couplings corresponding to 
the bulk masses for the hypermultiplets. 
In this case, (\ref{FinalExpr}) is reduced to 
\bea
 \Omglp \eql \frac{\abs{\phi_C}^2}{\cN^{2/3}}\sbk{
 \frac{(n_V-n_H-3)Q}{8\pi^2}
 +\int_0^\infty\frac{d\lmd}{16\pi^2}\;\lmd\ln\det\brc{2e^{\frac{T_R}{2}}
 e^{-\lmd}\sinh\brkt{\sqrt{1+\frac{T_R^2}{4\lmd^2}}\lmd}}} \nonumber\\
 &&+\cO(\hat{\chi}^2), 
\eea
where $\cN=\cN(\bdm{0}_{n_{V_{\rm e}}},2\Re T^{\Io})$, 
$T_R=\Re T^{\Io}\check{t}_{\Io}$, and 
$Q\equiv -\int_0^\infty\dr\lmd\;\lmd\ln(2e^{-\lmd}\sinh\lmd)\simeq 0.30$. 
The bulk mass parameters for the hypermultiplets are contained 
in the definition of the generators~$\check{t}_{\Io}$. 
The overall dependence on the moduli through $\cN^{-2/3}$ comes from 
the volume suppression of the extra dimension, and the nontrivial dependence 
through $T_R$ is induced by the gaugings accompanied by 
the hypermultiplet bulk mass parameters. 
}

\section{Summary} \label{summary}
We derived one-loop contributions to the K\"ahler potential 
in 4D effective theory of 5D SUGRA on $S^1/Z_2$ 
{\it with a generic form of the prepotential} and arbitrary boundary-localized terms. 
Our work is regarded as an extension of 
Refs.~\cite{Gherghetta:2001sa,Rattazzi:2003rj,Buchbinder:2003qu,
Gregoire:2004nn} to more general cases, and 
the result is applicable to a wide class of 5D SUGRA models, 
in which various isometries are gauged by arbitrary number 
of $Z_2$-odd vector multiplets (\ie, moduli multiplets). 
The calculations are performed by means of the $N=1$ superfield 
formalism~\cite{Abe:2011rg,Sakamura:2012bj}, which is  
based on the superconformal formulation of 
5D SUGRA~\cite{Kugo:2000af}-\cite{Kugo:2002js}. 
Since the off-shell formulation of SUGRA contains unphysical modes, 
such as the compensator multiplet, some projection operators appear 
in the calculations. 
This makes the procedure somewhat complicated. 
Especially, due to the projection operator~$\cP_V$, 
the ordinary Kaluza-Klein expansion of the vector superfields~$V^I$ cannot be performed 
in a way that the $N=1$ superfield structure is preserved~\cite{Abe:2008an,Abe:2006eg}. 
Instead, corresponding procedure becomes possible 
by changing the coordinate~$y$ with $V_s$ defined in (\ref{def:V_s}). 

The one-loop effective K\"ahler potential~$\Omglp$ is relevant to 
the brane-to-brane communication of SUSY-breaking effects and 
the moduli stabilization by the Casimir effect. 
Our result makes it possible to discuss these issues 
in much wider class of 5D SUGRA models than ever. 
Although the explicit forms of $\tl{\cF}_F(z)$, 
$\tl{\cF}_F^{\rm asp}(z)$ ($F=U,V,{\rm ch}$), 
and $C_{\rm rs}$ in our formula~(\ref{expr:Omg_eff^2}) are highly model-dependent, 
we can easily find them once a model is specified. 
As an illustrative example, 
we provided an explicit expression of $\Omglp$ 
in the case of 5D flat spacetime. 
In the case of a warped geometry, the expression becomes more complicated, 
and may not be expressed in an analytic form 
except in the Randall-Sundrum spacetime.\footnote{
As we have pointed out in Ref.~\cite{Abe:2011rg}, we have to require 
a fine-tuning among the gauge couplings and the vacuum expectation values 
of the moduli in order to obtain the Randall-Sundrum spacetime 
when there are more than one moduli. } 
Still, we expect that some properties can be extracted by means of a technique 
used in Ref.~\cite{Falkowski:2006vi}. 

The one-loop K\"ahler potential is also relevant to 
gauge symmetry breaking by the Wilson line phase~\cite{Hosotani:1983xw}. 
For example, 
we can discuss the gauge-Higgs unification scenario 
at the grand unification scale~\cite{Nomura:2006pn}-\cite{Brummer:2009ug} 
in the context of 5D SUGRA after extending our result to non-Abelian gauge groups.  

There are several ways to proceed. 
We plan to discuss the moduli stabilization 
and the SUSY-breaking mediation 
in 5D SUGRA models with a generic form of 
the prepotential by making use of our result, 
and derive useful information for the phenomenological model-building. 
An extension of our result to higher-dimensional SUGRA is another direction 
for future works. 
Notice that an $N=1$ superfield description of the action should be exist 
although such theories do not have a full off-shell formulation. 
Since our derivation in this paper is systematic, it can easily be extended to 
higher-dimensional SUGRA once we obtain the $N=1$ superfield description.

\subsection*{Acknowledgements}
This work was supported in part by 
Grant-in-Aid for Young Scientists (B) No.22740187 
from Japan Society for the Promotion of Science.

\appendix

\section{Superconformal transformations} \label{trf:SC}
Here we list the 5D superconformal transformation laws expressed 
in terms of the $N=1$ superfields. 
For the purpose of constructing the action up to linear in the gravitational 
superfields, it is enough to keep the transformations 
at the zeroth order in them. 

The $N=1$ part~$\dscp$ is given by 
\bea
 \dscp\Phi_{\rm odd} \eql \brkt{-\frac{1}{4}\bar{D}^2L^\alp D_\alp
 -i\sgm_{\alp\dalp}^\mu\bar{D}^{\dalp}L^\alp\der_\mu
 -\frac{1}{8}\bar{D}^2D^\alp L_\alp}\Phi_{\rm odd}, \nonumber\\
 \dscp\Phi_{\rm even} \eql \brkt{-\frac{1}{4}\bar{D}^2L^\alp D_\alp
 -i\sgm_{\alp\dalp}^\mu\bar{D}^{\dalp}L^\alp\der_\mu
 -\frac{1}{8}\bar{D}^2D^\alp L_\alp}\Phi_{\rm even}, \nonumber\\
 \dscp V \eql \brkt{-\frac{1}{4}\bar{D}^2L^\alp D_\alp
 -\frac{i}{2}\sgm_{\alp\dalp}^\mu\bar{D}^{\dalp}L^\alp\der_\mu+\hc}V, 
 \nonumber\\
 \dscp\Sgm \eql \brkt{-\frac{1}{4}\bar{D}^2L^\alp D_\alp
 -i\sgm_{\alp\dalp}^\mu\bar{D}^{\dalp}L^\alp\der_\mu}\Sgm, \nonumber\\
 \dscp U^\mu \eql \frac{1}{2}\sgm^\mu_{\alp\dalp}
 \brkt{\bar{D}^{\dalp}L^\alp-D^\alp\bar{L}^{\dalp}}, \;\;\;\;\;
 \dscp\Psi^\alp = -\der_y L^\alp, \;\;\;\;
 \dscp U^y = 0,  \nonumber\\
 \dscp V_E \eql \brkt{-\frac{1}{4}\bar{D}^2L^\alp D_\alp
 -\frac{i}{2}\sgm_{\alp\dalp}^\mu\bar{D}^{\dalp}L^\alp\der_\mu
 +\frac{1}{24}\bar{D}^2D^\alp L_\alp+\hc}V_E,  \label{dscp}
\eea
where a complex spinor superfield~$L^\alp$ is the transformation parameter. 
The remaining transformations~$\dscq$ are given by 
\bea
 \dscq\Phi_{\rm odd} \eql \frac{Y}{\vev{V_E}}\der_y\Phi_{\rm odd}
 -\frac{i}{4}\bar{D}^2\brc{\tl{N}(e^{-V})^t\bar{\Phi}_{\rm even}}, \nonumber\\
 \dscq\Phi_{\rm even} \eql \frac{Y}{\vev{V_E}}\der_y\Phi_{\rm even}
 +\frac{i}{4}\bar{D}^2\brc{\tl{N}e^V\bar{\Phi}_{\rm odd}}, \nonumber\\
 \dscq e^V \eql \frac{Y+\bar{Y}}{2\vev{V_E}}\der_y e^V
 +\frac{i\tl{N}}{\vev{V_E}}\brkt{\Sgm e^V-e^V\Sgm^\dagger}, \nonumber\\
 \dscq \Sgm \eql \der_y\brkt{\frac{Y\Sgm}{\vev{V_E}}}
 -\frac{i\vev{V_E}}{8}\bar{D}^2\brkt{D^\alp\tl{N}D_\alp e^Ve^{-V}}, \nonumber\\
 \dscq U^\mu \eql 0, \;\;\;\;\;
 \dscq V_E = \frac{1}{2}\der_y\brkt{Y+\bar{Y}}, \nonumber\\
 \dscq\Psi^\alp \eql \frac{i\vev{V_E}}{2}D^\alp\tl{N}, \;\;\;\;\;
 \dscq U^y = \frac{N}{\vev{V_E}},  \label{dscq}
\eea
where a chiral and real superfields~$Y$ and $N$ are the transformation parameters, 
and 
\be
 \tl{N} \equiv N-\frac{i}{2}\brkt{Y-\bar{Y}}. 
\ee
The components of $L_\alp$, 
\bea
 \xi^\mu \defa -\left.\Re\brkt{i\sgm_{\alp\dalp}^\mu\bar{D}^{\dalp}L^\alp}
 \right|_0, \;\;\;\;\;
 \ep_\alp \equiv -\frac{1}{4}\bar{D}^2L_\alp|_0, \nonumber\\
 \lmd_{\mu\nu} \defa -\frac{1}{2}\left.\Re\brc{\brkt{\sgm_{\mu\nu}}_\bt^{\;\;\alp}
 D_\alp\bar{D}^2L^\bt}\right|_0, \;\;\;\;\;
 \vph_D \equiv \left.\Re\brkt{\frac{1}{4}D^\alp\bar{D}^2L_\alp}\right|_0, 
 \nonumber\\
 \vth_A \defa \left.\Im\brkt{-\frac{1}{6}D^\alp\bar{D}^2L_\alp}\right|_0, 
 \;\;\;\;\;
 \eta_\alp \equiv -\frac{1}{32}D^2\bar{D}^2L_\alp|_0, 
\eea
where the symbol~$|_0$ denotes the lowest component of the superfield, 
are identified with the transformation parameters 
for the translation~$\bdm{P}$, the supersymmetry~$\bdm{Q}$, 
the Lorentz transformation~$\bdm{M}$, the dilatation~$\bdm{D}$, 
the R symmetry~$U(1)_A$ and the conformal supersymmetry~$\bdm{S}$, 
respectively. 
The components of $Y$ and $N$ are identified with 
the other transformation parameters that are $Z_2$-odd~\cite{Sakamura:2012bj}. 

In order to determine the kinetic terms for the gravitational 
superfields~$\cL_{\rm kin}^{{\mathbb E}_W}$, we need to extend the above 
transformations including linear order terms in the gravitational superfields. 
Since $\cL_{\rm kin}^{{\mathbb E}_W}$ is independent of 
the quantum fluctuation of the matter superfields, 
it is enough to focus on the background parts of the matter superfields 
in the extended parts of $\dscp$ and $\dscq$. 
We find the $U^\mu$-dependent part in the transformations as follows. 
The $\dscp$ does not receive any corrections at this order, 
but $\dscq$ is modified as 
\bea
 \dscq\Phi_{\rm odd} \eql 
 -\frac{i}{4}\bar{D}^2\brc{\brkt{\frac{\tl{N}}{3}\Dlt_\mu U^\mu
 -Y\der_\mu U^\mu}\Lvev{(e^{-V})^t\bar{\Phi}_{\rm even}}}
 +\cdots, \nonumber\\
 \dscq\Phi_{\rm even} \eql 
 \frac{i}{4}\bar{D}^2\brc{\brkt{\frac{\tl{N}}{3}\Dlt_\mu U^\mu
 -Y\der_\mu U^\mu}\Lvev{e^V\bar{\Phi}_{\rm odd}}}+\cdots, \nonumber\\
 \dscq V^I \eql 
 -i\der_\mu U^\mu\frac{Y\vev{\Sgm^I}-\bar{Y}\vev{\bar{\Sgm}^I}}{\vev{V_E}}
 +\cdots, \label{md:dscq}
\eea
where the ellipses denote terms shown in (\ref{dscq}). 
The other transformations are unchanged up to this order. 
Here we have considered in the Abelian case, for simplicity. 
Requiring the invariance of the action under this modified transformation, 
we can determine $\cL_{\rm kin}^{{\mathbb E}_W}$ as (\ref{cL_kin^EW}).

\section{Projectors in superspace} \label{superspinPi}
The chiral and anti-chiral projection operators are defined 
as~\cite{Wess:1992cp}  
\be
 P_+ \equiv -\frac{\bar{D}^2D^2}{16\Box_4}, \;\;\;\;\;
 P_- \equiv -\frac{D^2\bar{D}^2}{16\Box_4}.  \label{def:bP}
\ee

We can divide a vector superfield~$V$ into 
a chiral and a transverse parts by the following projectors. 
\be
 P_C \equiv P_++P_-, \;\;\;\;\;
 P_T \equiv \frac{D^\alp\bar{D}^2D_\alp}{8\Box_4}.  \label{def:P_CT}
\ee
These satisfy 
\be
 P_T+P_C = 1, \;\;\;\;\;
 P_T^2 = P_T, \;\;\;\;\;
 P_C^2 = P_C, \;\;\;\;\;
 P_TP_C = P_CP_T = 0. 
\ee

Similarly, the gravitational superfield~$U^\mu$ can be divided by 
the following superspin projectors 
as (\ref{U:decompose})~\cite{Gregoire:2004nn,Gates:1983nr,Gates:2003cz}. 
\bea
 \Pi_0^{\mu\nu} \defa \Pi_L^{\mu\nu}P_C, \nonumber\\
 \Pi_{1/2}^{\mu\nu} \defa \frac{1}{3}Q^{\mu\nu}+\Pi_L^{\mu\nu}P_T
 +\frac{1}{3}\Pi_L^{\mu\nu}P_C, \nonumber\\
 \Pi_1^{\mu\nu} \defa \Pi_T^{\mu\nu}P_C, \nonumber\\
 \Pi_{3/2}^{\mu\nu} \defa -\frac{1}{3}Q^{\mu\nu}+\eta^{\mu\nu}P_T
 -\Pi_L^{\mu\nu}+\frac{2}{3}\Pi_L^{\mu\nu}P_C,  \label{def:Pis}
\eea
where
\bea
 Q^{\mu\nu} \defa \frac{1}{16}\sgm^\mu_{\alp\dalp}\sgm^\nu_{\bt\dbt}
 \frac{[D^\alp,\bar{D}^{\dalp}][D^\bt,\bar{D}^{\dbt}]}{\Box_4}, \nonumber\\
 \Pi_T^{\mu\nu} \defa \eta^{\mu\nu}-\frac{\der^\mu\der^\nu}{\Box_4}, \;\;\;\;\;
 \Pi_L^{\mu\nu} \equiv \frac{\der^\mu\der^\nu}{\Box_4}. 
\eea
These projectors satisfy 
\bea
 &&\Pi_0^{\mu\nu}+\Pi_{1/2}^{\mu\nu}+\Pi_1^{\mu\nu}+\Pi_{3/2}^{\mu\nu} 
 = \eta^{\mu\nu}, \nonumber\\
 &&\Pi^{\mu\rho}_s\Pi_{r\rho}^{\;\;\;\;\nu} = \dlt_{rs}\Pi_s^{\mu\nu}, 
\eea
where $r,s=0,1/2,1,3/2$, and 
\be
 \der_\mu\Pi_0^{\mu\nu} = \der^\nu P_C, \;\;\;\;\;
 \der_\mu\Pi_{1/2}^{\mu\nu} = \der^\nu P_T, \;\;\;\;\;
 \der_\mu\Pi_1^{\mu\nu} = \der_\mu\Pi_{3/2}^{\mu\nu} = 0. 
\ee
Furthermore, $Q^{\mu\nu}$ satisfies 
\bea
 Q^{\mu\rho}Q_\rho^{\;\;\nu} \eql Q^{\mu\nu}\brkt{-4P_C+3}, \nonumber\\
 Q^{\mu\nu}P_C \eql P_CQ^{\mu\nu} = -\Pi_L^{\mu\nu}P_C = -\Pi_0^{\mu\nu}, 
 \nonumber\\
 \der_\mu Q^{\mu\nu}P_T \eql \der_\mu P_T Q^{\mu\nu}.  
\eea

The supertrace integrand~$\Istr$ in (\ref{def:Istr}) satisfies 
the following relations. 
\be
 \Istr\,\id = 0, \;\;\;\;\;
 \Istr\brkt{\bar{D}^2D^2} = \Istr\brkt{D^2\bar{D}^2} =
 \Istr\brkt{D^\alp\bar{D}^2D_\alp} = 16, 
\ee
and thus, 
\bea
 \Istr\, P_\pm \eql -\frac{1}{\Box_4}, \;\;\;\;\;
 \Istr\, P_T = \frac{2}{\Box_4}, \;\;\;\;\;
 \Istr\, Q^{\mu\nu} = \tr\brkt{\frac{2}{\Box_4}\eta^{\mu\nu}}
 = \frac{8}{\Box_4}, \nonumber\\
 \Istr\,\Pi_0^{\mu\nu} \eql -\frac{2}{\Box_4}, \;\;\;\;\;
 \Istr\,\Pi_{1/2}^{\mu\nu} = \frac{4}{\Box_4}, \;\;\;\;\;
 \Istr\,\Pi_1^{\mu\nu} = -\frac{6}{\Box_4}, \;\;\;\;\;
 \Istr\,\Pi_{3/2}^{\mu\nu} = \frac{4}{\Box_4}.  \label{Istr:formulae}
\eea

\section{Tree-level effective action} \label{tree:action}
In this section, we briefly review the derivation 
of the 4D effective action at tree level. 
We have developed a systematic method to derive it in Ref.~\cite{Abe:2006eg}. 
Explicit calculations in the flat and the warped spacetimes are 
performed in Refs.~\cite{Abe:2008an,Abe:2011rg}. 

The basic strategy is as follows. 
First, we drop the kinetic terms for $Z_2$-odd superfields 
because they do not have zero-modes that are dynamical below 
the compactification scale. 
Then the $Z_2$-odd superfields play a role of Lagrange multipliers, 
and their equations of motion extract zero-modes from the $Z_2$-even superfields. 

Since we are interested in the 4D effective action for the matter superfields, 
we neglect the gravitational superfields in this section. 
Namely, the 5D Lagrangian~(\ref{5D_action2}) reduces to 
\bea
 \cL \eql -\int\dr^4\tht\;3\cN^{1/3}(\cV)\brc{\Phi_{\rm odd}^\dagger\tl{d}
 (e^V)^t\Phi_{\rm odd}+\Phi_{\rm even}^\dagger\tl{d}e^{-V}\Phi_{\rm even}}^{2/3} 
 \nonumber\\
 &&+\sbk{\int\dr^2\tht\;\brc{
 2\Phi^t_{\rm odd}\tl{d}\brkt{\der_y-\Sgm}\Phi_{\rm even}
 +W_{\rm v}}+\hc} 
 +2\sum_{y_*=0,L}\cL_{\rm bd}^{(y_*)}\dlt(y-y_*),  \label{tree:cL}
\eea
where we have performed the partial integral. 

\subsection{Gauge kinetic functions and superpotential}  \label{tree:fW}
First, we divide $V$ into the $Z_2$-odd part~$V_{\rm o}$ and 
the $Z_2$-even part~$V_{\rm e}$ as 
\be
 e^V \equiv e^{V_{\rm e}/2}e^{V_{\rm o}}e^{V_{\rm e}/2}. 
\ee
Before dropping the kinetic terms for the $Z_2$-odd superfields, 
we eliminate $\Sgm$ from the bulk action  
by means of the supergauge transformation~(\ref{trf:gauge}) 
with the transformation parameter, 
\be
 e^{-\Lmd(y)} = \exp\brc{-\Lmd_\Sgm(y)} 
 \equiv \cP\exp\brc{\int_0^y\dr y'\;\Sgm(y')}, \label{def:Lmd_Sgm}
\ee
where $\cP$ denotes the path-ordering operator. 
Namely, this is a solution to $\der_y e^{-\Lmd} = \Sgm e^{-\Lmd}$. 
Although the $Z_2$-odd superfields~$\Sgm^{\Ie}$ are completely gauged away, 
the zero-modes of the $Z_2$-even superfields~$\Sgm^{\Io}$ remain in the theory 
as we will explain below. 
We define 4D superfields~$T$ and $S$ as 
\bea
 e^{S}e^{T} \defa \lim_{y\to L}\exp\brc{-\Lmd_\Sgm(y)} 
 =\cP\exp\brc{\int_0^L\dr y\;\Sgm(y)}, \nonumber\\
 T \eql \sum_{\Io}T^{\Io}\hat{t}_{\Io}, \;\;\;\;\;
 S = \sum_{\Ie}S^{\Ie}\hat{t}_{\Ie},   \label{def:T}
\eea
where $\hat{t}_I\equiv 2igt_I$ are hermitian generators, 
and the limits are taken from the bulk region~($0<y<L$). 
Then, the gauge-transformed vector superfields have the following 
boundary conditions. 
\be
%
 \lim_{y\to 0}e^V = \brkt{e^{V'_{\rm e}}}_{y=0}, \;\;\;\;\;
 \lim_{y\to L}e^V 
 = e^{-T}e^{-S}\brkt{e^{V'_{\rm e}}}_{y=L}e^{-S^\dagger}e^{-T^\dagger}. 
 \label{bdcd:tlV}
\ee
where $V'_{\rm e}$ and $V'_{\rm o}$ denote the vector superfields 
before the gauge transformation by (\ref{def:Lmd_Sgm}). 

\ignore{
Notice that there is a residual gauge symmetry that preserves $\Sgm=0$, 
\ie, the 4D gauge transformation whose transformation 
parameter~$\Lmd_0=\sum_{\Ie}\Lmd_0^{\Ie}\hat{t}_{\Ie}$ is independent of $y$. 
Under such residual gauge transformation, $T$, $S$, $V'_{\rm o}$ 
and $V'_{\rm e}$ transform as 
\bea
 T \toa e^{\Lmd_0}Te^{-\Lmd_0}, \;\;\;\;\;
 S \to e^{\Lmd_0}Se^{-\Lmd_0}, \nonumber\\
 e^{V'_{\rm o}} \toa e^{-\Lmd_0^\dagger}e^{V'_{\rm o}}e^{-\Lmd_0}, 
 \;\;\;\;\;
 e^{V'_{\rm e}/2} \to e^{\Lmd_0}e^{V'_{\rm e}/2}e^{\Lmd_0^\dagger}. 
 \label{trf:4Dgauge1}
\eea
}
Since $V_{\rm e}$ corresponds to the gauge superfield 
for the 4D unbroken gauge group, it should vanish 
in $\cN^{1/3}(\cV)$ in (\ref{tree:cL}) because 
there is no corresponding term in 4D gauge theories. 
This implies that 
\be
 \der_y V_{\rm e}=0.  \label{flat:V_e}
\ee
Then, $\cN(\cV)$ reduces to 
\be
 \cN(\cV) = \cN\brkt{e^{V_{\rm o}}\der_y e^{-V_{\rm o}}}, 
\ee
and the boundary conditions for $V_{\rm o}$ in (\ref{bdcd:tlV}) becomes 
\be
 V_{\rm o}|_{y=0} = 0, \;\;\;\;\;
 \lim_{y\to L}V_{\rm o} = \bar{V}_{\rm o}
 \equiv -T-\bar{T}^\dagger+\frac{1}{2}\sbk{V_{\rm e},T-T^\dagger}
 +\cdots,  \label{def:barV_s}
\ee
where $\bar{V}_{\rm o}$ is defined so that 
$e^Te^{V_{\rm e}}e^{\bar{V}_{\rm o}}e^{V_{\rm e}}e^{T^\dagger}$ 
belongs to the unbroken gauge group. 
Notice that $V_{\rm o}$ is discontinuous at $y=L$ since it is $Z_2$-odd. 
This discontinuity stems from the discontinuous 
gauge transformation~(\ref{def:Lmd_Sgm}). 
(See (\ref{limits:Lmd_Sgm}).)

Now we impose constraints~$D_\alp V^{\Io}=0$ to drop the kinetic terms 
for $V^{\Io}$. 
To illustrate the procedure of deriving the gauge kinetic functions 
in (\ref{abelian:W_v}), 
we consider a case that the gauge group is Abelian. 
Then, since $\Sgm$ has been gauged away, $W_{\rm v}$ becomes 
\bea
 W_{\rm v} \eql \frac{c^3}{16g^3}\tr\brc{\frac{1}{12}\bar{D}^2
 \brkt{\der_y V_{\rm o}D^\alp V_{\rm e}}\cW_{{\rm e}\alp}}
 -\frac{c^3}{48g^3}\der_y\tr\brc{\Lmd_\Sgm\cW_{\rm e}^2} \nonumber\\
 \eql \frac{c^3}{48g^3}\der_y\tr\brc{\frac{1}{4}\bar{D}^2
 \brkt{V_{\rm o}D^\alp V_{\rm e}}\cW_{{\rm e}\alp}
 -\Lmd_\Sgm\cW_{\rm e}^2}, 
 \label{tree:W_v}
\eea
where we have used (\ref{flat:V_e}) at the second equality, and 
\be
 \cW_{{\rm e}\alp} = \frac{1}{4}\bar{D}^2\brkt{e^{V_{\rm e}}
 D_\alp e^{-V_{\rm e}}} 
 = -\frac{1}{4}\bar{D}^2D_\alp V_{\rm e}. 
\ee
We have also used that $\tr\brkt{\brc{\hat{t}_{\Io},\hat{t}_{J_{\rm e}}}
\hat{t}_{K_{\rm e}}}=0$. 
(See the footnote~\ref{zero:C_IJK}.)
The last term in the first line of (\ref{tree:W_v}) 
is induced by the supergauge transformation with $\Lmd_\Sgm$. 
Thus, 
\bea
 \int_0^L\dr y\brc{\int\dr^2\tht\;W_{\rm v}+\hc} 
 \eql \frac{c^3}{48g^3}\int\dr^4\tht\sbk{
 \tr\brc{\brkt{-2V_{\rm o}-\Lmd_\Sgm-\Lmd_\Sgm^\dagger}
 D^\alp V_{\rm e}\cW_{{\rm e}\alp}}}_0^{L-\ep} \nonumber\\
 \eql \frac{c^3}{16g^3}\int\dr^4\tht\;\tr\brc{
 \brkt{T+T^\dagger}D^\alp V_{\rm e}
 \cW_{{\rm e}\alp}} \nonumber\\
 \eql \frac{c^3}{16g^3}\int\dr^2\tht\;\tr\brkt{T\cW_{\rm e}^2}+\hc. 
 \label{abelian:W_v}
\eea
We have performed the partial integrals, 
used the relations~$d^2\bar{\tht}=-\frac{1}{4}\bar{D}^2$, 
$D^\alp\cW_{{\rm e}\alp}=\bar{D}_{\dalp}\bar{\cW}^{\dalp}_{\rm e}$, 
$\tr(\hat{t}_{\Ie}\hat{t}_{J_{\rm e}}\hat{t}_{K_{\rm e}})=0$, and 
\bea
 \lim_{y\to 0}\Lmd_\Sgm \eql -S, \;\;\;\;\;
 \lim_{y\to 0}V_{\rm o} = 0, \nonumber\\
 \lim_{y\to L}\Lmd_\Sgm \eql -T-S, \;\;\;\;\;
 \lim_{y\to L}V_{\rm o} = -T-T^\dagger.  \label{limits:Lmd_Sgm}
\eea
The expression~(\ref{abelian:W_v}) is also valid in the non-Abelian case. 
In fact, it is invariant under the unbroken 4D gauge transformation,\footnote{
This gauge transformation preserves the gauge in which $\Sgm=0$. } 
\be
 T \to e^{\Lmd_0}T e^{-\Lmd_0}, \;\;\;\;\;
 e^{V_{\rm e}} \to e^{\Lmd_0}e^{V_{\rm e}}e^{\Lmd_0^\dagger}, 
\ee
where $\Lmd_0=\sum_{\Ie}\Lmd_0^{\Ie}\hat{t}_{\Ie}$ is $y$-independent. 

Next we drop the kinetic terms for $\Phi_{\rm odd}$ in the first line 
of (\ref{tree:cL}). 
Then, from the equation of motion for $\Phi_{\rm odd}$, we obtain
\be
 \der_y\Phi_{\rm even} = 0, 
\ee
which means that $\Phi_{\rm even}$ is $y$-independent for $0\leq y<L$. 

Recall that the gauge transformation parameter~$e^{\Lmd_\Sgm}$ 
is discontinuous at $y=L$, 
\be
 \left.e^{\Lmd_\Sgm}\right|_{y=L} = e^{-S}, \;\;\;\;\;
 \lim_{y\to L}e^{\Lmd_\Sgm} = e^{-T}e^{-S},  
\ee
since the $Z_2$-odd generators~$\hat{t}_{\Io}$ vanish there.\footnote{
Note that $\hat{t}_{\Io}$ include the $Z_2$-odd gauge couplings. }
Hence, the boundary values of $\Phi_{\rm even}$ and $e^{-V_{\rm e}}$ 
that appear in $\cL_{\rm bd}^{(L)}$ are related to their bulk values as 
\bea
 \left.\Phi_{\rm even}\right|_{y=L} 
 \eql e^T\Phi_{\rm even}, \nonumber\\
 V_{\rm e}|_{y=L} \eql V_{\rm e}^{(L)} 
 \equiv \ln\brkt{e^Te^{V_{\rm e}/2}e^{\bar{V}_{\rm o}}
 e^{V_{\rm e}/2}e^{T^\dagger}} 
 = V_{\rm e}-\frac{1}{2}\sbk{T,T^\dagger}+\cdots,  \label{rel:bulk-bd}
\eea
where $\Phi_{\rm even}$ and $V_{\rm e}$ in the left-hand side 
denote the values in the bulk~($0<y<L$). 

Therefore, we obtain the expression of the 4D effective Lagrangian, 
\bea 
 \cL_{\rm eff} = \int_0^L\dr y\;\cL 
 \eql -\int\dr^4\tht\;\int_0^L\dr y\;3\cN^{1/3}\brkt{e^{V_{\rm o}}\der_y
 e^{-V_{\rm o}}}\brkt{\hat{\Phi}_{\rm even}^\dagger\tl{d}e^{-V_{\rm o}}
 \hat{\Phi}_{\rm even}}^{2/3} \nonumber\\
 &&+\sbk{\int\dr^2\tht\;\frac{c^3}{16g^3}\tr\brkt{T\cW_{\rm e}^2}+\hc} 
 \nonumber\\
 &&+\cL_{\rm bd}^{(0)}\brkt{e^{-V_{\rm e}},\Phi_{\rm even}}
 +\cL_{\rm bd}^{(L)}\brkt{e^{-V_{\rm e}^{(L)}},
 e^T\Phi_{\rm even}}, \label{expr:cL_eff}
\eea
where $\hat{\Phi}_{\rm even}\equiv e^{V_{\rm e}/2}\Phi_{\rm even}$ 
is independent of $y$. 
From this expression, we can read off the gauge kinetic functions 
and the superpotential in the effective theory.

\subsection{K\"ahler potential} \label{tree:Kahler}
In (\ref{expr:cL_eff}), the only $y$-dependent superfield is $V_{\rm o}$. 
Since we have dropped its kinetic term, we can integrate it out 
by using its equation of motion. 

In the following derivation, we focus on a subset of 
$\brc{V^{\Io}\hat{t}_{\Io}}$, 
in which every generator commutes with each other. 
We also consider a single compensator case ($n_C=1$), and the generators 
have the following form. 
\be
 \hat{t}_{\Io} = \begin{pmatrix} -3k_{\Io} & \\
 & -3k_{\Io}\id_{n_H}+\check{t}_{\Io} \end{pmatrix},  \label{form:generator}
\ee
where $\check{t}_{\Io}$ are $n_H\times n_H$ matrices. 
Then the effective K\"ahler potential~$\Omg_{\rm eff}\equiv -3e^{-K_{\rm eff}/3}$ 
at tree level is rewritten as 
\be
 \Omg^{\rm tree}_{\rm eff} = -\int_0^L\dr y\;3\abs{\phi_C}^2
 \hat{\cN}^{1/3}(-\der_y V_{\rm o})
 e^{2k\cdot V}\brkt{1-\chi^\dagger e^{-\check{V}_{\rm o}}
 \chi}^{2/3}, 
\ee
where $k\cdot V\equiv\sum_{\Io}k_{\Io}V^{\Io}$, and 
\be
 \phi_C \equiv \brkt{\hat{\Phi}_{\rm even}^1}^{2/3}, \;\;\;\;\;
 \chi^a \equiv \frac{\hat{\Phi}_{\rm even}^{a+1}}{\hat{\Phi}_{\rm even}^1}, 
 \;\;\;\;\;
 \check{V}_{\rm o} \equiv \sum_{\Io}V^{\Io}\check{t}_{\Io}. 
\ee
Then, from the equation of motion for $V^{\Io}$, we obtain 
\be
 \brc{\der_y\brkt{\frac{\cN_{\Jo}}{\cN^{2/3}}}
 +6k_{\Jo}\cN^{1/3}
 +\frac{2\cN^{1/3}\chi^\dagger e^{-\check{V}_{\rm o}}\check{t}_{\Jo}\chi}
 {1-\chi^\dagger e^{-\check{V}_{\rm o}}\chi}}(\cP_V)^{\Jo}_{\;\;\Io} = 0, 
 \label{EOM:V_o}
\ee
the arguments of the norm function and its derivative are 
$(\bdm{0}_{n_{V_{\rm e}}}-\der_y V_{\rm o})$, 
and the projection operator~$(\cP_V)^{\Jo}_{\;\;\Io}$ 
is defined by (\ref{def:cP_V}). 
The presence of $(\cP_V)^{\Jo}_{\;\;\Io}$ indicates that the number of 
independent equations is less than that of $V^{\Io}$. 
Thus we cannot solve $V^{\Io}$ as functions of $y$. 
Hence we need another method to integrate them out. 

Let us define 
\be
 V_s \equiv s_{\Io}V^{\Io}, \;\;\;\;\;
 v^{\Io} \equiv \frac{\der_y V^{\Io}}{\der_y V_s},  \label{def:v^Io}
\ee
where $s_{\Io}$ are arbitrarily chosen constants, 
and $V_s$ satisfies the boundary conditions, 
\be
 \lim_{y\to 0}V_s = 0, \;\;\;\;\;
 \lim_{y\to L}V_s = \bar{V}_s \equiv -2s_{\Io}\Re T^{\Io}. 
 \label{bdcd:V_s}
\ee
Then (\ref{EOM:V_o}) is rewritten as 
\be
 \brc{\der_y v^{\Jo}a_{\Jo\Ko}(v)+\brkt{3k_{\Ko}
 +\frac{\chi^\dagger e^{-\check{V}_{\rm o}}\check{t}_{\Ko}\chi}
 {1-\chi^\dagger e^{-\check{V}_{\rm o}}\chi}}\der_y V_s}(\cP_V)^{\Ko}_{\;\;\Io}(v) 
 = 0.  \label{derva}
\ee
From (\ref{def:v^Io}), $v^{\Io}$ satisfies $s_{\Io}v^{\Io}=1$, and thus, 
$s_{\Io}(dv^{\Io}/dV_s)=0$. 
Therefore, (\ref{derva}) is rewritten as 
\be
 \frac{dv^{\Io}}{dV_s} = \cG^{\Io\Jo}(v)
 \brkt{3k_{\Jo}+\frac{\chi^\dagger e^{-\check{V}_{\rm o}}\check{t}_{\Jo}\chi}
 {1-\chi^\dagger e^{-\check{V}_{\rm o}}\chi}}, \label{dv/dV}
\ee
where $\cG^{\Io\Jo} = -\brkt{\dlt^{\Io}_{\;\;\Ko}-v^{\Io}s_{\Ko}}a^{\Ko\Jo}$. 
Notice that these equations are solvable 
in contrast to (\ref{EOM:V_o}). 
Once $v^{\Io}(V_s)$ are obtained, $V^{\Io}$ are also expressed 
as functions of $V_s$ through 
\be
 V^{\Io} = \int_0^y\dr y'\;\der_y V^{\Io} 
 = \int_0^y\dr y'\;v^{\Io}\der_y V_s 
 = \int_0^{V_s}\dr V'_s\;v^{\Io}(V'_s).  \label{fcn:Vv}
\ee
In the limit of $y\to L$, this becomes 
\be
 -2\Re T^{\Io} = \int_0^{\bar{V}_s}\dr V_s\;v^{\Io}(V_s), 
\ee
which determines the integral constants for solutions of (\ref{dv/dV}). 
Therefore, $\Omg_{\rm eff}^{\rm tree}$ can be calculated 
as an integral for $V_s$, instead of $y$. 
\be
 \Omg_{\rm eff}^{\rm tree} = \int_0^{\bar{V}_s}\dr V_s\;3\abs{\phi_C}^2
 \cN^{1/3}(v(V_s))e^{2k\cdot V(V_s)}
 \brkt{1-\chi^\dagger e^{-\check{V}_{\rm o}(V_s)}\chi}^{2/3}. 
\ee

We can solve (\ref{dv/dV}) order by order 
in the matter chiral superfields~$\chi^a$. 
Here we consider a case of $k_{\Io}=0$, which means that 
the background 5D spacetime is flat.\footnote{
We calculated $\Omg_{\rm eff}^{\rm tree}$ 
in the case of $k_{\Io}\neq 0$ in Ref.~\cite{Abe:2011rg}. }
In this case, we find that
\be
 v^{\Io}(V_s) = \bar{v}^{\Io}
 -\chi^\dagger\cG^{\Io\Jo}(\bar{v})\check{\bar{v}}^{-1}
 \brkt{e^{-\check{\bar{v}}V_s}-\frac{(\Re\check{T})^{-1}(e^{2\Re\check{T}}-1)}{2}}
 \check{t}_{\Jo}\chi+\cO(\chi^4),  \label{fcn:v}
\ee
where 
\be
 \bar{v}^{\Io} \equiv \frac{\Re T^{\Io}}{s\cdot\Re T}, \;\;\;\;\;
 \check{\bar{v}} \equiv \sum_{\Io}\bar{v}^{\Io}\check{t}_{\Io}, \;\;\;\;\;
 \Re\check{T} \equiv \sum_{\Io}(\Re T^{\Io})\check{t}_{\Io}.  \label{def:barv}
\ee
Since $\check{t}_{\Io}$ commute with each other, 
they can be diagonalized simultaneously. 
\be
 U\check{t}_{\Io}U^{-1} = -\diag(c_{1\Io},c_{2\Io},\cdots,c_{n_H\Io}), \;\;\;\;\;
 \hat{\chi} \equiv U\chi. 
\ee
After some calculations, we obtain~\cite{Abe:2008an,Abe:2011rg} 
\be
 \Omg_{\rm eff}^{\rm tree} = \abs{\phi_C}^2\cN^{1/3}\brc{
 -3+\sum_a 2Y(c_a\cdot\Re T)\abs{\hat{\chi}^a}^2
 +\sum_{a,b}\Omg_{ab}^{(4)}\abs{\hat{\chi}^a}^2\abs{\hat{\chi}^b}^2
 +\cO(\abs{\chi}^6)},  \label{tree:Omg_eff}
\ee
where $Y(x)\equiv\frac{1-e^{-2x}}{2x}$, and~\footnote{
The definitions of the moduli~$T^{\Io}$ and the gauge couplings~$c_a$ ($a=1,\cdots,n_H$) 
are different from those of 
Ref.~\cite{Abe:2008an,Abe:2011rg} by a factor 2. } 
\bea
 \Omg_{ab}^{(4)} \defa 
 -\frac{(c_a\cdot\cP_V a^{-1}\cdot c_b)\brc{Y((c_a+c_b)\cdot\Re T)
 -Y(c_a\cdot\Re T)Y(c_b\cdot\Re T)}}{(c_a\cdot\Re T)(c_b\cdot\Re T)} 
 \nonumber\\
 &&+\frac{Y((c_a+c_b)\cdot\Re T)}{3}. 
\eea
The arguments of $\cN$ and $\cP_V$ are $(\bdm{0}_{n_{V_{\rm e}}},2\Re T^{\Io})$. 
Notice that the $s_{\Io}$-dependences are caancelled 
in the final result~(\ref{tree:Omg_eff}).

\section{Quadratic terms for fluctuation superfields} \label{quad_flct}
Here we show the detailed derivation of 
the quadratic terms for the fluctuation superfields~(\ref{quad:cL_bulk}). 
\subsection{Gravitational sector} \label{quad_flct:grv}
Notice that $\Psi_\alp$ appears in the action only through 
$\bar{D}_{\dalp}\Psi_\alp$ and its derivatives. 
Thus we define the following two real superfields, 
\be
 V_+^\mu \equiv \frac{i}{2}\sgm_{\alp\dalp}^\mu
 \brkt{\bar{D}^{\dalp}\Psi^\alp+D^\alp\bar{\Psi}^{\dalp}}, \;\;\;\;\;
 V_-^\mu \equiv \frac{1}{2}\sgm_{\alp\dalp}^\mu
 \brkt{\bar{D}^{\dalp}\Psi^\alp-D^\alp\bar{\Psi}^{\dalp}}, 
\ee
to describe the degree of freedom for $\Psi_\alp$. 
Since
\be
 E_2 = -U_\mu\Box_4\brkt{\Pi_{3/2}^{\mu\nu}
 -\frac{2}{3}\Pi_0^{\mu\nu}}U_\nu, 
\ee
up to total derivatives, we can expand the integrand in (\ref{5D_action2}) as 
\bea
 &&\Lvev{\Omg_{\rm v}^{1/3}\Omg_{\rm h}^{2/3}}E_2
 -\Lvev{\Omg_{\rm v}^{-1/3}\Omg_{\rm h}^{4/3}}
 \brkt{\cC^\mu\cC_\mu+\bar{D}^{\dalp}\Psi^\alp 
 D_\alp\bar{\Psi}_{\dalp}}
 -3\brkt{1+\frac{\Dlt_\mu U^\mu}{3}}
 \Omg_{\rm v}^{1/3}\Omg_{\rm h}^{2/3}  \nonumber\\
 \eql -\Lvev{\Omg_{\rm v}^{1/3}\Omg_{\rm h}^{2/3}}
 \brc{U_{3/2}^\mu\brkt{\Box_4+\cD_U}U_{3/2\mu}
 -\frac{2}{3}U_0^\mu\Box_4 U_{0\mu}} \nonumber\\
 &&-\Lvev{\Omg_{\rm v}^{-1/3}\Omg_{\rm h}^{4/3}}
 \brkt{\der_y\bar{U}^\mu\der_y\bar{U}_\mu
 +2\der_y\bar{U}^\mu V_{-\mu}
 +\frac{1}{2}V_-^\mu V_{-\mu}-\frac{1}{2}V_+^\mu V_{+\mu}} 
 \nonumber\\
 &&-\Lvev{\Omg_{\rm v}^{1/3}\Omg_{\rm h}^{2/3}}\left\{
 2iU^\mu\der_\mu\brkt{\cT+\tl{\Phi}_C-\bar{\cT}-\bar{\tl{\Phi}}_C}
 +\frac{3}{2}\brkt{V_+^\mu\der_\mu-V_-^\mu\Dlt_\mu}V_{\cT} \right.
 \nonumber\\
 &&\hspace{25mm}\left.
 -3\der_y U^\mu\Dlt_\mu V_{\cT}
 +U^\mu\Dlt_\mu\brkt{\tl{V}_{\rm v}+\tl{V}_{\rm h}}+\cdots\right\}, 
\eea
where we have performed the partial integrals, and 
$\cT$, $V_{\cT}$, $\tl{V}_{\rm v}$ and $\tl{V}_{\rm h}$ are defined 
in (\ref{def:cT}). 

Since $V_\pm^\mu$ do not have kinetic terms, 
they are integrated out as 
\bea
 V_+^\mu \eql \Lvev{\frac{3\Omg_{\rm v}^{2/3}}{2\Omg_{\rm h}^{2/3}}}
 \der^\mu V_{\cT}, \;\;\;\;\;
 V_-^\mu = -2\der_y\bar{U}^\mu
 +\Lvev{\frac{3\Omg_{\rm v}^{2/3}}{2\Omg_{\rm h}^{2/3}}}
 \Dlt^\mu V_{\cT}. 
\eea
After eliminating $V_\pm^\mu$, the 5D Lagrangian becomes
\bea
 \cL \eql \int\dr^4\tht\;\Lvev{\Omg_{\rm v}^{1/3}\Omg_{\rm h}^{2/3}}
 \left[-U_{3/2}^\mu\brkt{\Box_4+\cD_U}U_{3/2\mu}
 +\frac{2}{3}U_0^\mu\Box_4 U_{0\mu}
 +\Lvev{\frac{\Omg_{\rm h}^{4/3}}{\Omg_{\rm v}^{1/3}}}
 \der_y\bar{U}^\mu\der_y\bar{U}_\mu \right. \nonumber\\
 &&\hspace{20mm}\left.
 -2iU_0^\mu\der_\mu\brkt{\cT+\tl{\Phi}_C-\bar{\cT}-\bar{\tl{\Phi}}_C}
 -\bar{U}^\mu\Dlt_\mu\brkt{\tl{V}_{\rm v}+\tl{V}_{\rm h}}\right]
 \nonumber\\
 &&+\int\dr^4\tht\;\frac{9\Lvev{\Omg_{\rm v}}}{8}V_{\cT}
 \brkt{\Dlt^\mu\Dlt_\mu+\Box_4}V_{\cT}+\cdots. 
\eea
Adding the gauge-fixing term~(\ref{cL_gf^sc}), 
the cross terms between $U_\mu$ and the other superfields are canceled, 
and we obtain (\ref{quad:SUGRA}).

\subsection{Matter sector}
Since we have moved to the gauge where $\bdm{\Sgm}=0$ by the supergauge transformation 
for the background superfields, 
$W_{\rm v}$ in (\ref{def:Omgs}) is rewritten 
in terms of the gauge-transformed superfields as 
\be
 W_{\rm v} = \frac{c^3}{16g^3}\tr\brc{\tl{\Sgm}\tl{\cW}^2
 -\frac{1}{12}\bar{D}^2\brkt{V\der_yD^\alp\tl{V}
 -\der_y VD^\alp\tl{V}}\tl{\cW}_\alp}
 -\frac{c^3}{48g^3}\der_y\tr\brkt{\Lmd_\Sgm\tl{\cW}^2}, 
\ee
where 
\be
 \Lmd_\Sgm \equiv -\int_0^y\dr y'\;\bdm{\Sgm}(y').  \label{def:Lmd_Sgm2}
\ee
Note that $\bdm{V}^{\Io}$ and $\Lmd_\Sgm^{\Io}$ have nontrivial boundary conditions
at $y=L$ (see (\ref{limits:Lmd_Sgm})). 
\ignore{
\bea
 \lim_{y\to 0}\bdm{V}^{\Io}\hat{t}_{\Io} \eql 0, \;\;\;\;\;
 \lim_{y\to 0}\Lmd_\Sgm = 0, \nonumber\\
 \lim_{y\to L}\bdm{V}^{\Io}\hat{t}_{\Io} \eql -T-T^\dagger, \;\;\;\;\;
 \lim_{y\to L}\Lmd_\Sgm = -T.  \label{bdcd:V^Io}
\eea
}
The quadratic terms for $\tl{V}$ are read off as 
\bea
 &&\int\dr^2\tht\;W_{\rm v}+\hc \nonumber\\
 \eql -\int\dr^4\tht\;\frac{c^3}{16g^3}\tr\brc{
 \frac{1}{12}\brkt{\bdm{V}\der_yD^\alp\tl{V}
 -\der_y\bdm{V}D^\alp\tl{V}}\bar{D}^2D_\alp\tl{V}+\hc} \nonumber\\
 &&-\int\dr^4\tht\;\frac{c^3}{192g^3}
 \der_y\tr\brkt{\Lmd_\Sgm D^\alp\tl{V}\bar{D}^2D_\alp\tl{V}+\hc}+\cdots 
 \nonumber\\
 \eql -\int\dr^4\tht\;\frac{c^3}{16g^3}\tr\left[
 \frac{1}{4}\der_y\bdm{V}\tl{V}D^\alp\bar{D}^2D_\alp\tl{V}
 -\frac{1}{12}\der_y\brc{\brkt{\bdm{V}-\Lmd_\Sgm-\Lmd_\Sgm^\dagger}
 \tl{V}D^\alp\bar{D}^2D_\alp\tl{V}}\right] \nonumber\\
 &&+\cdots \nonumber\\ 
 \eql -\int\dr^4\tht\;\frac{c^3}{8g^3}\sbk{\tr\brkt{
 \der_y\bdm{V}\tl{V}\Box_4P_T\tl{V}}}+\cdots \nonumber\\
 \eql -\int\dr^4\tht\;\frac{\cN_{IJ}(\vev{\cV})}{2}\tl{V}^I\Box_4P_T\tl{V}^J
 +\cdots, \label{int:W_v}
\eea
where we have dropped total derivatives, and used (\ref{limits:Lmd_Sgm}). 
Combining this with the last term in (\ref{quad:SUGRA}), 
we find the kinetic terms for $\tl{V}$ as (\ref{cL_kin^vec}). 

Next we consider kinetic terms for the chiral superfields. 
We can expand $\Omg_{\rm v}^{1/3}\Omg_{\rm h}^{2/3}$ as 
\bea
 &&\Omg_{\rm v}^{1/3}\Omg_{\rm h}^{2/3} \nonumber\\
 \eql \Lvev{\Omg_{\rm v}^{1/3}\Omg_{\rm h}^{2/3}}
 \left[\frac{\cN_{IJ}}{6\cN}\brc{\der_y\tl{V}^I\der_y\tl{V}^J
 -2\brkt{\tl{\Sgm}+\bar{\tl{\Sgm}}}^I\der_y\tl{V}^J
 +\brkt{\tl{\Sgm}+\bar{\tl{\Sgm}}}^I
 \brkt{\tl{\Sgm}+\bar{\tl{\Sgm}}}^J} \right. \nonumber\\
 &&\hspace{20mm}
 -\brkt{\tl{V}_{\rm v}+\cT+\bar{\cT}}^2
 +\Lvev{\frac{\der_I\der_J\Omg_{\rm h}}{3\Omg_{\rm h}}}\tl{V}^I\tl{V}^J
 +\frac{2}{3}\tl{V}^I\brkt{\Ups_I^\dagger\tl{\Phi}_{\rm even}+\hc} \nonumber\\
 &&\hspace{20mm}
 +\frac{2}{3\Lvev{\Omg_{\rm h}}}\brkt{\tl{\Phi}_{\rm odd}\tl{d}
 \brkt{e^{\sbdm{V}}}^t\tl{\Phi}_{\rm odd}
 +\tl{\Phi}_{\rm even}\tl{d}e^{-\sbdm{V}}\tl{\Phi}_{\rm even}}
 -\frac{1}{4}\brkt{\tl{V}_{\rm h}+\tl{\Phi}_C+\bar{\tl{\Phi}}_C}^2
 \nonumber\\
 &&\hspace{20mm}\left. 
 +\brkt{\tl{V}_{\rm v}+\cT+\bar{\cT}}
 \brkt{\tl{V}_{\rm h}+\tl{\Phi}_C+\bar{\tl{\Phi}}_C}\right]+\cdots. 
\eea
Thus the cross terms between $\tl{V}$ and the chiral superfields are 
\be
 \cL_{\rm cross} = \int\dr^4\tht\;\tl{V}^I\brkt{\Xi_I+\bar{\Xi}_I}, 
 \label{L_cross}
\ee
where
\bea
  \Xi_I \defa \der_y\brc{2\Lvev{\Omg_{\rm v}^{1/3}\Omg_{\rm h}^{2/3}}
 (a\cdot\cP_V)_{IJ}\tl{\Sgm}^J}
 -\frac{2}{3}\Lvev{\frac{\Omg_{\rm h}}{\Omg_{\rm v}}}^{2/3}\Ups_I^\dagger
 \cN_J\tl{\Sgm}^J \nonumber\\
 &&-\der_y\brkt{\frac{2}{3}\Lvev{\frac{\Omg_{\rm h}}{\Omg_{\rm v}}}^{2/3}
 \cN_I\Ups^\dagger\tl{\Phi}_{\rm even}}
 -2\Lvev{\Omg_{\rm v}^{1/3}\Omg_{\rm h}^{2/3}}\Ups_I^\dagger
 \brkt{1-\frac{\bdm{\Phi}\Ups^\dagger}{3}}\tl{\Phi}_{\rm even}. 
 \label{def:Xi_I}
\eea
Adding the gauge-fixing term~(\ref{GF:gg}), these cross terms are canceled, 
and we obtain 
\bea
 &&\cL+\cL_{\rm gf}^{\rm sg} \nonumber\\
 \eql \int\dr^4\tht\;\tl{V}^I\left[\Lvev{\Omg_{\rm v}}a_{IJ}
 \Box_4\brkt{P_T+\frac{1}{\xi_{\rm sg}}P_C}\tl{V}^J
 -\der_y\brc{\Lvev{\Omg_{\rm v}^{1/3}\Omg_{\rm h}^{2/3}}
 (a\cdot\cP_V)_{IJ}\der_y\tl{V}^J} \right. \nonumber\\
 &&\hspace{20mm}
 -\der_y\brc{\Lvev{\Omg_{\rm v}^{1/3}\Omg_{\rm h}^{2/3}}
 \frac{\cN_I}{3\cN}\Ups_J^\dagger\bdm{\Phi}\tl{V}^J}
 +\Lvev{\Omg_{\rm v}^{1/3}\Omg_{\rm h}^{2/3}}
 \frac{\cN_J}{3\cN}\Ups_I^\dagger\bdm{\Phi}\der_y\tl{V}^J \nonumber\\
 &&\left.\hspace{20mm}
 -3\Lvev{\Omg_{\rm v}^{1/3}\Omg_{\rm h}^{2/3}}\brkt{
 \Lvev{\frac{\der_I\der_J\Omg_{\rm h}}{3\Omg_{\rm h}}}
 -\frac{\Ups_I^\dagger\bdm{\Phi}\bdm{\Phi}^\dagger\Ups_J}{9}}
 \tl{V}^J \right] \nonumber\\
 &&+\int\dr^4\tht\;\Lvev{\Omg_{\rm v}^{1/3}\Omg_{\rm h}^{2/3}}
 \left[2(a\cdot\cP_V)_{IJ}\tl{\Sgm}^I\bar{\tl{\Sgm}}^J
 -\frac{2}{\Lvev{\Omg_{\rm h}}}\tl{\Phi}_{\rm odd}^\dagger\tl{d}
 (e^{\sbdm{V}})^t\tl{\Phi}_{\rm odd} \right. \nonumber\\
 &&\left.\hspace{25mm}
 -\tl{\Phi}_{\rm even}^\dagger\brkt{
 \frac{2\tl{d}e^{-\sbdm{V}}}{\Lvev{\Omg_{\rm h}}}
 -\frac{2}{3}\Ups\Ups^\dagger}\tl{\Phi}_{\rm even}
 -\brkt{\frac{2\cN_I}{3\cN}\Ups^\dagger\bar{\tl{\Sgm}}^I\tl{\Phi}_{\rm even}
 +\hc}\right] 
 +\cO(\xi_{\rm sg}) \nonumber\\
 &&+\sbk{\int\dr^2\tht\;\brkt{
 \tl{\Phi}_{\rm odd}^t\tl{d}\der_y\tl{\Phi}_{\rm even}
 -\tl{\Phi}^t_{\rm even}\tl{d}\der_y\tl{\Phi}_{\rm odd}
 -2\tl{\Phi}_{\rm odd}^t\tl{d}\tl{\Sgm}\bdm{\Phi}}
 +\hc}+\cdots. \label{quad:chiral}
\eea
From (\ref{quad:SUGRA}), (\ref{int:W_v}) and (\ref{quad:chiral}), 
the quadratic terms for the fluctuation superfields 
in the bulk Lagrangian are summarized as (\ref{quad:cL_bulk}).

\section{Boundary conditions for bulk fluctuation modes} \label{BC:flct_md}
Here we derive the boundary conditions for the fluctuation modes of 
the bulk superfields, which are determined by the orbifold parities 
and the boundary actions. 

First, let us consider the boundary conditions of $U^\mu$ and $\tl{V}^I$. 
Since we have chosen the gauge~$\xi_{\rm sc}=\xi_{\rm sg}
=\zeta_{\rm sc}^{(y_*)}=\zeta_{\rm sg}^{(y_*)}=0$, 
only the transverse modes of $U^\mu$ and $\tl{V}^I$
(\ie, $U_{3/2}^\mu$ and $\tl{V}_T^I\equiv P_T\tl{V}^I$) can propagate.\footnote{
In fact, $U_{3/2}^\mu$ and $\tl{V}_T^I$ are gauge-invariant 
under $\dscp$ and $\dgt$, respectively. 
} 
From (\ref{quad:cL_bulk}) and (\ref{quad:cL_bd}), 
the equations of motion for them are 
\bea
 &&\brc{\Lvev{\Omg_{\rm v}^{1/3}\Omg_{\rm h}^{2/3}}\brkt{-p^2+\cD_U}
 -2\sum_{y_*=0,L}\dlt(y-y_*)\abs{\bdm{\phi}_C}^2h^{(y_*)}p^2}U^\mu_{3/2} = 0, 
 \nonumber\\
 &&\Lvev{\Omg_{\rm v}}a_{IK}\brc{-\dlt^K_{\;\;J}p^2+(\cD_V)^K_{\;\;J}}\tl{V}_T^J
 \nonumber\\
 &&-2\sum_{y_*=0,L}\dlt(y-y_*)\brkt{\Re f_{\Ie\Je}^{(y_*)}p^2
 +\frac{3}{2}\abs{\bdm{\phi}_C}^2h^{(y_*)}_{\Ie\Je}}\tl{V}_T^{\Je} = 0. 
 \label{EOM:UV}
\eea
By integrating these over infinitesimal intervals~$[y_*-\ep,y_*+\ep]$ ($y_*=0,L$), 
we obtain 
\bea
 &&\left.\Lvev{\Omg_{\rm v}^{-1/3}\Omg_{\rm h}^{4/3}}
 \der_y U_{3/2}^\mu\right|_{y=y_*+\eta_{y_*}\ep}
 +\left.\eta_{y_*}\abs{\bdm{\phi}_C}^2h^{(y_*)}p^2 U_{3/2}^\mu\right|_{y=y_*} = 0, 
 \nonumber\\
 &&\left.\Lvev{\Omg_{\rm v}^{1/3}\Omg_{\rm h}^{2/3}}
 a_{\Ie\Je}\der_y\tl{V}_T^{\Je}\right|_{y=y_*+\eta_{y_*}\ep}
 +\eta_{y_*}\left.\brkt{\Re f_{\Ie\Je}^{(y_*)}p^2
 +\frac{3}{2}\abs{\bdm{\phi}_C}^2h_{\Ie\Je}^{(y_*)}}\tl{V}_T^{\Je}
 \right|_{y=y_*} = 0,  \nonumber\\
 &&\left.\tl{V}_T^{\Io}\right|_{y=y_*+\eta_{y_*}\ep} = 0, 
 \label{bdcd:UV}
\eea
where $y_*=0,L$, $\eta_0=1$ and $\eta_L=-1$, and 
we have used that 
\be
 \cN_{\Ie} = 0, \;\;\;\;\;
 (a\cdot\cP_V)_{\Ie\Je} = a_{\Ie\Je}, \;\;\;\;\;
 (a\cdot\cP_V)_{\Ie\Jo} = 0, 
\ee
which follow from the fact that $\bdm{V}^{\Ie}$ are independent of $y$. 
(See Appendix~\ref{tree:action}.)

Next we derive the boundary conditions for the chiral superfields. 
Since $\tl{\phi}_C$ and some of $\tl{\chi}^a$ are expressed 
in terms of $\tl{\Phi}_{\rm even}$ as 
\be
 \begin{pmatrix} \tl{\phi}_C \\ \tl{\chi}^a \end{pmatrix} 
 = \begin{pmatrix} \frac{2}{3}\bdm{\phi}_C^{-1/2} & 0 \\
 -\bdm{\chi}^a/\bdm{\phi}_C^{3/2} & 1/\bdm{\phi}_C^{3/2} 
 \end{pmatrix}\tl{\Phi}_{\rm even}+\cO(\tl{\Phi}_{\rm even}^2), 
\ee
the boundary Lagrangians~(\ref{quad:cL_bd}) are rewritten as 
\bea
 \cL_{\rm bd}^{(y_*)} \eql 
 \int\dr^4\tht\;2\tl{\Phi}_{\rm even}^\dagger\cM_{\rm bd}^{(y_*)}\tl{\Phi}_{\rm even}
 +\sbk{\int\dr^2\tht\;\tl{\Phi}_{\rm even}^tW_{\rm bd}^{(y_*)}\tl{\Phi}_{\rm even}
 +\hc}+\cdots,  \label{cL_boundary}
\eea
where
\bea
 \cM_{\rm bd}^{(y_*)} \defa \frac{1}{2\abs{\bdm{\phi}_C}}\begin{pmatrix} 
 h_{c\bar{d}}^{(y_*)}\bdm{\chi}^c\bar{\bdm{\chi}}^d
 -\frac{2}{3}\brkt{h_c^{(y_*)}\bdm{\chi}^c
 +h_{\bar{c}}^{(y_*)}\bar{\bdm{\chi}}^c}+\frac{4}{9}h^{(y_*)} & 
 -h_{\bar{c}b}^{(y_*)}\bar{\bdm{\chi}}^c-\frac{2}{3}h_b^{(y_*)} \\
 -h^{(y_*)}_{\bar{a}c}\bdm{\chi}^c-\frac{2}{3}h_{\bar{a}}^{(y_*)} & 
 h_{a\bar{b}}^{(y_*)} \end{pmatrix},  \nonumber\\
 W_{\rm bd}^{(y_*)} \defa \frac{1}{2}\begin{pmatrix}
 P_{cd}^{(y_*)}\bdm{\chi}^c\bdm{\chi}^d-4P_c^{(y_*)}\bdm{\chi}^c
 +\frac{8}{3}P^{(y_*)} & -P_{cb}^{(y_*)}\bdm{\chi}^c+2P_b^{(y_*)} \\
 -P_{ac}^{(y_*)}\bdm{\chi}^c+2P_a^{(y_*)} & 
 P_{ab}^{(y_*)} \end{pmatrix}. 
\eea
Thus the equations of motion for the chiral superfields are read off 
from (\ref{quad:cL_bulk}) and (\ref{cL_boundary}) as 
\be
 -\frac{1}{4}\cM D^2\vph+\bar{W}\bar{\vph}
 +\sum_{y_*=0,L}\brc{-\frac{1}{4}\cM_{\rm bd}^{(y_*)}D^2\tl{\Phi}_{\rm even}
 +\bar{W}_{\rm bd}^{(y_*)}\bar{\tl{\Phi}}_{\rm even}}\cdot 2\dlt(y-y_*)
 = 0.  \label{EOM:chiral}
\ee
By integrating this over $[y_*-\ep,y_*+\ep]$, we obtain 
\be
 -\left.\eta_{y_*}\tl{d}\bar{\tl{\Phi}}_{\rm odd}\right|_{y=y_*+\eta_{y_*}\ep} 
 +\left.\brc{-\frac{1}{4}\cM^{(y_*)}_{\rm bd}D^2\tl{\Phi}_{\rm even}
 +\bar{W}_{\rm bd}^{(y_*)}\bar{\tl{\Phi}}_{\rm even}}\right|_{y=y_*} = 0. 
 \label{BC:chiral1}
\ee
Multiplying (\ref{EOM:chiral}) by $\cM^{-1}$ from the left and 
taking limits~$y\to y_*$ from the fundamental region~$0<y<L$, we also obtain
\bea
 \brc{-\frac{1}{4}D^2\tl{\Sgm}^I
 -\frac{\bdm{\Phi}^\dagger\tl{d}\brkt{a^{IJ}\hat{t}_J
 -\Lvev{\cV^I}\der_y}}
 {\Lvev{\Omg_{\rm v}^{1/3}\Omg_{\rm h}^{2/3}}}\bar{\tl{\Phi}}_{\rm odd}
 }_{y=y_*+\eta_{y_*}\ep} = 0,  \nonumber\\
 \brc{-\frac{1}{4}D^2\tl{\Phi}_{\rm odd}
 +\Lvev{\frac{\Omg_{\rm h}}{\Omg_{\rm v}}}^{1/3}(e^{-\sbdm{V}})^t
 \brkt{\hat{t}^t_I\bar{\bdm{\Phi}}\bar{\tl{\Sgm}}^I
 -\der_y\bar{\tl{\Phi}}_{\rm even}}}_{y=y_*+\eta_{y_*}\ep} = 0. 
 \label{BC:chiral2}
\eea
If we denote an eigenvalue of the differential operator~$\cM^{-1}\bar{W}$ 
as $\mu_{\rm ch}$, the equation of motion in the bulk can be expressed as 
$\frac{1}{4}D^2\vph=\cM^{-1}\bar{W}\bar{\vph}=\mu_{\rm ch}\bar{\vph}$. 
Hence the boundary conditions~(\ref{BC:chiral1}) and (\ref{BC:chiral2}) 
are rewritten as 
\be
 \brc{\cA_{\rm ch}^{(y_*)}\der_y\vph
 +\cB_{\rm ch}^{(y_*)}\vph}_{y=y_*+\eta_{y_*}\ep} = 0, 
 \label{bdcd:chiral2}
\ee
where 
\bea
 \cA_{\rm ch}^{(y_*)} \defa \begin{pmatrix} 
 0 & 0 & -\frac{\vev{\cV^I}\bdm{\Phi}^t\tl{d}}
 {\Lvev{\Omg_{\rm v}^{1/3}\Omg_{\rm h}^{2/3}}} \\
 0 & 0 & 0 \\ 
 0 & \Lvev{\frac{\Omg_{\rm h}}{\Omg_{\rm v}}}^{1/3}e^{-\sbdm{V}} & 0 
 \end{pmatrix}_{y=y_*+\eta_{y_*}\ep},  \nonumber\\
 \cB_{\rm ch}^{(y_*)} \defa \begin{pmatrix} 
 \mu_{\rm ch}\dlt^I_{\;\;J} & 0 & 
 \frac{a^{IJ}\bdm{\Phi}^t\tl{d}\hat{t}^t_J}{\Lvev{\Omg_{\rm v}^{1/3}
 \Omg_{\rm h}^{2/3}}} \\ 
 0 & \brkt{\bar{\cM}_{\rm bd}^{(y_*)}\mu_{\rm ch}
 -W_{\rm bd}^{(y_*)}}\id_{n_C+n_H} 
 & \eta_{y_*}\tl{d} \\
 -\Lvev{\frac{\Omg_{\rm h}}{\Omg_{\rm v}}}^{1/3}e^{-\sbdm{V}}\hat{t}_J
 \bdm{\Phi} & 0 & \mu_{\rm ch}\id_{n_C+n_H} 
 \end{pmatrix}_{y=y_*+\eta_{y_*}\ep}. 
 \label{def:AB}
\eea

\section{Bosonic component expression of one-loop action} \label{BS_component}
Here we provide an explicit expression of the one-loop Lagrangian 
in terms of the bosonic components in a simple case 
where 5D spacetime is flat and the boundary terms are absent. 
In this case, the one-loop K\"ahler potential~(\ref{FinalExpr}) is reduced to 
\be
 \Omglp = \frac{\abs{\phi_C}^2}{\cN^{2/3}}
 \brc{\tl{Q}_1-Q_2\sum_a\brkt{c_a\cdot\Re T}^3}+\cO(\hat{\chi}^2), 
\ee
where $\tl{Q}_1\equiv (n_V-n_H+1)Q_1/(8\pi^2)$, 
and $c_a\cdot\Re T\equiv c_{a\Io}\Re T^{\Io}$. 
Thus the one-loop Lagrangian is written as 
\bea
 \Dlt^{\rm 1loop}\cL \eql \int\dr^4\tht\;\Omglp+\cdots \nonumber\\
 \eql \frac{1}{\cN^{2/3}}\brc{\abs{F_{\phi_C}}^2L_{\bar{\phi}\phi}
 +\brkt{\bar{F}_{\phi_C}F_{T^{\Io}}\phi_C L_{\bar{\phi}T^{\Io}}+\hc}
 +\bar{F}_{T^{\Io}}F_{T^{\Jo}}\abs{\phi_C}^2L_{\bar{T}^{\Io}T^{\Jo}}} \nonumber\\
 &&+\cO(F_{\chi},D_{V^{\Ie}})+\cdots, 
\eea
where $F_{\vph}$ ($\vph=\phi_C,\chi,T^{\Io}$) and $D_{V^{\Ie}}$ denote 
the $F$-component of a superfield~$\vph$ and the $D$-component of $V^{\Ie}$ 
respectively, and 
\bea
 L_{\bar{\phi}\phi} \defa \tl{Q}_1-Q_2\sum_a\brkt{c_a\cdot\Re T}^3, \nonumber\\
 L_{\bar{\phi}T^{\Io}} \defa 
 -\frac{2\cN_{\Io}}{3\cN}\brc{\tl{Q}_1-Q_2\sum_a\brkt{c_a\cdot\Re T}^3}
 -\frac{3Q_2}{2}\sum_a\brkt{c_a\cdot\Re T}^2c_{a\Io}, \nonumber\\
 L_{\bar{T}^{\Io}T^{\Jo}} \defa 
 -\frac{3\cN\cN_{\Io\Jo}-5\cN_{\Io}\cN_{\Jo}}{3\cN^2}
 \brc{\tl{Q}_1-Q_2\sum_a\brkt{c_a\cdot\Re T}^3} \nonumber\\
 &&+\frac{Q_2}{\cN}\sum_a\brkt{c_a\cdot\Re T}^2
 \brkt{\cN_{\Io}c_{a\Jo}+\cN_{\Jo}c_{a\Io}} 
 -Q_2\sum_a\brkt{c_a\cdot\Re T}c_{a\Io}c_{a\Jo}. 
\eea
Here the arguments of $\cN$ and its derivatives are 
$(\bdm{0}_{n_{V_{\rm e}}},2\Re T^{\Io})$, and $\phi_C$ and $T^{\Io}$ denote 
the lowest components of the corresponding superfields.


\end{document}